\documentclass[pdflatex,sn-basic]{sn-jnl}
\usepackage{aas_macros}
\usepackage[normalem]{ulem}
\usepackage{titlesec}
\usepackage{array}
\usepackage{booktabs}
\usepackage{adjustbox}
\usepackage{amssymb}

\def\Rs{${\rm{R_S}}$}  
\def\Rss{${\rm{R_S}}\,$} 

\jyear{2021}%

\theoremstyle{thmstyleone}%

\theoremstyle{thmstyletwo}%

\theoremstyle{thmstylethree}%

\begin{document}

\title[Saturn Rings Composition]{The Composition of Saturn's Rings}

\author[1]{\fnm{Kelly E.} \sur{Miller}}\email{kmiller@swri.edu}

\author*[2]{\fnm{Gianrico} \sur{Filacchione}}\email{gianrico.filacchione@inaf.it}

\author[3]{\fnm{Jeffrey} \sur{Cuzzi}}\email{jeffrey.cuzzi@nasa.gov}

\author[4]{\fnm{Philip D.} \sur{Nicholson}}\email{pdn2@cornell.edu}

\author[5]{\fnm{Matthew M.} \sur{Hedman}}\email{mhedman@uidaho.edu}

\author[6]{\fnm{Kevin} \sur{Bailli\'e}}\email{kevin.baillie@obspm.fr}

\author[7]{\fnm{Robert E.} \sur{Johnson}}\email{rej@virginia.edu}

\author[8]{\fnm{Wei-Ling} \sur{Tseng}}\email{wltseng@ntnu.edu.tw}

\author[3]{\fnm{Paul R.} \sur{Estrada}}\email{paul.r.estrada@nasa.gov}

\author[9]{\fnm{Jack Hunter} \sur{Waite}}\email{hunterwaite@gmail.com}

\author[2]{\fnm{Mauro} \sur{Ciarniello}}\email{mauro.ciarniello@inaf.it}

\author[10]{\fnm{Cécile} \sur{Ferrari}}\email{ferrari@ipgp.fr}

\author[11]{\fnm{Zhimeng} \sur{Zhang$^{11}$}}\email{zzm19881204@gmail.com}

\author[12]{\fnm{Amanda} \sur{Hendrix}}\email{ahendrix@psi.edu}

\author[13]{\fnm{Julianne I.} \sur{Moses}}\email{jmoses@spacescience.org}

\affil[1]{\orgdiv{Space Science}, \orgname{Southwest Research Institute}, \orgaddress{\street{9503 W. Commerce}, \city{San Antonio}, \state{TX}, \postcode{78227}, \country{USA}}}

\affil*[2]{\orgdiv{INAF}, \orgname{IAPS Institute for Space Astrophysics and Planetology}, \orgaddress{\street{via del Fosso del Cavaliere, 100}, \city{Rome}, \postcode{00133}, \country{Italy}}}

\affil[3]{\orgdiv{Space Sciences Div.}, \orgname{NASA Ames Research Center}, \orgaddress{\street{MS 245-3}, \city{Moffett Field}, \state{CA}, \postcode{94035}, \country{USA}}}

\affil[4]{\orgdiv{Department of Astronomy}, \orgname{Cornell University}, 
\city{Ithaca}, \state{NY}, \postcode{14853}, \country{USA}}

\affil[5]{\orgdiv{Department of Physics}, \orgname{University of Idaho}, 
\city{Moscow}, \state{ID}, \postcode{83844}, \country{USA}}

\affil[6]{\orgdiv{IMCCE}, \orgname{Observatoire de Paris, PSL Research University, CNRS, Sorbonne Universités, UPMC Univ Paris 06, Univ. Lille}, \orgaddress{\street{77 Av. Denfert-Rochereau}, \city{Paris}, \postcode{75014}, \state{}, \country{France}}}

\affil[7]{\orgdiv{Engineering Physics}, \orgname{University of Virgina}, \orgaddress{ \city{Charlottesville},\state{VA}, \postcode{22902}, \country{USA}} }

\affil[8]{\orgdiv{Department of Earth Sciences}, \orgname{National Taiwan Normal University}, \orgaddress{ \city{Taipei}, \postcode{116}, \country{Taiwan}}}

\affil[9]{ \orgname{Waite Science, LLC}, \orgaddress{\street{16284 Narwhal Drive}, \city{Pensacola}, \state{FL}, \postcode{32507}, \country{USA}}}

\affil[10]{ \orgname{Université de Paris Cité, Institut de physique du globe de Paris}, \orgaddress{\street{CNRS}, \city{Paris}, \postcode{F-75005}, \country{France}}}

\affil[11]{ \orgdiv{Division of Geological and Planetary Sciences}, \orgname{California Institute of Technology}, \orgaddress{ \city{Pasadena},\state{CA}, \postcode{91125}, \country{USA}} }

\affil[12]{ \orgname{Planetary Science Institute}, \orgaddress{ \city{Tucson},\state{AZ}, \postcode{85719}, \country{USA}} }

\affil[13]{ \orgname{Space Science Institute}, \orgaddress{\city{Boulder}, \state{CO},\postcode{80301}, \country{USA}}}

\abstract{The origin and evolution of Saturn's rings is critical to understanding the Saturnian system as a whole. Here, we discuss the physical and chemical composition of the rings, as a foundation for evolutionary models described in subsequent chapters. We review the physical characteristics of the main rings, and summarize current constraints on their chemical composition. Radial trends are observed in temperature and to a limited extent in particle size distribution, with the C ring exhibiting higher temperatures and a larger population of small particles. The C ring also shows evidence for the greatest abundance of silicate material, perhaps indicative of formation from a rocky body. The C ring and Cassini Division have lower optical depths than the A and B rings, which contributes to the higher abundance of the exogenous neutral absorber in these regions. Overall, the main ring composition is strongly dominated by water ice, with minor silicate, UV absorber, and neutral absorber components. Sampling of the innermost D ring during Cassini's Grand Finale provides a new set of in situ constraints on the ring composition, and we explore ongoing work to understand the linkages between the main rings and the D ring. The D ring material is organic- and silicate-rich and water-poor relative to the main rings, with a large population of small grains. This composition may be explained in part by volatile losses in the D ring, and current constraints suggest some degree of fractionation rather than sampling of the bulk D ring material.}


\maketitle

\setcounter{secnumdepth}{4}

\section{Introduction}\label{Intro}
The rings are a prominent and unique feature in the Saturnian system, and they serve both as a witness plate and a  driver for the evolution of the system via interactions with both the moons and Saturn itself. 
In this chapter we review current knowledge of the physical and chemical composition of Saturn's rings, with a focus on the main rings and the innermost D ring. 
We begin in Section \ref{Intro} with an overview of the ring structure and main evolutionary processes affecting the rings. 
In Section \ref{mainRingPhys}, we describe the physical properties of the main rings, including their brightness, particle size distribution and temperature. 
Remote sensing constraints on the main rings' chemical composition are then reviewed in Sections \ref{mainRingRegolith}, \ref{RRTModels} and \ref{bulk}.
Section \ref{dring} reviews our current knowledge of the D ring, which is thought to be the immediate source for the inflowing ring material that was measured in situ by Cassini and is described in Section \ref{inflowIntro}. 
Current constraints suggest this material includes a gas-phase component, and we review our knowledge of the ring atmosphere and possible connections to the inflowing material in Section \ref{ringAtmosphere}. 
Finally, we briefly explore the relationship between the main ring composition and the ring moons in Section \ref{ringMoon}, before summarizing our conclusions in Section \ref{conclusions}.
\subsection{Background on ring structure}\label{Strucure}

\subsubsection{Terminology and dimensions}\label{Terminology} 
\par
\begin{figure}
    \centering
    \resizebox{4.5in}{!}{\includegraphics[angle=0]{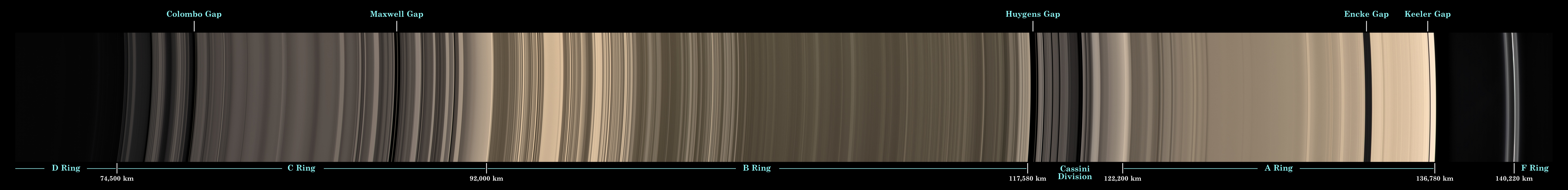}}
    \caption{An optical mosaic of the dark side of Saturn's rings, obtained by the narrow-angle camera of the Cassini ISS instrument. 
    From left to right, we see the broad C, B and A rings at increasing radii, flanked by the narrow F ring at the extreme right. Note the numerous narrow gaps, appearing here as black bands in the C ring, the Cassini Division and in the outer part of the A ring. Courtesy: NASA/JPL-Caltech/Space Science Institute}
    \label{fig:ring_mosaic}
\end{figure}

\par
Saturn's dense rings (or ``main rings") are traditionally divided into three major components known as the A, B and C rings (see Fig.~\ref{fig:ring_mosaic}). The outermost or A ring extends from a radius of 136,770~km relative to Saturn's center in to $\sim122,000$~km; the outer edge at $\sim2.25$ Saturn radii (\Rs, where Saturn’s equatorial radius is 60,268 km) 
essentially marks the Roche Limit for self-gravitating icy particles. 
The middle or B ring extends from 117,570~km in to $\sim92,000$~km, and is both the brightest and most opaque of the main rings, containing approximately $2/3$ of the system's total mass.
The inner or C ring extends from the B ring in to $74,490$~km, or $\sim1.25$ \Rs. 
Located between the A and B rings is the 4500~km wide Cassini Division ("CD"), which Voyager data showed to be comparable in brightness and optical depth to the C ring and which is now considered to be a distinct fourth component of the main rings.
\par
In comparison to their radial width of $\sim62,000$~km, or almost 10 Earth radii, Saturn's rings are extremely thin. 
Theoretical estimates based on collisional equilibrium suggest that their vertical scale height $H$ is likely to be of order 10~m, a value that is supported by a limited number of experimental estimates ranging from $<30$~m in the C ring and Cassini Division to $\sim6$~m in the A ring \citep{Jerouseketal2011, Hedmanetal2007a}. 

\par
Although it is less opaque and almost certainly contains much less material than the main rings, we include in this chapter the nearby D ring, primarily because of the {\it in situ} sampling by Cassini on its final orbits that observed material hypothesized to originate from this region. 
The D ring extends inwards from the C ring almost to the outer reaches of the planet's atmosphere at a radius of $\lt 68,000$~km. 
Invisible in Earth-based images and in most optical depth profiles, the D ring is dominated by a series of narrow, dusty ringlets and is most readily seen in images taken in forward-scattering geometries (see Section \ref{dring}).
\par
Limited {\it in situ} sampling near the F ring region outside of the A ring was also performed by Cassini. This ring, which is shepherded by the ring moons Pandora and Prometheus, is discussed further in Chapter 11 along with the dusty G ring that is associated with the satellite Aegaeon.
\par
Not included in this volume is the very tenuous and isolated  E ring, which is associated with the satellite Enceladus, or the even more distant `Phoebe ring' discovered by the Spitzer space telescope \citep{Verbisceretal2009}. 
For information on these rings the interested reader is directed to Chapters 12 and 13 in the book `Planetary Ring Systems' \citep{TiscarenoMurray2018}.

\subsubsection{Optical depths and ring structure}\label{opticalDepth} 

The density of a planetary ring is usually expressed in terms of its normal optical depth $\tau$, which in general is a function of the wavelength of observation $\lambda$ and the particle size distribution $n(a)$:
\begin{equation}
\tau(\lambda) = \pi\int_{a_{\rm min}}^{a_{\rm max}} Q_{\rm ext}(a,\lambda)\,a^2\,n(a)\,da
    \label{eq:optical_depth}
\end{equation}   
\noindent where $n(a)da$ is the number of ring particles per unit ring area with particle radii between $a$ and $a+da$ and $Q_{\rm ext}(a,\lambda)$ is their extinction efficiency at wavelength $\lambda$. 
For quasi-spherical particles, $Q_{\rm ext}$ approaches $0$ when the size parameter $x = 2\pi a/\lambda$ is much les than $1$, but approaches an asymptotic value between 1 and 2 for $x \gg 1$ \citep{Cuzzi1985, FrenchNicholson2000}.

Detailed radial profiles of optical depth for Saturn's rings are available from stellar occultations observed by the Ultraviolet Imaging Spectrometer \citep[UVIS;][]{Espositoetal2004} at $\lambda \simeq 0.15~\mu$m and by the Visual and Infrared Mapping Spectrometer \citep[VIMS;][]{Brownetal2004} at $\lambda = 2.9~\mu$m, for both of which $Q_{\rm ext}\simeq1$,
as well as radio occultations observed by the Radio Science Subsystem \citep[RSS;][]{Klioreetal2004} at Ka-band ($\lambda = 0.94$~cm), X-band (3.4~cm) and S-band (13~cm) \citep{Cuzzietal2009}.
\footnote{For most of the regions in the main rings, $a\gg\lambda$ for the UVIS and VIMS occultation data and $Q_{\rm ext}\simeq1$, but for the RSS occultation data, $a\simeq\lambda$ and $Q_{\rm ext}\simeq2$ due to the particles' broad diffraction lobes at cm-wavelengths, as well as the coherent nature of the spacecraft's radio signal.}
\par
A second useful quantity is the surface mass density of the ring for a given distance ({\it i.e.,} the mass per unit ring area), expressed as:
\begin{equation}
\Sigma = \frac{4\pi}{3}\int_{a_{\rm min}}^{a_{\rm max}} \rho\,a^3\,n(a)\,da
    \label{eq:surface_density}
\end{equation}  
\noindent where $\rho$ is the average internal density of the ring particles.
For a ring composed of spherical particles with a narrow range of sizes and $Q_{\rm ext}\simeq1$, the ratio $\Sigma/\tau$ is approximately equal to $\frac{4}{3}\rho\bar{a}$, where $\bar{a}$ is the mean particle radius. The inverse quantity $\tau/\Sigma$ is known as the opacity of the ring, and is a measure of its optical cross section per unit mass.

\subsection{Relationship of Composition to Ring Age} \label{ringAge} 
\par
 Evolutionary processes alter the physical and chemical composition of the rings, and an understanding of the types and timescales of these forms of alteration is essential for a determination of the ring age. 
 Here, we summarize the primary processes known to drive evolution of the main rings. 
 For an in-depth discussion of the ring age, see Chapter 12 of this volume (Crida et al.).
 
\subsubsection{Physical evolution}\label{dynamics}
The physical evolution of the rings is controlled by interparticle collisions, which determine the rings' viscosity and thickness as described below, as well as the rings' self-gravity and gravitational interactions with distant and embedded satellites. 
In order to estimate the dynamical age of the rings, the two critical parameters are the surface mass density $\Sigma$ and the effective kinematic viscosity of the rings $\nu$, which control the rate at which keplerian shear within the rings leads to their radial spreading over time. 
The latter rate is determined by the outward viscous angular momentum flux across the rings
\begin{equation}
F_v(r) = \frac{3}{2}\,\nu\Sigma\Omega r,
    \label{eq:ang_mom_flux}
\end{equation}   
\noindent where $\Omega$ is the local keplerian angular velocity at a radius $r$ from Saturn in the rings \citep{LyndenBellPringle1974, Longaretti2018}.
\par
Theoretical models suggest that there are at least three important sources of viscosity: the normal `local viscosity' due to random velocities and the resulting collisions between individual ring particles,\footnote{Akin to molecular viscosity in a fluid medium.}
a `nonlocal viscosity' due to transmission of momentum across large particles in densely-packed rings, and the transfer of momentum by gravitational interactions associated with self-gravity wakes \citep{Daisakaetal2001, Longaretti2018}. 
These self-gravity wakes are 100~m to km scale temporary aggregations of particles that form within the Roche limit but are subsequently sheared out and disrupted by tidal forces and the keplerian angular velocity gradient within the rings.
The relative importance of these processes in different parts of the rings is not well-known, but gravitational interactions are expected to be dominant in the A ring, where self-gravity wakes are ubiquitous \citep{Colwelletal2006, Hedmanetal2007b}.
\par
Empirical estimates of $\nu$ come from two sources: the damping of density and bending waves, and the balancing of the viscous torque $2\pi rF_v$ with gravitational torques exerted at satellite resonances. 
For the A ring, estimates of the former lead to $\nu = 40$ to 200~cm$^2$~s$^{-1}$ \citep{Tiscarenoetal2007}, comparable to the self-gravity-wake viscosity of $\sim100$~cm$^2$~s$^{-1}$.
Balancing the viscous and satellite torques, on the other hand, \cite{Tajeddineetal2017} find somewhat lower values for the A ring of $\nu = 10-50$~cm$^2$~s$^{-1}$ and a similar value for the B ring of $\sim 30$~cm$^2$~s$^{-1}$. 
There are few reliable estimates in the C ring or Cassini Division.
\par
Under the (probably erroneous) assumption that the dominant contribution to viscosity is local, the measured value of $\nu$ can be used to estimate an upper limit on the root mean square (rms) random velocity of particles in the ring plane $c$ via the expression \citep{GoldreichTremaine1978a}:
\begin{equation}
c^2 = 2\nu\Omega\,\frac{1+\tau^2}{\tau}.
    \label{eq:vel_disp}
\end{equation}   
\noindent If the velocity ellipsoid is isotropic, then the ring's vertical scale height is simply $H \simeq c/\Omega$.  
The lower values quoted above for $\nu$ imply that $c\leq 0.05-0.25$~cm~s$^{-1}$ and $H\leq 4-20$~m in the A ring and that $c\leq0.2$~cm~s$^{-1}$ and $H\leq10$~m in the B ring, consistent with the more direct estimates given above. 
In the Cassini Division (and perhaps also the C ring), where self-gravity wakes are absent, the same argument leads to $H\simeq 3-5$~m \citep{Colwelletal2009a}. In all cases, the ring thickness appears to be comparable to the size of the largest ring particles (see Section \ref{particleSize} below).
\par
A lower limit on the velocity dispersion comes from the requirement that the rings be stable against axisymmetric instabilities due to their self-gravity. Known as the Toomre criterion, this specifies a minimum value for the dimensionless ratio
\begin{equation}
Q = \frac{c\Omega}{\pi G\Sigma}
    \label{eq:ToomreQ}
\end{equation}
\noindent of about 1.
Numerical simulations show that stable rings with self-gravity wakes typically have $Q\simeq 2$ \citep{Schmidtetal2009}, leading to estimates of $c\simeq 0.10$ and 0.15~cm~s$^{-1}$ in the A and B rings, respectively.\footnote{Note that the keplerian angular velocity $\Omega \simeq 1.29\times10^{-4}$~s$^{-1}$ in the outer A ring and $\sim1.66\times10^{-4}$~s$^{-1}$ in the middle B ring.} 
For further discussion of the dynamical and structural evolution timescales of the rings, see Chapter 12 of this volume by Crida et al..

\subsubsection{Compositional evolution}\label{compositionalEvolution} 

Apart from their dynamical evolution, ring particles also undergo compositional evolution because they are continuously exposed to exogenous processes such as photolysis and meteoritic particle bombardment, which are capable of altering their original state and chemical makeup. The composition of the rings is one of the key measurements that is useful in constraining their age. This topic is discussed in detail in Chapter 12 of this volume, and is briefly discussed here as context for the ring composition as a whole.

\par
While the majority of compositional data for the rings comes from remote sensing at UV to infrared wavelengths, and therefore represents the ring particles' surface materials, mixing effectively homogenizes the rings such that these data are still useful constraints for the bulk composition of the rings. The continuous low-velocity collisions among nearby ring particles, and mixing of their subsurfaces due to extrinsic bombardment by meteoroids (known as ``impact gardening") is commonly thought to result in a very efficient mixing of the exogenous and original ring material over timescales of 1~Myr or so, even at depths comparable to the size of the largest ring particles \citep[][]{ElliottEsposito2011}. Similarly, the ring layer itself is well-mixed, as ring particles move vertically on their inclined orbits, colliding and bouncing around, fragmenting and re-forming, spending time at the faces as well as in the midplane and getting equally exposed on average \citep{Morishimaetal2010}. Comparison of results from visible and near-IR studies with those obtained from microwave observations that sample the bulk of the ring material \citep[see also Section \ref{bulk}] {Zhangetal2017AB,Zhangetal2017C, Zhangetal2019} generally supports this mixing framework. Interestingly though, microwave data identify a unique part of the mid-C Ring where there is evidence favoring particles having icy mantles and rocky cores. 

Owing to their huge surface-area-to-mass ratio (see Table \ref{tbl:ring_mass}), the rings are highly susceptible to the effects of micrometeoroid bombardment and the resulting ballistic transport of their impact ejecta. In particular, ring composition (which is predominantly icy, $\gtrsim 95\%$ by mass) is sensitive to the continuous infall of non-icy meteoroidal material that may accumulate over time. The CDA team has recently published a final analysis of an extensive set of in-situ observations collected over the more than a dozen years of the Cassini tour that provides the total mass flux of extrinsic, non-icy material impinging on the rings \citep[and chapter 12 by Crida et al]{Kempfetal2023}. The main conclusions from this study indicate that the magnitude of the measured flux entering the Saturn system, \(6.9 \times 10^{-17}\) kg m\(^{-2}\) s\(^{-1}\) to \(2.7 \times 10^{-16}\) kg m\(^{-2}\) s\(^{-1}\), is roughly a factor of two smaller than previously derived estimates.
Moreover, the dynamical origin of these extrinsic micrometeoroids is not from the Oort Cloud as previously assumed, but predominantly from the Edgeworth-Kuiper Belt \citep{Kempfetal2023}. The latter dynamical population is characterized by much lower entry speeds upon crossing Saturn's Hill sphere, and as a result is more strongly affected by gravitational focusing by the planet. As a result, the rate at which the rings are polluted is also faster. This polluting material is linked to the neutral absorbers, discussed further in Section \ref{neutralAbsorber}.
  
\section{Physical Properties of the Main Rings}\label{mainRingPhys}

The physical properties of the rings include characteristics such as their brightness and optical depth profiles, particle size distribution, surface mass density, temperature and thermal behavior.
\par
\begin{figure}
    \centering
    \resizebox{4.5in}{!}{\includegraphics[angle=90]{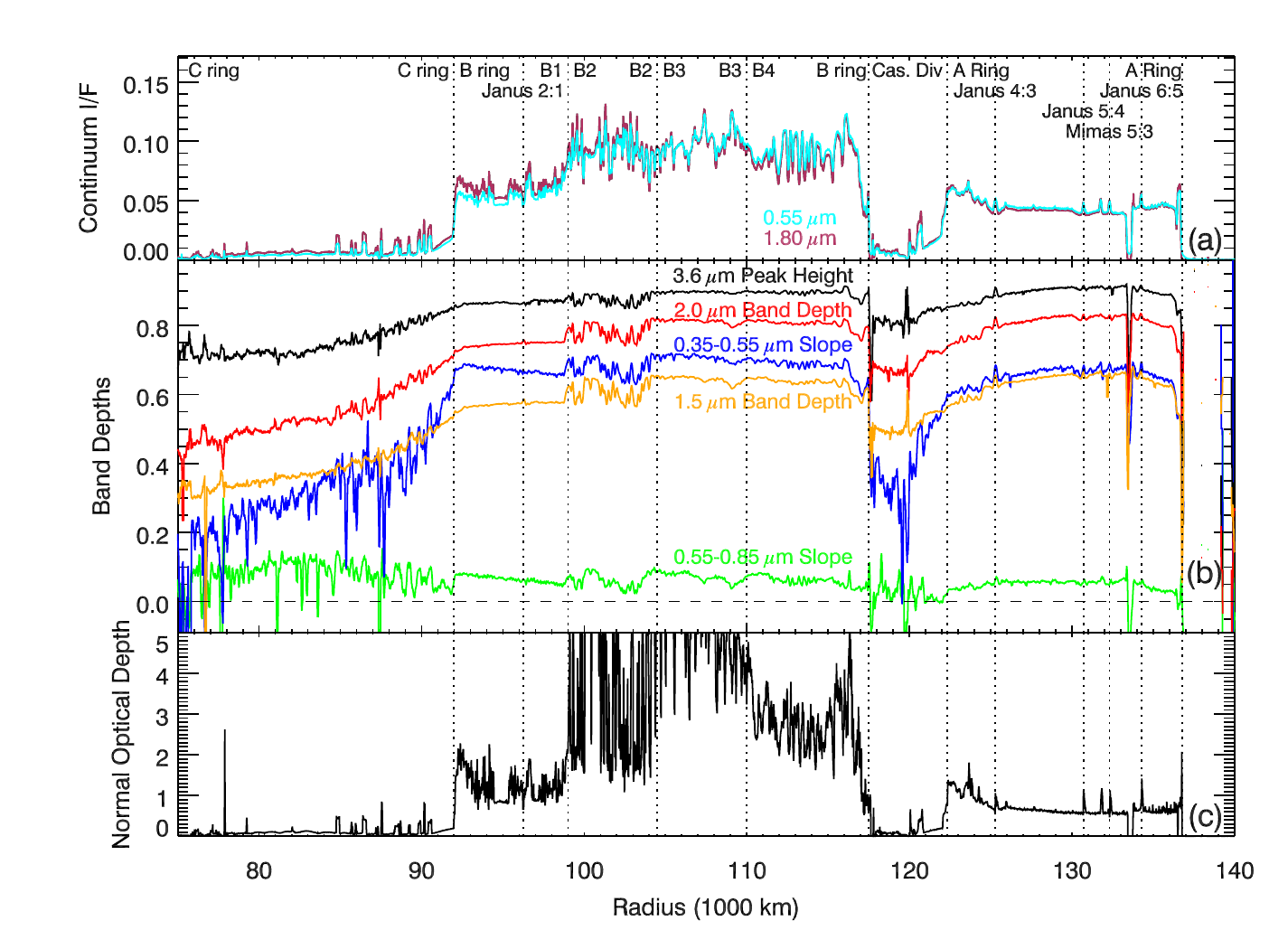}}
    \caption{Radial profiles of the main rings of Saturn derived from Cassini-VIMS observations. 
    The upper panel shows the reflectivity of the rings at two continuum wavelengths of 0.55 and 1.8~$\mu$m, with the locations of the major ring boundaries and the strongest density waves identified by vertical dotted lines. 
    The middle panel shows the fractional depths of the water ice bands at 1.55 and 2.0~$\mu$m, as well as the peak height at 3.6~$\mu$m and the blue and red slopes, rebinned to a uniform sampling resolution of 20~km and smoothed to 100 km.  
    The lower panel shows an optical depth profile of the rings obtained from an occultation of the star $\gamma$~Crucis, binned to 10~km resolution. \citep[From][]{Tiscarenoetal2019}.}
    \label{fig:ring_profiles}
\end{figure}
\subsection{Brightness and optical depth profiles} \label{brightness}
Figure~\ref{fig:ring_profiles} compares two radial brightness profiles of the main rings in reflected light (I/F) as measured in the visible and near-infared (top), as well as a profile of optical depth (bottom) measured by the VIMS instrument on the Cassini spacecraft. The spectral profiles in the middle panel of Figure \ref{fig:ring_profiles} are discussed further in Section \ref{mainRingRegolith} below.
In the A ring, both the optical depth $\tau$ and I/F profiles are relatively flat, with an average $\tau \simeq 0.5$ in both the UV \citep{Colwelletal2010} and near-IR \citep{Hedmanetal2013}. 
There are two relatively narrow gaps in its outer part, known as the Encke and Keeler gaps.
\par
The B ring is somewhat brighter than the A ring and much more structurally complex, with $\tau_{\rm UV}$ and $\tau_{\rm IR}$ ranging from a minimum of $\sim0.9$ in the inner parts (known as the B1 region; see labels at top of Figure \ref{fig:ring_profiles}) to $\gt5$ in parts of the central (or B3) region. 
The microwave optical depths exceed the radio science detection threshold in the densest parts of the B Ring, partly because $Q_{\rm ext}\simeq2$.
\par
The C ring, on the other hand, is relatively transparent. It is dominated by an undulating profile with an average $\tau \simeq 0.05 - 0.10$, punctuated by a series of narrow, more opaque features known as ``plateaux" where $\tau\simeq0.5$, as well as several narrow gaps, some of which contain optically-thick ringlets. 
\par
Located between the A and B rings, the Cassini Division shows many similarities to the C ring, with an average $\tau \simeq 0.10-0.20$, comprised of eight narrow gaps and three isolated ringlets \citep{Nicholsonetal2018}.
\par
The sharp outer edges of the A and B rings are controlled by strong satellite resonances \citep{Nicholsonetal2018}, but their more gradual inner edges take the form of `ramps', where $\tau$ decreases more-or-less linearly towards the much less opaque Cassini Division and C ring.
These inner edges, and the inner edge of the D Ring, appear to be unconfined. 
In fact, many physical parameters change gradually across these ramps, which are believed to be generated by the process of ballistic transport \citep{Durisenetal1989, Durisenetal1996, Estradaetal2015}. They are thus important sites for assessing the degree of long-term evolution of the rings (see Chapter 12 in this volume, by Crida, et al.)

\subsection{Surface mass density}\label{surfaceDensity} 
\par
\par
\begin{table}[htbp]
    \centering
   
    \caption{Radii, surface areas, surface mass densities, and estimated masses for the principal Saturnian rings.}
    \label{tbl:ring_mass}
    \begin{adjustbox}{width=\textwidth}
    \begin{tabular}{lrrccccl}
    \toprule
    Ring & $R_{\text{in}}$ & $R_{\text{out}}$ & Ring surface & $\Sigma$ & Mass & Mass & Ref. \\ region & (km) & (km) & area (km$^2$) & (kg m$^{-2}$) & (kg) & (M$_{\text{Mimas}}$) \\[.2em]
    \midrule
    A ring & 122,340 & 136,770 & $1.18\times 10^{10}$ & 150-400 & $3.7\times 10^{18}$ & 0.10 & a \\
    Cass. Div. & 117,930 & 122,340 & $3.3 \times 10^9$ & 15--190 & $1.7\times 10^{17}$ & 0.004 & b \\
    B ring & 92,000 & 117,570 & $1.68\times 10^{10}$ & 600 & $1.01\times 10^{19}$ & 0.27 & c \\ 
    C ring & 74,490 & 92,000 & $9.2\times 10^9$ & 10--60 & $3.4\times 10^{17}$ & 0.009 & d \\ [0.4em]
    Total & 74,490 & 136,770 & $4.11\times 10^{10}$ &  & $1.43\times 10^{19}$ & 0.38 \\[0.2em]
    \bottomrule
    \end{tabular}
    \end{adjustbox}
    \begin{tablenotes}\footnotesize
        \item[a]\, 3-component model \citep{Tiscarenoetal2007, TiscarenoHarris2018}
        \item[b]\, 2-component model \citep{Colwelletal2009a, Tiscarenoetal2013b}
        \item[c]\, average value used \citep{HedmanNicholson2016}
        \item[d]\, 3-component model \citep{Baillieetal2011, HedmanNicholson2014}
    \end{tablenotes}

\end{table}

The surface mass densities of the main ring regions (Table \ref{tbl:ring_mass}) have been estimated from the radial wavelengths of the many density and bending waves in the rings that are driven by external satellites and planetary internal oscillations \citep{Tiscarenoetal2007, Colwelletal2009a, Baillieetal2011, HedmanNicholson2014, HedmanNicholson2016}. 
Numerous estimates for the A ring show that $150 < \Sigma < 400$~kg~m$^{-2}$ \citep{Tiscarenoetal2007, TiscarenoHarris2018}, with an average value in the central A ring of $350$~kg~m$^{-2}$.
In the Cassini Division $\Sigma$ is much lower but varies substantially, ranging from $\sim15$ in the inner part to almost $200$~kg~m$^{-2}$ in the outer ramp \citep{Colwelletal2009a, Tiscarenoetal2013b}. 
Reliable data are much scarcer in the B ring, but one study finds that $\Sigma \simeq 600$~kg~m$^{-2}$ over much of this region \citep{HedmanNicholson2016}, with surprisingly little radial variation.
For the C ring $\Sigma$ ranges from $\sim10$ to $60$~kg~m$^{-2}$, with the highest values in the central undulating region \citep{Baillieetal2011, HedmanNicholson2014}. 
\par
Based on the above surface mass densities, we can estimate the mass of the main ring components, as listed in Table~\ref{tbl:ring_mass}. 
In units of the mass of the satellite Mimas ($M_{\rm Mimas} = 3.75\times10^{19}$~kg), we find that the overall mass of the rings is $\sim0.38~M_{\rm Mimas}$. This value was confirmed when direct gravity measurements were made by Cassini on its final orbits, yielding $M_{\rm rings} = (0.41\pm0.13)~M_{\rm Mimas}$ \citep{Iessetal2019}.
\par
Indirect estimates of the average particle size in different ring regions can be made by combining the optical depth with the local surface mass density. 
As noted in Section \ref{opticalDepth}, $\Sigma/\tau \simeq \frac{4}{3}\rho\bar{a}$, where $\rho$ is the internal mass density of a particle and $\bar{a}$ is the mean particle radius. Assuming an average particle density of $500$~kg~m$^{-3}$, suitable for porous water ice, we find that $\bar{a}\simeq 100$~cm in the A ring, $25-60$~cm in the B ring and $10-20$~cm in the C ring.

An unexpected result of such comparisons is that within both the B and C rings there appears to be little local correlation between $\Sigma$ and $\tau$, with $\tau$ varying much more than does $\Sigma$. This lack of correlation suggests that variations in the particle size distribution may be the cause of much of the observed variability in $\tau$, rather than variations in the total amount of ring material per unit area.
This is not currently well-understood. In the B ring, moreover, some local brightness variations may reflect variations in the albedo or phase function of the ring particles, rather than variations in either $\tau$ or $\Sigma$ \citep{EstradaCuzzi1996}.
\par

\subsection{Particle size distribution}\label{particleSize}

\begin{table}
\begin{center} 
\caption{Particle size distributions in the main rings. Values are averaged by region for differential indices for model power-law particle size distributions in Saturn's rings. Fits are derived from Voyager RSS data \citep{Zebkeretal1985}, Earth-based 28 Sgr occultation in 1989 \citep{FrenchNicholson2000} and Cassini UVIS, VIMS \& RSS data \citep{Jerouseketal2020, Jerousek2018}.}
\label{tbl:size_distribution} 
\begin{tabular}{lccccc}\\
\toprule
Ring Region & Radial location & Avg. q	& Min. $a_{\rm min}$ &	Max. $a_{\rm max}$ \\
& (km) && (cm) & (m) & \\
\midrule
C Ring background 	&	74500 - 91980	& 3.16$^a$	& 0.42$^a$	& 11$^a$\\
C Ring Plateaux		&					& 3.08$^a$	& 0.36$^a$	& 13$^a$\\
C Ring ramp			&	90620 - 91980	& 3.16$^a$	& 0.38$^a$	& 13$^a$\\
B Ring$^b$				&	91980 - 117500	& $\sim2.8^{c,d}$	& 0.5$^d$	& 7$^e$\\
Cassini Division background		&	117500 - 122100	& 2.85$^f$	& 0.41$^a$	& 8.9$^a$\\
Cassini Division Triple Band		&	120550 - 120800 & 2.94$^a$	& 0.61$^a$	& 7.6$^a$\\
Cassini Division ramp				&	120900 - 122100 & 3.07$^a$	& 0.33$^a$	& 20$^a$\\
A Ring inside Encke Gap & 122100 - 133410 & $2.9 \pm 0.1^{d,g}$ & $\sim 2.5^d$	& $5-10^{d,g}$\\
A Ring outside Encke Gap				&	133410 - 136800 & $3.05 \pm 0.15^{d,g}$	& $0.4^d$	& $5-10^{d,g}$\\
\botrule
\end{tabular} 
\end{center}
\begin{tablenotes}\footnotesize
        \item[a]\, \cite{Jerouseketal2020}
        \item[b]\, Constraints are limited to regions of the B ring with lower optical depth
        \item[c]\, \cite{FrenchNicholson2000}
        \item[d]\, \cite{Jerousek2018}
        \item[e]\, inner B ring, \cite{Jerousek2018}
        \item[f]\, mean of values from \cite{Zebkeretal1985} and \cite{Jerouseketal2020}
        \item[g]\, \cite{Zebkeretal1985}
    \end{tablenotes}
\end{table}

The most direct constraints on the particle size distribution come primarily from the multi-wavelength RSS occultation data from Voyager \citep{Zebkeretal1985}, estimates of the near-forward scattering cross-section of the rings at optical and near-IR wavelengths from Earth-based stellar occultation \citep{FrenchNicholson2000}, and comparisons of the measured values of optical depth $\tau$ in the Cassini UVIS (UV) and VIMS (near-IR) data with those obtained from microwave (RSS) data \citep{Jerouseketal2020, Jerousek2018}.

The areal density $n(a)$ of particles of radius $a$ can be modeled as a power law, $n(a) \propto a^{-q}$ for $a_{\rm min}<a<a_{\rm max}$. The power-law index $q$ and the upper size limit for particles $a_{\rm max}$ are both found to vary across the rings, as estimated using various methods. 
Representative estimates by ring location are provided in Table \ref{tbl:size_distribution}. In the A and C rings and in the Cassini Division, \cite{Zebkeretal1985} and \cite{Cuzzietal2009} report a differential index $q = 2.7 - 3.2$ and $a_{\rm max} \simeq 2-10$~m.
Because of its large optical depth, estimates of $a_{\rm max}$ and $q$ have a much greater uncertainty for the B ring.

\begin{figure}
    \centering
    \resizebox{5in}{!}{\includegraphics{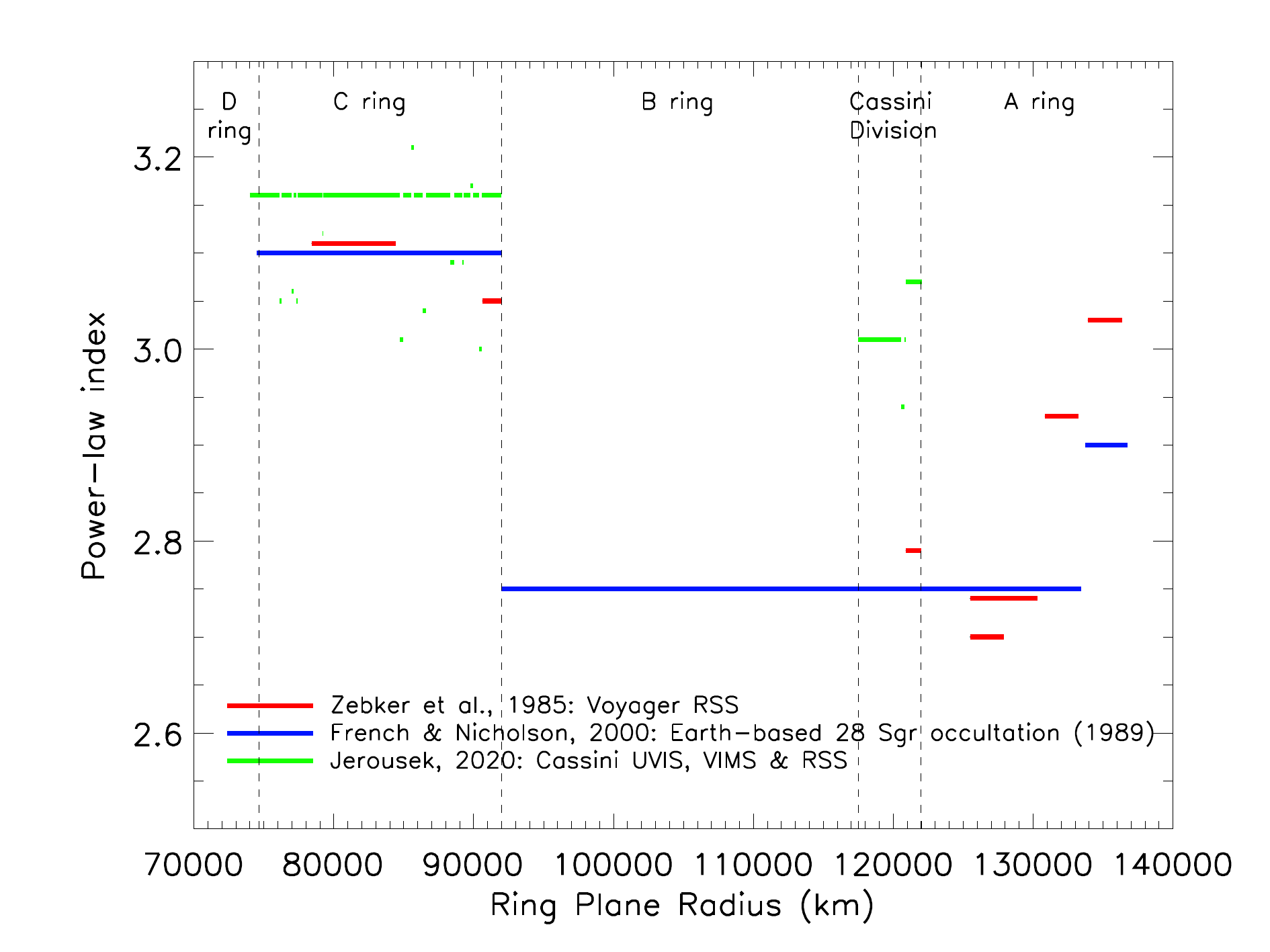}}
    \caption{Particle size distributions across Saturn's main rings: power-law indices inferred from Voyager RSS data \citep{Zebkeretal1985}, 28 Sgr Earth-based occultation \citep{FrenchNicholson2000}, and Cassini UVIS, VIMS and RSS \citep{Jerouseketal2020}.}
    \label{fig:partsizedistr_powerlaw}
\end{figure}

\begin{figure}
    \centering
    \resizebox{2.3in}{!}{\includegraphics{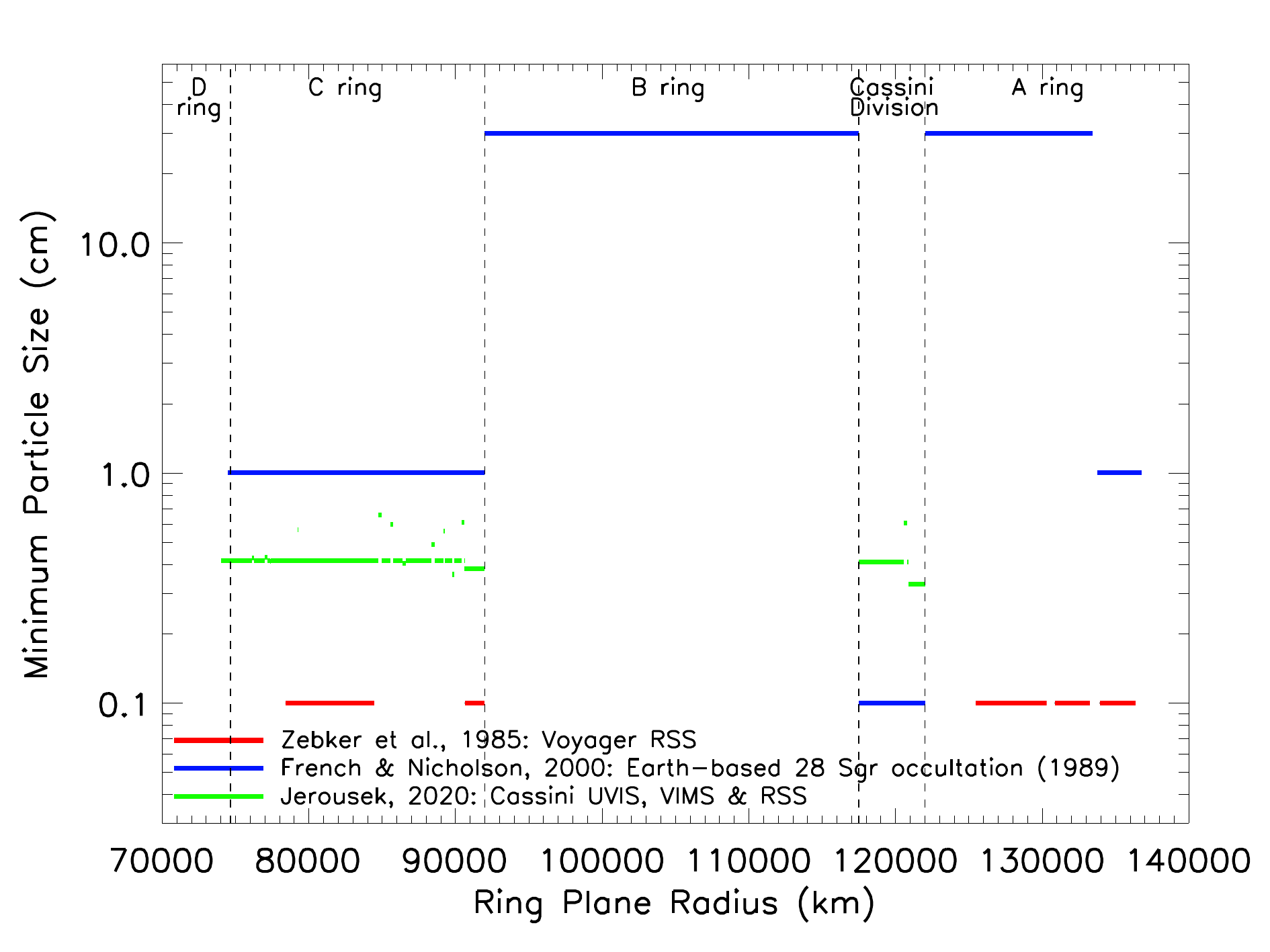}}
    \resizebox{2.3in}{!}{\includegraphics{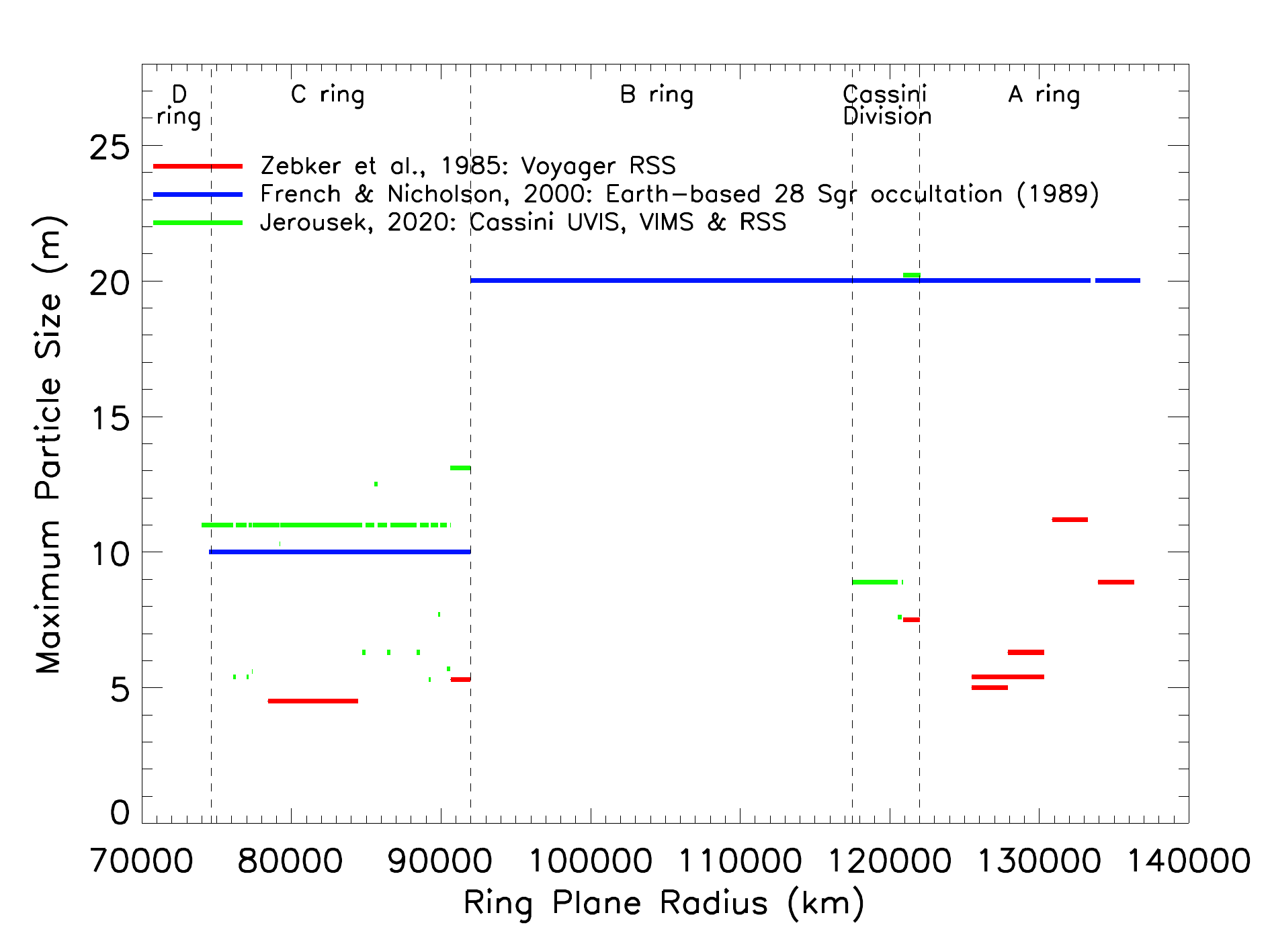}}
    \caption{Particle size distributions across Saturn's main rings: minimal (left) and maximal (right) limits for power-law distributions inferred from Voyager RSS data \citep{Zebkeretal1985}, 28 Sgr Earth-based occultation \citep{FrenchNicholson2000}, and Cassini UVIS, VIMS and RSS \citep{Jerouseketal2020}.}
    \label{fig:partsizedistr_limits}
\end{figure}

The value of  $a_{\rm min}$ is more poorly defined, but occultations in the near-IR suggest that it varies from a few mm in the outer A ring and C ring to as much as 30~cm in the B and inner A rings \citep{FrenchNicholson2000, Cuzzietal2009, Harbisonetal2013}. Figure \ref{fig:partsizedistr_powerlaw} summarizes the published estimates of power-law indices across the rings, while Figure \ref{fig:partsizedistr_limits} shows the corresponding values of $a_{\rm min}$ and $a_{\rm max}$.

\par

Inverting the near-forward scattered signal from RSS, \cite{Zebkeretal1985} derived a power-law index between $q = 2.7$ in the A ring and $q = 3.11$ in the C ring, with particles ranging from 1 mm to $\sim10$ m. They also estimated that the Cassini Division has a slightly shallower ({\it i.e.,} smaller $q$) particle size distribution than the C ring, with $q = 2.79$, and maximum sizes of $4.5-7.5$~m.
Following \cite{Zebkeretal1985}'s method, \cite{Jerouseketal2020} determined separate power-law indices in the background C ring, in the C ring plateaux, in the background Cassini Division, and in the Triple Band that connects the inner Cassini Division outward to the ramp to the inner edge of the A ring.

\par
\cite{Zebkeretal1985} also found that the size distribution steepens from the inner A ring ($q = 2.7$) to the outer A ring ($q = 3.03$), with the largest values of $q$ being found outside the Encke Gap at a radius of 133,500~km. This result was confirmed by subsequent studies (see Fig.~\ref{fig:partsizedistr_powerlaw}).
An abrupt jump in $q$ has also been observed by \cite{Beckeretal2016} in diffraction signatures at the edges of the Encke gap: they measured $q = 2.9$ and $a_{\rm min} = 15$ mm at the inner edge and $q = 3.1$ and $a_{\rm min} = 9.3$ mm at its outer edge. This trend continues to the outer edge of the A ring, where the minimum size is 4.4~mm with a power-law index of 3.2. This change in the size distribution may be related to an abrupt change in the A ring's photometric properties (color and phase function) at the Encke gap \citep{Tiscarenoetal2019}. The relative lack of small particles in the inner A ring is consistent with the analyses by \cite{Maroufetal1986} and \cite{Jerouseketal2016}. 
\cite{Becker2016PhD} suggests that this depletion in sub-mm particles may indicate that the collisions in this region are too gentle to release enough energy to remove micron-sized dust from the surfaces of larger aggregates \citep{AlbersSpahn2006}.

From the analysis of Cassini UVIS stellar occultation scans crossing the edges of the Encke and Keeler gaps, R6, Huygens and Titan ringlets, as well as the outer edges of the A and B rings, \cite{Eckertetal2021} found minimum particle sizes ranging from 4.5~mm to 66~mm and average power law indices ranging from 3.0 to 3.2. 
Consistent with the results of \cite{Beckeretal2016} and the RSS multi-wavelength profiles, this indicates that edges perturbed by satellite resonances tend to host more numerous sub-cm particles than the sharp edges of other ringlets. 
This result appears consistent with the predator-prey model detailed in \cite{Espositoetal2012}, based on the competition between aggregation and fragmentation mechanisms in regions perturbed by satellite resonances. 
The kinetic model of \cite{Brilliantovetal2015} confirmed that the observed distributions could be consistent with the steady-state of such an aggregation/fragmentation mechanism.

Particles are much smaller in the D and F rings, with photometric and spectrophotometric data suggesting average sizes of a few tens of microns \citep{Hedmanetal2011, Vahidiniaetal2011, HedmanStark2015}, although some regions of the F ring may contain a substantial population of cm-sized particles \citep{Cuzzietal2014a}. 
See Section \ref{dringcomp} for further discussion of particle sizes in the D ring.

\subsection{Embedded moonlets}

There is also evidence for a small population of much larger ring particles that are up to a few kilometers in size, distributed across the main rings.
Daphnis (3.8 km in radius) and Pan (14.1 km in radius) define the upper limit of this population, followed by a population of $\sim100$ to 1000~m size objects, mainly observed by Cassini-ISS in the A and B rings \citep{SpahnSremcevic2000, Tiscarenoetal2006, Tiscarenoetal2008, Tiscarenoetal2010}. 
These 100-1000 m objects are known as ``propellers" because of the shape of the denser regions they produce in the rings surrounding depletion zones located ahead of the object (which is itself unseen) on its interior and behind the object on its exterior.

Propellers form when the orbiting objects are too small to clear a full azimuthal gap but are large enough to clear small azimuthal regions of the rings in their immediate vicinity. 
More massive moons result in perturbations that can open a gap all the way around the rings, as with Pan and Daphnis, which clear the Encke and Keeler gaps in the outer A ring \citep[][; see Section \ref{ringMoon} for further discussion]{Nicholsonetal2018}. 
Propellers in the A ring are primarily detected in three narrow belts between radii of 126,750 and 132,000 km \citep{Tiscarenoetal2008}, and their differential size distribution may be modeled as a steep power-law of index $q \simeq6$ \citep{Tiscarenoetal2010}.

Finally, from the analysis of Cassini UVIS stellar occultations, \cite{Baillieetal2013} reported the detection of a population of ``ghosts" (small regions of low local optical depth, or `holes') in the C ring and in the Cassini Division, which they believe are consistent with the depletion zones surrounding propellers. 
Their estimate of the upper limit of the particle size distribution is consistent with the existence of a population of icy particles of up to 12 meters in the C ring and $\sim 50$ m in the Cassini Division. 
These objects do not contribute significantly to the measured normal optical depth but may result in the ``streaky texture" identified in the C ring by \cite{Tiscarenoetal2019} in Cassini ISS images.

From the distribution of the ghost widths, \cite{Baillieetal2013} were able to model the size distribution of these boulders as a power-law with a differential index q = 1.6 in the C ring and q = 1.8 in the Cassini Division assuming the density of water ice. These power-law indices for objects between a few tens of meters and a few hundreds of meters are thus quite different from the estimated values for propellers in the A ring or for smaller ring particles. It is uncertain at present how the assumption of pure water ice for the ghost objects affects the retrieved size distribution, and further work in this area is needed. Ghost objects in the C ring and Cassini Division may represent the coherent shards of one or more consolidated objects that broke up to form the rings, as they exist well inside the Roche limit for solid water ice and are therefore unlikely to have accreted {\it in situ} as is supposed to have happened for the A ring propellers \citep{CanupEsposito1996,LewisStewart2009}. The increased abundance of non-icy material in the central C ring has been cited as additional evidence for the break up of a Centaur or other solid body \citep[][; see also Section \ref{bulk}]{Zhangetal2017C}.
\par

\subsection{Ring brightness temperatures measured by CIRS}\label{brightnessTemp} 

The numerous radial scans of the main rings made with the Composite Infrared Spectrometer \citep[CIRS; ][]{Flasaretal2004}
FP1 detector have provided radial profiles of brightness temperatures, both on their lit and unlit sides \citep[see][for a review]{Spilkeretal2018}.
These temperatures are a complex function of the ring's 3-dimensional structure, particle properties such as spin state, 
Saturn's season, the radius from Saturn (which is a supplemental heating source), and viewing geometry.
On the lit side of the rings, the brightness temperature generally decreases from the C ring to the A ring except for the Cassini Division, reflecting the decreasing flux from the planet. Saturn remains a significant source of heating for particles within the close, thin C ring \citep{Spilkeretal2013}.
The higher temperatures seen in the C ring and Cassini Division are additionally attributed to the albedos of the particles being lower in these regions, as well as to the lower optical depths in these regions that results in reduced particle mutual shadowing.
\par
On the unlit side, the optically-thin C ring and Cassini Division rings appear hotter than the more opaque A ring, while the optically thick B ring is the coldest of all, reflecting the effects of mutual shadowing between the particles as well as the difficulty with which heat may be transferred across the ring plane \citep[][]{Spilkeretal2006, Morishimaetal2010}. Any deconvolution of the properties of the ring particles and regolith grains from the brightness temperature profiles is therefore ring-model dependent. 
\par
\cite{Morishimaetal2010} have derived the ring particle bolometric Bond albedo $A_V$ as a function of Saturn distance, assuming a multi-particle-layer model that mixes slow and fast rotating particles. 
The albedo for the A and B rings is roughly constant, with $A_V \sim 0.55$, while the C ring albedo seems to decrease towards the planet from 0.4 to 0.1. In this study, the infrared emissivity was fixed to unity. Were the true infrared emissivity to be a bit lower, the fitted temperatures would favor lower albedos to reproduce a given observed temperature. Other thermal models for the C ring provide average values of about 0.2 \citep[see][]{Spilkeretal2018}, similar to but slightly higher than the value of $\sim0.15$ derived from visible and NIR observations \citep{Smithetal1981,Cooke1991, Cuzzietal2009}.
In comparison, the empirical slab model of \cite{Altobellietal2014} yields an albedo $A_V \sim 0.45$ that is approximately constant with distance to the planet, albeit with large fluctuations.
\par
If the apparently increasing albedo from the C ring to the A and B rings is real, a correlation between local optical depth and Bond albedo should exist. Such a correlation would be compatible with a darkening pollutant, such as one delivered by meteoroid bombardment, comprising a larger portion of the optically thin C ring \citep{Cuzzietal2009}.
At this stage, no thermal model has taken into account this variation of albedo in terms of pollutant fraction.

Radial variations in ring brightness temperatures may also originate from changes in emissivity that can stem from compositional or structural causes, such as a change in the regolith grain sizes covering the ring particles. \cite{Morishimaetal2012} have investigated radial variations in regolith properties of particles, assuming grains are made of pure water ice. Any variation of ring emissivity versus distance has then been interpreted as due to variations in the size distribution of grains with distance (see Section \ref{mainRingsWater}).

\subsection{Comparison to ring temperatures measured by VIMS}\label{TempCompare}

The differences in albedo, optical depth, proximity to Saturn and regolith grain size discussed in the previous sections are the main cause of the temperature variation observed across the rings. While Cassini/CIRS data directly measure the rings' brightness temperature, as discussed in Section \ref{brightnessTemp}, VIMS data permit an indirect measurement based on the spectral properties of water ice. 
Specifically, the position of the 3.6 $\mu$m continuum peak exhibited by crystalline water ice is temperature-dependent and can be determined from optical constants measured in the laboratory. As \cite{Mastrapaetal2009} have demonstrated, 
the imaginary part of the refractive index of crystalline water ice shows a change in its minimum
around 3.6~$\mu$m for a range of temperatures between 20 and 150 K. 
\par
\begin{figure}
    \centering
    \resizebox{5in}{!}{\includegraphics{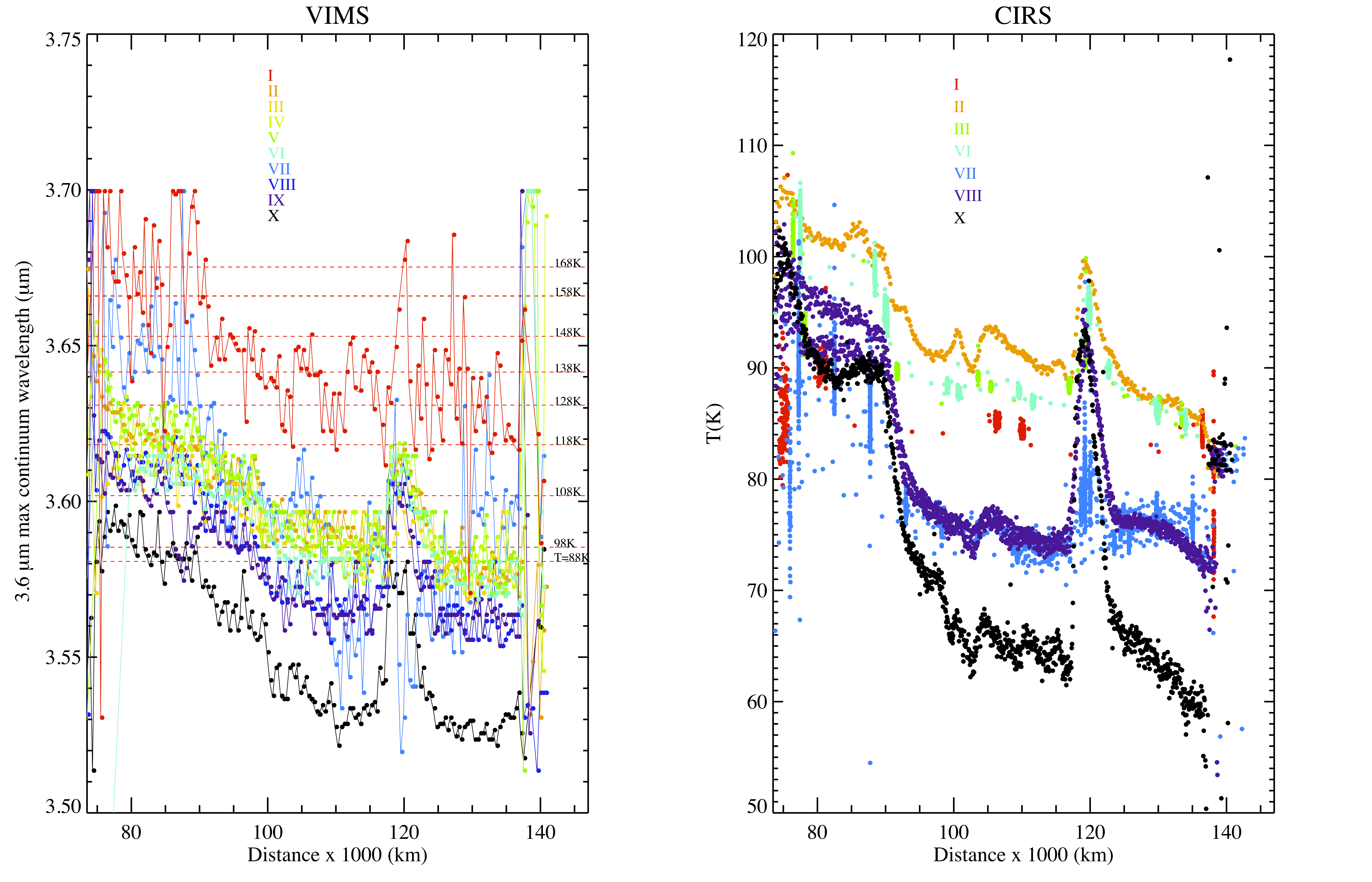}}
    \caption{Left panel: VIMS-retrieved 3.6 $\mu$m continuum peak wavelength and corresponding water ice temperature scale (horizontal dashed lines) for the 10 observations considered in \cite{Filacchioneetal2014}. 
    From observation (I) to (X) the sub-solar latitude, or solar elevation angle, decreases from -23.5$^\circ$ to 2.6$^\circ$. When the sub-solar latitude is negative VIMS observes the south face of the ring plane, (I) to (IX); when sub-solar latitude is positive, VIMS observes the north face, (X). 
    Right panel: CIRS-retrieved ring’s temperature during 7 of the 10 observations. 
    For some CIRS observations, like in (VIII), the data shown are taken across two consecutive radial scans; this explains the differences observed across the C ring.}
    \label{fig:rings_temp_vims}
\end{figure}

A similar behaviour has been observed by \cite{Clarketal2012} in reflectance measurements of small grains of pure water ice at standard illumination conditions (phase = 30$^\circ$) for sample temperatures varying between 88 K and 172 K.
By determining the wavelength of the 3.6~$\mu$m peak in reflectance, it is therefore possible to derive the temperature of the ice grains that are the main constituent of the ring particles. 
The 3.6~$\mu$m peak shifts towards shorter wavelengths when the ice is cooled, moving from about 3.675~$\mu$m at T = 172~K to about 3.581~$\mu$m at T = 88~K \citep{Filacchioneetal2016}. 
By applying this method to ten VIMS radial mosaics of the rings sampled at 400 km/bin in the radial direction, \cite{Filacchioneetal2014} derived radial profiles of the rings' temperature on the sunlit side as shown in Fig.~\ref{fig:rings_temp_vims}. 
The observations were taken between 2004 near the southern solstice when the solar elevation angle $B_0$ was -23.5$^\circ$, and 2010 near equinox ($B_0 = 2.6^\circ$), allowing tracking of the diurnal temperature changes over one half of the seasonal cycle.

Before equinox the south side of the ring plane was illuminated, while after equinox the sun shone on the north face. 
The maximum temperature was measured in 2004 when the solar elevation angle, B$_0$, was at a maximum (observation I, red curve in Fig. \ref{fig:rings_temp_vims}). Ring particles cool down moving from southern summer to equinox. 
At equinox the solar rays are parallel to the ring plane and the particles receive the minimum solar illumination. 
As a consequence of this effect the peak’s position shifts to shorter wavelengths and the minimum temperatures are observed across the rings (observation X, black curve in Fig. \ref{fig:rings_temp_vims}). 
When both solar and antisolar ansae observations\footnote{Here ''solar ansa" and ''antisolar ansa" refer to the ansa closer to the subsolar and antisolar points respectively. The antisolar ansa is therefore closer to Saturn's shadow, but both acquisitions are from the illuminated portion of the rings.} are available, such as in the two paired observations (V)–(VI) and (VIII)–(IX), it is possible to distinguish diurnal variations: on the two solar ansa (V and  VIII), higher temperatures are measured than for the corresponding antisolar observations, (VI and IX). 
This effect is particularly evident in the C ring inside of 92,000 km.
Since Saturn’s thermal emission is isotropic, a similar thermal behavior is probably the consequence of the cooling of the ring particles during the eclipse period.
\par
While substantial agreement between VIMS
and CIRS results across the A and B rings is observed, VIMS systematically measures higher temperatures than CIRS across the C ring and Cassini Division. 
Several effects could explain this discrepancy. 
One possibility is that the deviation of the peak’s position in the presence of contaminants, in particular organics that have absorption bands in the 3.6~$\mu$m peak's short-wavelength wing (see Section \ref{organicsSpectral}), may change the apparent position of the peak towards longer wavelengths. 
This would result in higher apparent temperatures. Notably, the C ring and Cassini Division are the regions of the rings where the amount of contaminants is larger, and where the assumption of pure water ice made to derive the temperature is less robust.
\par
Alternatively, the difference in temperatures reported by the two instruments may be a consequence of the different skin depths $l_{\lambda}$ sampled by the two instruments as a result of the wavelengths at which they operate.
While VIMS is sensitive to very shallow depths a few microns from the outer surface of the particles, CIRS measures temperature at a greater depth,perhaps a few millimeters or even a centimeter in clean ice. In this case, VIMS may be measuring surface temperatures while CIRS retrieves volume temperatures. Indeed, the diurnal thermal inertia for the main rings is estimated from eclipse cooling data to be $\sim 10~ \, \text{J m}^{-2} \, \text{K}^{-1} \, \text{s}^{-0.5}$, with the lowest values in the Cassini Division. The corresponding thermal skin depth is on the order of 1 mm, greater than the characteristic VIMS penetration but less than that for CIRS \citep{Ferrarietal2005,Morishimaetal2011}. 
\par
Overall, the comparison between VIMS' ``indirect" temperature measurements and CIRS brightness temperatures reveals that they show the same radial trends, with temperatures on the sunlit side increasing towards Saturn.

\section{UV/Optical/IR Observations of the Main Ring Composition}\label{mainRingRegolith}  
\par
Cassini UV and near-IR spectral data show that Saturn's ring particles are predominantly made of crystalline water ice with minor contamination from non-icy materials \citep{Cuzzietal2009, Cuzzietal2018book}.
While water ice is original to the rings, part of the non-icy materials that is responsible for the darkening and red-colored appearance at visible wavelengths is exogenous in origin. 
In this respect, when comparing the current composition of the rings with their original composition, one has to formulate some hypothesis about the ab-initio conditions proposed by the different formation scenarios, whether the rings are remnants of the Saturn protoplanetary nebula or fragments of a destroyed moon or comet \citep{Harris1984,EspositoDeStefano2018}.
For example, rings originating in the protoplanetary nebula might be expected to contain some silicates and some volatile ices such as NH$_3$ or CO$_2$, while material tidally stripped from a differentiated icy satellite \citep{Canup2010} might be almost pure water ice. A disrupted comet might have a very different composition from a satellite's mantle.
\par
In Section \ref{mainRingsWater} we discuss evidence for crystalline water ice as the dominant ring component, before turning in Section \ref{nonWater} to a description of endmember materials added to water ice (silicates and rocky material, carbon-bearing material and other ices) and their distribution across ring regions as derived from a synergistic analysis and modeling of Cassini data.
In Sections \ref{UVAbsorber} to \ref{otherIces} below, we describe spectral features that are observed in the data, as well as how these features have been interpreted in the context of radial trends and other associations with regions of the main rings. Modeling of the data is discussed separately in Sections \ref{subsec:caveats} and \ref{regolithModels}. The data and interpretations described in Sections \ref{UVAbsorber} to \ref{regolithModels} are associated with spectra at nm to µm wavelengths, and constrain the surface composition of the ring particles. Regolith gardening processes may homogenize the particles on timescales of $~1$ Myr \citep[][]{ElliottEsposito2011}. Data and interpretation related to the ring particle interiors are discussed in Section \ref{bulk}.

\subsection{Water}\label{mainRingsWater} 
\begin{figure}
    \centering
    \resizebox{4.5in}{!}{\includegraphics[angle=0]{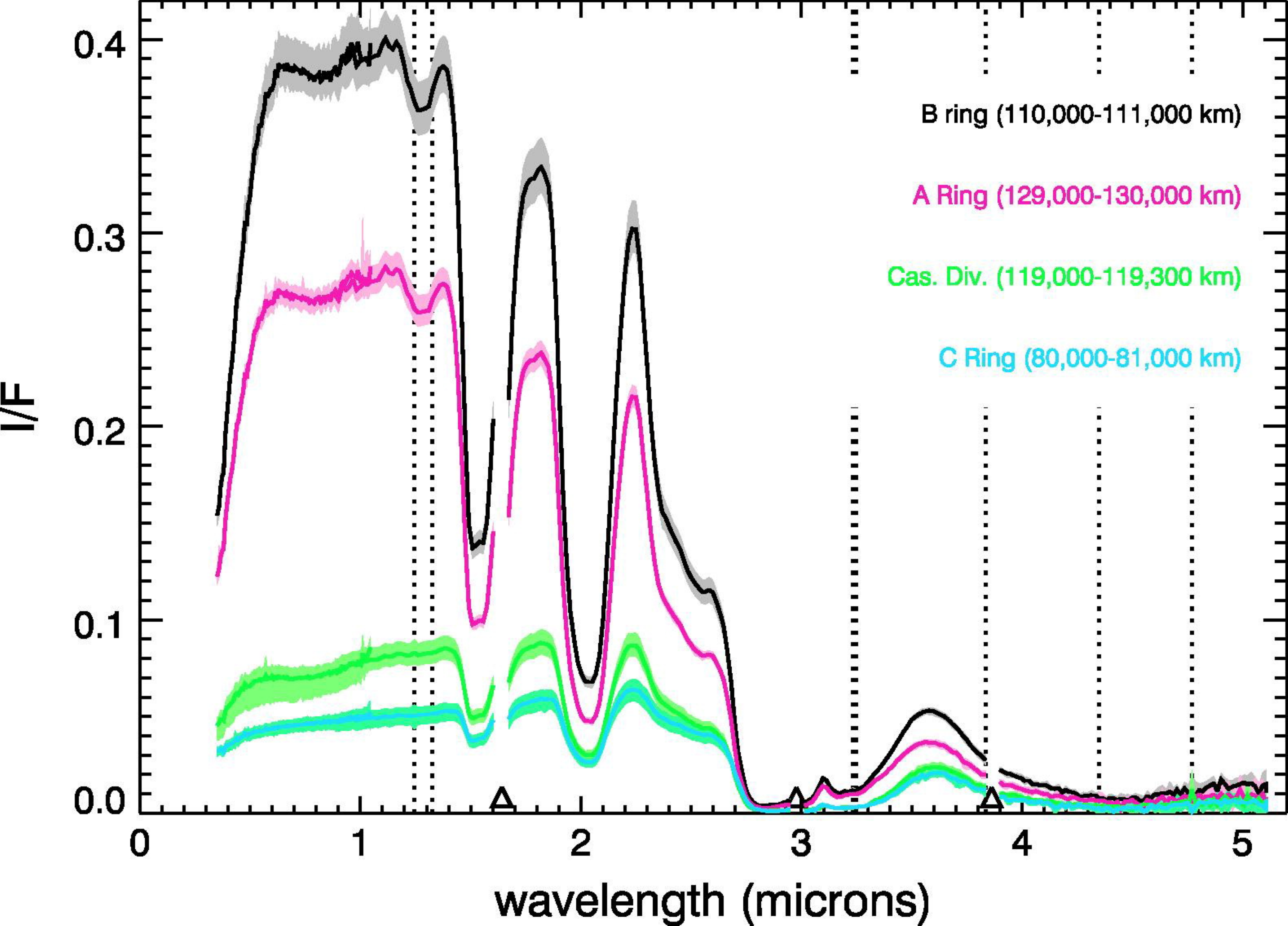}}
    \caption{Reflectance spectra of the sunlit side of the main rings of Saturn derived from Cassini-VIMS observations in 2005 at phase angles of $13-41^\circ$.
    Vertical dotted lines mark the locations of “hot pixels” in the detector with higher than average dark current, while triangles mark the locations of gaps in the order-sorting filter, where the spectral data are less secure and thus not plotted. 
    The solid curves show the mean I/F spectrum for each region, while the shaded bands indicate the range of values observed in each of these locations. 
    These ranges primarily reflect variations in the absolute brightness of the ring, rather than statistical uncertainties in the spectra themselves. 
    Note the strong water–ice absorption bands at 1.5, 2.0 and $3.0~\mu$m, as well as weaker ice bands in the A and B ring spectra at 1.25 and $1.05~\mu$m, and the steep slope at wavelengths below $0.6~\mu$m in the A and B rings.
    Also note the variations in the spectral slope at continuum wavelengths between 0.6 and $2.3~\mu$m among the different spectra. \citep[From][]{Hedmanetal2013}. }
    \label{fig:ring_IRspectra}
\end{figure}

The reflectance spectrum of the main rings from the near-UV ($\lambda = 0.30~\mu$m) to the near-IR ($\lambda = 5.2~\mu$m) is dominated by crystalline H$_2$O frost (Fig.~\ref{fig:ring_IRspectra}), with typical grain sizes of a few tens to $100~\mu$m \citep{ClarkMcCord1980, Pouletetal2003, Cuzzietal2009, Cuzzietal2018, Filacchioneetal2012, Filacchioneetal2014}. 
The comparative purity of the ice is supported by the very high albedo of the rings in the visible region, where I/F $\simeq0.5$ at low phase angles, as well as an extremely low albedo in the $2.7-3.1~\mu$m region where water ice is strongly absorbing, and with the very strong absorption edge in the rings' far-UV spectrum at $\sim0.17~\mu$m that is characteristic of fine-grained H$_2$O ice \citep{Bradleyetal2010, Bradleyetal2013}. 
\par
\par

\par

\cite{Spilkeretal2005} derived emissivity spectra for the A, B and C rings from CIRS data. 
Particles much larger than the observing wavelengths (10 to 600 cm$^{-1}$, or 16.7 to 1000 $\mu$m) were found to dominate the spectrum for most of the FP1 spectral range, 

since the observed emissivity spectra appeared free of features within the error bars across this range. 
Of most interest, however, they found a clear roll-off in the averaged ring particle emissivity at frequencies below 50~cm$^{-1}$ ($\lambda\gt200~\mu$m) that they attributed to a combination of particle size effects along with the low absorption of pure water ice in this spectral region (see their figure 5 in particular).
\par
The shape of the roll-off is sensitive to the size distribution and upper and lower size limits of ring particles, as well as the nature of the regolith that covers them. With a radiative transfer model making use of a four-stream calculation along with the delta-M method on a single layer with randomly-distributed regolith-free ring particles, \cite{Spilkeretal2005} found that a size distribution with power-law index $q=3.4$ reasonably fits the spectrum of the B ring. However, no robust analysis has been performed on the possible range of these parameters, nor on the fraction of contaminants in the water ice. Near the wavelength of the roll-off, the transparency of water ice makes the influence of sub-millimeter-sized water-ice particles apparent, but seems to preclude the presence of grains smaller than a few tens of micrometers. The fact that the CIRS results show the influence of sub-cm size particles so clearly, and that Cassini radio occultation results that sense only free-floating particles have larger minimum-particle-size cutoffs (Section \ref{bulk}), suggests that the spectral signature of sub-cm size particles seen in the CIRS data might actually be a manifestation of regolith emissivity, which \cite{Spilkeretal2005} did not model.

\par
\cite{Morishimaetal2012} derived the spectral emissivity from a large number of spectra accumulated by the CIRS FP1 channel at various seasons and viewing geometries before the end of 2010. Good fits were obtained using pure water-ice and a power-law distribution of grain sizes ranging from $\sim1~\mu$m to 1~cm. For the main rings, they found that emissivity varies with ring temperature range (i.e. radial distance, season or phase angle). This observed emissivity is interpreted to be that of the regolith covering ring particles.

Despite many searches, there appear to be no unambiguous spectral indicators of other materials in the UV to near-IR spectral range \citep{Cuzzietal2009, Cuzzietal2018}, or in the mid-IR spectrum measured by the CIRS instrument \citep{Edgingtonetal2008}.
However, the steep decrease in the main rings' albedo shortward of $\sim0.6~\mu$m --- which is especially apparent in the A and B rings in Fig.~\ref{fig:ring_IRspectra}, and leads to their distinctly reddish color at visible wavelengths \citep{EstradaCuzzi1996, Cuzzietal2002} --- is {\it not} characteristic of pure water ice. 
This is discussed further in Sections \ref{UVAbsorber} and \ref{neutralAbsorber}.
\par
Data from the Cassini-VIMS instrument have been used by several investigators to spatially map the variations in the purity, regolith grain size and even the temperature of the water ice.
This is done using the relative depths of the water-ice bands at 1.55 and $2.0~\mu$m, the height and exact wavelength of the spectral peak at $\sim3.6~\mu$m, and the average spectral slope between 0.35 and $0.55~\mu$m (referred to below as the `UV slope').\footnote{The band depths are defined as $B_\lambda = (I_0 - I_\lambda)/I_0$, where $I_0$ is the average continuum intensity level outside the band, while the spectral slope is defined by $S = (I_{\lambda2} - I_{\lambda1})/I_0$. Here $I_0$ is the average intensity level over the interval [$\lambda1,\lambda2$], and $I_\lambda$ is the calibrated $I/F$ at wavelength $\lambda$.}
Examples of such measurements are shown in the middle panel of Fig.~\ref{fig:ring_profiles} above, where it is apparent that the strengths of the water ice bands closely track the steepness of the UV slope.
This has led to the interpretation that the material responsible for the UV absorption is embedded within the icy grains ({\it i.e.,} it is an `intra-mix' in spectral parlance; see Section \ref{RRTModels}).
However, the observed variations in band depth could be due predominantly to variations in the mixing ratio of the non-icy material or to systematic variations in the water ice grain size in the regolith \citep{Cuzzietal2009}. 
\par
Overall, the strongest water-ice bands are found in the outer A and middle B rings, and the weakest in the C ring and Cassini Division. However, the tranisitions between these regions are much more gradual than might be expected, given the sharpness of the underlying transitions in optical depth. 
Across the C ring and in the part of the A ring beyond $\sim136,000$~km there is a steady weakening of the ice bands and of the UV slope \citep{Nicholsonetal2008, Tiscarenoetal2019}.

\subsection{Non-Water Components}\label{nonWater}
\par
The primary spectral features that have been ascribed to non-water components are referred to as the ''ultraviolet absorber" (Section \ref{UVAbsorber}) and the ''neutral absorber" (Section \ref{neutralAbsorber}). Since organic material is a significant component in the Cassini Grand Finale in situ ring measurements (Section \ref{eqCompositon}) and different carbon-based materials have been hypothesized as the cause of these two sets of spectral characteristics, we discuss spectral evidence for organic material in the main rings further in Section \ref{organicsSpectral}. Finally, in Section \ref{otherIces} we briefly address the lack of evidence in the main rings for volatile ices such as CH$_4$ or CO$_2$, which were both observed in the in situ data in the D ring region (Section \ref{eqCompositon}).

\subsubsection{Ultraviolet absorber}\label{UVAbsorber}
Two competing explanations have been offered for the steep decrease in the main rings' albedo shortward of $\sim0.5~\mu$m (Figure \ref{fig:ring_IRspectra}): the presence of small amounts of carbon or carbon-bearing material such as tholins within the ice grains \citep{CuzziEstrada1998, Pouletetal2003}, or the presence of nano-phase inclusions of metallic iron or iron minerals such as hematite (Fe$_2$O$_3$) or troilite (FeS) \citep{Clarketal2012}.
These two models have very different cosmogonic implications (see further discussion in Section \ref{organicsSpectral} below). 
\par
Iron metal or hematite, perhaps in the form of extremely tiny nano-grains, has been proposed as a darkening and/or reddening agent \citep{Clarketal2012}. This suggestion has really only been made for the C Ring, which is far less red than the A and B rings (section \ref{regolithModels}). Meteoroid bombardment and ballistic transport would predict that the C Ring is indeed more polluted by extrinsic meteoroids. However, radiative transfer models tend to find that carbon-tholin-silicate mixtures provide better fits to C ring spectra than iron metal or hematite \citep[Section  \ref{regolithModels}, ][and Chapter 10 in this volume]{Cuzzietal2018}.
While carbon-based materials therefore provide the better spectral fits, the Grand Finale data provide some new evidence for the presence of Fe-based materials in the rings. In analyses of CDA data from Cassini's Grand Finale (see Section \ref{inflowIntro}), \cite{Fischeretal2018} reported two distinct compositional populations: iron-rich grains (93\% of them sulfides) providing more than half of the detections; and iron-poor grains (most of them silicates) providing less than half.  A dynamical analysis of these detected grains, expanded upon by \cite{Fischeretal2023}, found that the silicate component plausibly derives from ejecta off the surfaces of Saturn's retrograde irregular moons, while the iron sulfide-rich component is plausibly exogenic to the Saturn system. Iron sulfides and silicates have not been found to provide optimal fits to spectra of  the C ring, and less so of the A and B rings. However, any measurement of the composition of non-icy material is of interest. 
\par    
More recently, \cite{Lintietalinprep} performed compositional analysis of silicate grains detected by CDA during the Grand Finale that were traced to the B, C, and D rings. They found that the compositions were homogeneous, independent of the source region, and that the grains were Fe-poor in comparison to chondritic materials, as well as in comparison to the \cite{Fischeretal2018} analysis. They suggest these results may point to chemical fractionation occurring in the rings that may lead to the small, Fe-poor silicate grains that were sampled, and larger, Fe-enriched grains that may not have been measured by in situ techniques. 

\subsubsection{Neutral absorber}\label{neutralAbsorber}
As noted above, the water-ice band depths in the C ring and Cassini Division are low in comparison with those in the A and B rings, as seen in Fig.~\ref{fig:ring_profiles}.  
At the same time, the UV slope is shallower, which makes the rings' visible colors less red. 
These changes in the UV, visible and near-IR spectrum are accompanied by a decrease in the visual albedo in these same spectral regions that has generally been attributed to the addition of a spectrally-neutral absorber from an external source \citep{CuzziEstrada1998}.
The higher concentrations of this material in the C ring and Cassini Division can then readily be explained in terms of these regions' lower mean surface mass densities and thus greater susceptibility to `contamination' by non-icy extrinsic material \citep[e.g. ][]{Filacchioneetal2014}.
\par
This neutral absorber has no specific spectral signature, but on cosmogonic grounds it has usually been identified either as  carbonaceous material such as in carbonaceous meteorites or comets, amorphous elemental carbon from the {\it in situ} breakdown of accreted organic material, or perhaps silicates such as in stony meteorites. While most recent modellers assume this material is carbonaceous in nature, silicates remain a possibility.
Future analyses of spectral data from the CIRS FP3 channel ($9-17~\mu$m) might permit detection of the characteristic emission or absorption features due to silicates in the $10-11~\mu$m region, but will be challenging due to the relatively low ring temperatures and poor signal-to-noise ratio of these data \citep{Edgingtonetal2008}.

\subsubsection{Spectral evidence for organic material}\label{organicsSpectral}

\par
The question remains whether the neutral absorber in the rings (and probably also the UV absorber) is predominantly some sort of carbonaceous material, perhaps in the form of tholins, polycyclic aromatic hydrocarbons (PAHs) or amorphous carbon; or a combination of silicates and oxidized iron.
It seems likely that this issue can only be finally settled by analysis of {\it in situ} data such as that discussed in Section \ref{inflowIntro} below, but there are some additional remote sensing observations that bear on this issue. 

\par
\begin{figure}
    \centering
    \resizebox{4.5in}{!}{\includegraphics{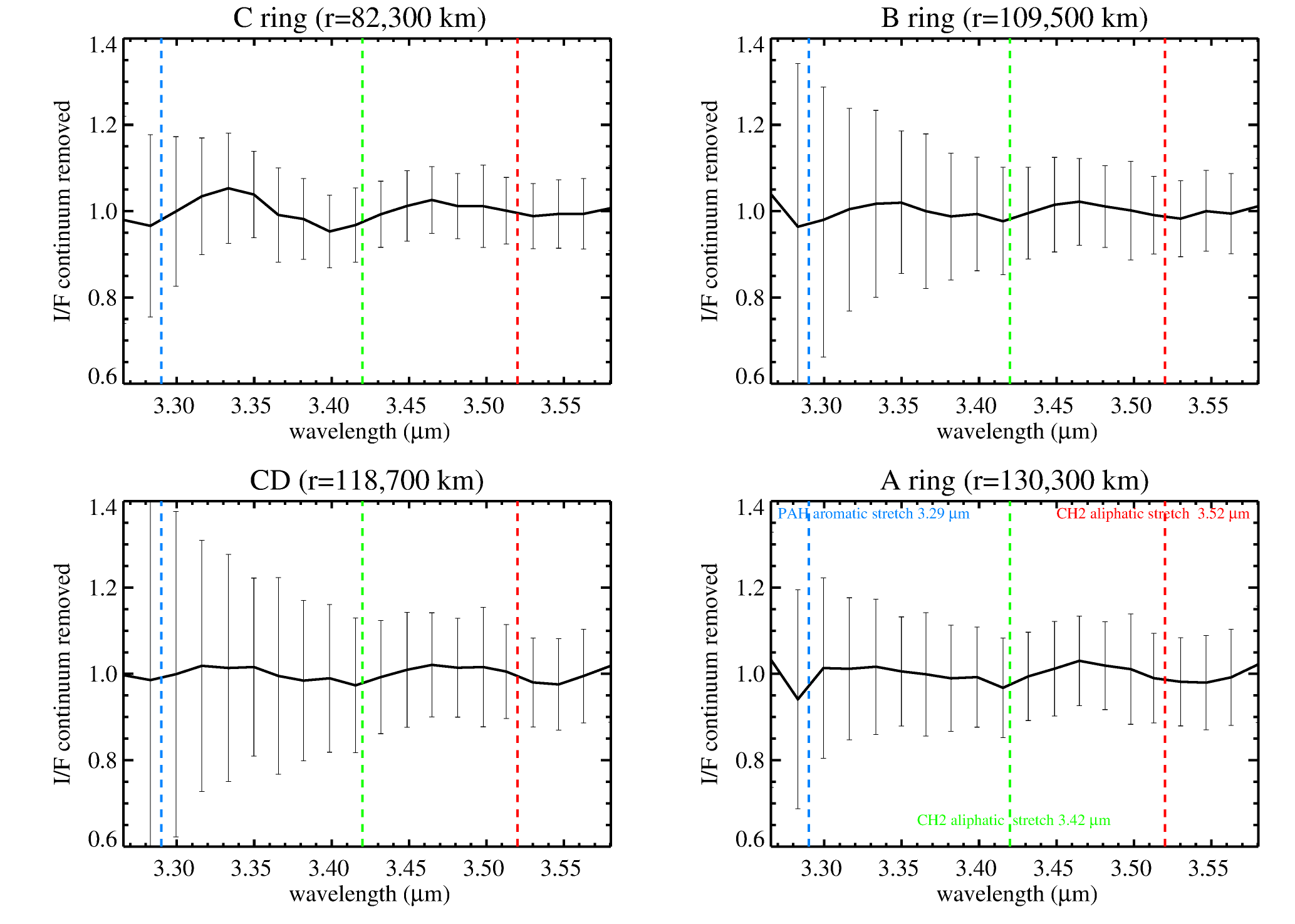}}
    \caption{Continuum-removed IR reflectance spectra by VIMS showing the two aliphatic hydrocarbon absorption features at 3.42–3.52 $\mu$m across the A, B, and C rings and CD. From \citep{Filacchioneetal2014}.}
    \label{fig:rings_aliphatic_bands}
\end{figure}

In their study of Cassini VIMS spectra of the rings, which covered a wide range of phase, incidence and emission angles, \cite{Filacchioneetal2014} reported the detection of the fundamental --- but very weak --- absorption bands due to CH bond stretching. 
The analysis, performed on a low phase ($\approx6^\circ$), high signal-to-noise ring mosaic containing a statistically significant dataset ($\gt$8,000 individual spectra), allowed the detection of faint aliphatic CH$_2$ bands at 3.42 and 3.52 $\mu$m (Fig. \ref{fig:rings_aliphatic_bands}). 
Conversely, the other main spectral feature associated with organic matter and particularly with C-C or aromatic bonds in polycyclic aromatic hydrocarbons (PAHs) at 3.29~$\mu$m\footnote{\cite{DalleOreetal2012} report the aromatic mode at 3.284 µm. The positions of the 4 aliphatic bands of CH$_2$-CH$_3$ symmetric-asymmetric are in fact not fixed, but have a small range of variability ($~20$ nm) depending on the specific molecule.}, which was detected by VIMS on the dark terrains of Iapetus, Hyperion and Phoebe \citep{Coradinietal2008, DalleOreetal2012, Cruikshanketal2007, Cruikshanketal2008}, is not recognizable in VIMS rings spectra. 
These aromatic features are thought to be produced in warmer, higher-energy environments such as the protosolar or proto-Saturnian nebula. This feature seems to be absent in the rings, though its proximity to the very strong water ice absorption at $\sim3.0~\mu$m makes any detection very difficult, if not impossible.
\par
\begin{figure}
    \centering
    \resizebox{4.5in}{!}{\includegraphics{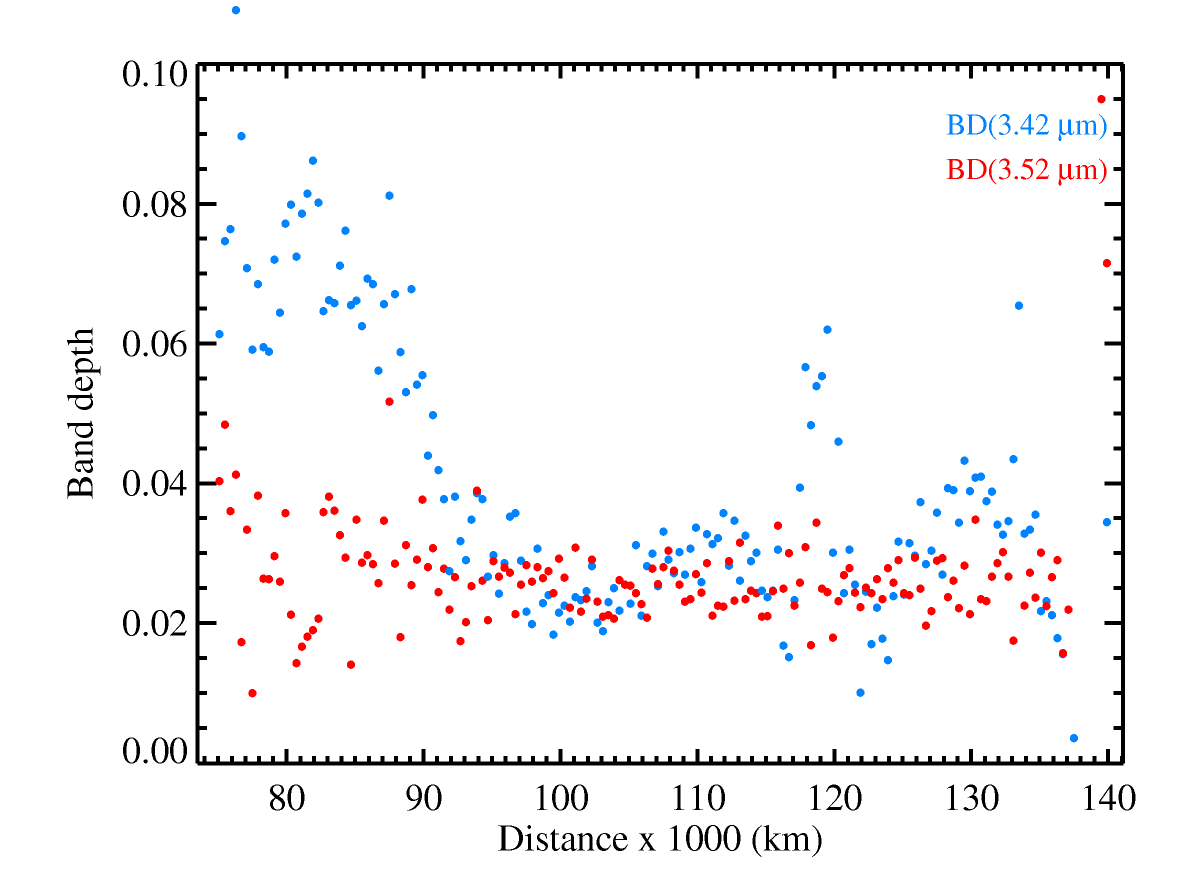}}
    \caption{Radial profiles showing the variability of the aliphatic hydrocarbon band depths at 3.42–3.52 $\mu$m. From \citep{Filacchioneetal2014}.}
    \label{fig:rings_aliphatic_profile}
\end{figure}

The detection of the aliphatic features across the rings and the absence of the aromatic band led  \cite{Filacchioneetal2014} to suggest an ultimate origin for this material in the diffuse interstellar medium where aliphatic bonds outnumber the aromatic ones \citep{PendletonAllamandola2002, Raponietal2020}. Although seen in all major ring regions, the 3.42~$\mu$m band was stronger in the C ring and in the Cassini Division (Fig. \ref{fig:rings_aliphatic_profile}), implying a likely association with accumulated exogenous, interplanetary debris that is now suspected to originate in the Kuiper Belt \citep{Kempfetal2023}. This spatial distribution matches very well with the quantitative spectral fits reported by \cite{Ciarnielloetal2019}, who found higher abundances of organics in the C ring and Cassini Division (see Tables \ref{tbl:ring_models} and \ref{tab:ring-comp-summary} below).

\par
A careful study of the B ring by \cite{Clarketal2019}, on the other hand, failed to find any evidence for an aliphatic band at the 0.2\% level, after the co-addition of almost 22,000 individual VIMS spectra from regions selected to be uncontaminated by saturnshine.\footnote{Reflected light from the planet is dominated by strong methane bands, and thus carries an imprint of a similar C-H stretch feature at $\sim3.3~\mu$m.}
One way to reconcile these two observations, of course, is to attribute the CH band to the `neutral absorber' brought in by interplanetary bombardment. As noted above, this aliphatic material is more visible in parts of the rings with lower optical depths such as the C ring or Cassini Division, and much less abundant in the A and B rings.

\subsubsection{Other ices}\label{otherIces}

Most common ices ({\it e.g.,} CO, CO$_2$, NH$_3$ and especially N$_2$ and CH$_4$) are too volatile to survive over geological timescales in Saturn's rings. Of course, merely being volatile is no guarantee of escape from Saturn's gravity field, but the gaseous species are likely to be photoionized and then effectively swept away into the magnetosphere on seasonal timescales \citep{Cuzzietal2009}.
The detection of other, less-volatile organic ices is again complicated by the presence of saturnshine, which imprints strong CH$_4$ bands in the rings' spectra except at large phase angles or at longitudes well beyond the terminator. 

Nevertheless, CO$_2$ or NH$_3$ ice should be readily detectable, if they are present.  
Models of the near-UV spectrum suggest that there is very little or no NH$_3$ ice in the rings \citep{Cuzzietal2018}, while near-IR spectra of the rings show no sign of the strong CO$_2$ feature at $4.2~\mu$m seen on Iapetus and several other mid-size icy satellites \citep{Clarketal2012, Filacchioneetal2014}.

\section{Regolith radiative transfer modeling of the main rings}\label{RRTModels}
\subsection{Overview of regolith radiative transfer modeling}\label{subsec:caveats}
Traditional regolith radiative transfer (RRT) modeling, following either \citet{Hapke1993, Hapke2013} or \citet{Shkuratovetal1999, Shkuratovetal2005, Shkuratovetal2012},  treats regolith grains on the surface of a moon or an individual ring particle as independent, isolated scattering objects with some ``single-scattering" albedo and phase function. The basic versions of these approaches presume the particles have no internal structure on scales of a wavelength; being internally uniform they are treated using some average real and imaginary refractive indices as a function of wavelength. 

\begin{figure}
    \centering
    \resizebox{4.5in}{!}{\includegraphics{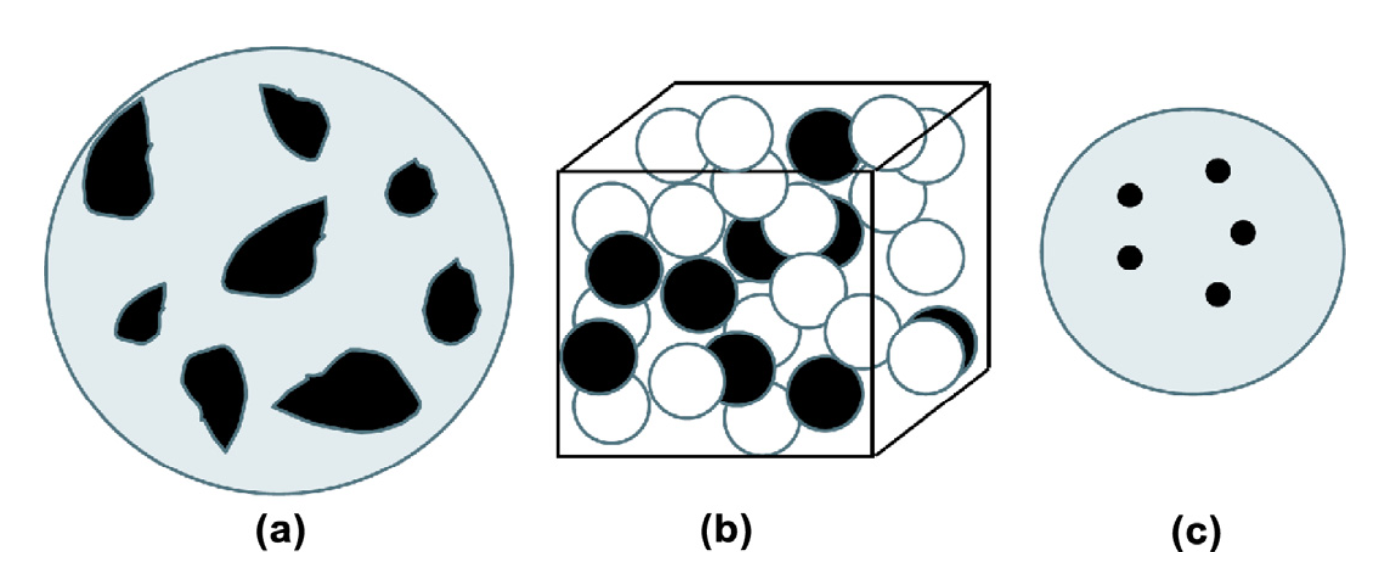}}
    \caption{Schematic diagram of different modalities for two-component mixtures. (a) areal mixture, blue circle represents instrument field of view; (b) intimate mixture; (c) intramixture, blue circle represents a single grain. From \cite{Ciarnielloetal2011}.}
    \label{fig:two-component-mixtures}
\end{figure}

This approach has frequently been adapted to two similar-sounding regolith structures, which however have important differences from the standpoint of compositional interpretation. One regolith structure is the so-called {\it intramixture} in which tiny particles of different materials - much smaller than the wavelength of interest - are embedded in regolith grains of some primary material (Figure \ref{fig:two-component-mixtures}). The refractive indices of these mixed grains are treated with Effective Medium Theory  \citep[see][for reviews and references]{BohrenHuffman1983, Cuzzietal2014b}. On the other hand, a physical mixture of grains having multiple different sizes and/or compositions is called an {\it intimate mixture}, like salt and pepper or grains of sand on a beach. Traditional Hapke theory provides standard ways to treat this heterogeneous mix \citep{Roush1994, HendrixHansen2008}, using an average grain albedo and phase function calculated and adopted for the entire mixture, with different components weighted by their cross sections. RRT models can even include combinations of both kinds - an intimate mixture of grains containing different kinds of intramixtures \citep{Cuzzietal2018}. In the even more advanced adaptations these regolith grains can themselves have {\it internal scatterers}, distinct from any internal intramixed composition \citep[Chapter 6]{Hapke1993}. The most basic assumption is that the regolith mixture has a single typical regolith grain size. However, multiple size distributions, different size distributions for different compositions, or other refinements are often incorporated, all at the cost of additional parameters \citep{Pouletetal2003, Clarketal2012}.
\par
A complication of realistic RRT, is that longer wavelengths may be insensitive to scattering by grains much smaller than the wavelength, instead selecting out grains (or clusters of grains) with size perhaps closer to the wavelength in question to interact most strongly with. This may be why RRT analyses often tend to infer ``optimum" regolith grain sizes that increase with wavelength. For example, compare VIMS  implications of particle size in figure 15.27 of \citet{Cuzzietal2009} with those from UVIS analyses by \citet{Bradleyetal2010}. This effect is likely to be most important in the analysis of observations covering wide spectral ranges, from perhaps fractions of a micron well into the IR. \citet{Pouletetal2003} dealt with this problem by assuming a broad regolith grain size distribution, not unrealistic in general. 

Intramixures tend to maximize the absorbing power per unit mass of nonicy material. Grains of even strongly colored material that is highly absorbing (large imaginary index) can become totally absorbing over wide ranges of wavelength at fairly small sizes, suppressing any wavelength dependence (i.e. becoming less strongly colored). Tholins are good examples of this. While their imaginary indices are strong functions of wavelength through the visible and NIR, grains of the pure material quickly become so opaque that their characteristic reddish intrinsic color caused by strong wavelength dependence is lost if they are more than several microns in size. This drove \citet{CuzziEstrada1998} to adopt a tholin-in-ice intramixing model to fit the very strong redness of the A and B rings in Voyager imaging data. The same approach was used by \citet{Pouletetal2003}, and \citet{Cuzzietal2018} still find that this approach provides the best fits to HST-STIS ring spectra.
\par
Even for material with strong but wavelength-independent intrinsic absorption, large grains become so opaque at even small sizes that their darkening power becomes proportional to their surface area, not their volume, weakening the absorption per unit mass as they get larger. An example of this might be seen in comparing the fraction of carbonaceous material in the rings (the so-called neutral absorber) inferred by \citet{Cuzzietal2018} using intramixtures (see Table \ref{tbl:cuzzietal_HST_models}) with the values inferred by \citet{Ciarnielloetal2019} using intimate mixtures assuming 10 $\mu$m radius grains (see Table \ref{tbl:ring_models}). The fraction of nonicy absorber inferred by \citet{Ciarnielloetal2019}  is considerably larger than by \citet{Cuzzietal2018}, probably because the assumed carbonaceous grain sizes of \cite{Ciarnielloetal2019} are larger. Because inferences about the age and origin of the rings depend sensitively upon the fraction of nonicy material, these model-based uncertainties should be kept in mind. 
\par
One possible way to distinguish these different RRT hypotheses is to compare the above  abundance inferences (both at visible and near-IR wavelengths, that sample only the very surfaces of ring particles) with cm-wavelength radiometry that samples the bulk of the ring particles. At the long wavelengths of microwave emission and scattering, intimate mixtures containing purely absorbing grains that are tens of microns or more in size behave like intramixtures. At present, the abundance of nonicy material over most of the rings inferred from 2.2 cm radiometry by \citet{Zhangetal2017AB, Zhangetal2017C, Zhangetal2019}  seems to be closer to the results of \citet{Cuzzietal2018} than those of \citet{Ciarnielloetal2019}, possibly providing support for the lower fractions of nonicy material obtained from the intramixing model.

Finally, a recent wrinkle has emerged regarding ring RRT, that the planetary surfaces community has long been aware of: unresolved macroscopic shadows on/in the surfaces being observed make a surface look darker. If this effect is ignored, one infers an abundance of darkening material that is too large. It would not be unreasonable for ring particles, if they have a grape-cluster aggregate nature, to be even more strongly affected by this shadowing effect than other smoother surfaces. \citet[Chapter 12]{Hapke1993} has advanced a first-principles, geometrically-inspired, forward model along these lines, with several parameters.  The role of shadowing in the rings context is also discussed by \citet[see also references therein]{Cuzzietal2017}, where a simple one-parameter model is advanced to modify a basically Lambertian underlying smooth particle phase function. This simpler model turns out to be fairly similar in properties to the more complicated forward model of \citet{Hapke1993}.\footnote{As the Hapke model is parameterized assuming shadows are formed by craters, one might not expect the quantitative results (slopes) to be meaningful for clumpy aggregate ring particles.} 

In the approach suggested by \citet{Cuzzietal2017}, the particle phase function is measured using observations over a wide range of viewing and illumination geometries. The shadowing behavior is manifested in the shape of the observed phase function, and this phase function defines the correction from the directly observed rough-surface, shadowed, effective ring particle albedo that rings analyses normally discuss, to a theoretical  smooth-particle albedo with a Lambertian phase function and composition unbiased by shadowing. This is most easily done in low-optical-depth regions where multiple scattering is not a complication, but can be incorporated into a proper multiple-(ring particle) scattering model of arbitrary optical depth. The smooth particle albedo can then be modeled in terms of its underlying composition by the simplest forms of Hapke or Shkuratov theory. Of course, all of the above caveats apply as well to ring moon surfaces as to ring particle surfaces. 

One prediction of this approach is that the shadowing parameter characterizing the observed particle phase function depends on the intrinsic, smooth-surface reflectance of the particle material - with particles appearing ``smoother" at wavelengths where they are intrinsically brighter because shadows are filled in by nearby lit facets. A look at recent phase functions from icy moons provides some support for this hypothesis \citep{Pitmanetal2010}, but more comparisons need to be done for the probably much craggier ring particles. 

\footnote{An undergraduate class could test this conjecture in a large, empty, blacked out room with a flashlight and a light meter.} 

Along these lines, \citet{Cooke1991} contains some carefully derived and highly relevant results, even if only in the Voyager clear filter. For example, the particle phase functions in all the outer C ring plateaux, and certain parts of the very inner C ring, are steeper (suggesting stronger shadowing, requiring a larger albedo correction) than the phase function in the mid-C Ring. Currently, the implications of this effect for the underlying intrinsic ring particle material are being explored by Cuzzi et al (2023, in prep) using C ring regions intentionally chosen to be complementary to those of \citet{Cooke1991} but covering Cassini UV3, BL1, and RED filters, as well as by \citet{ElliottEsposito2021} over all the rings using the basic Hapke approach. 

\subsection{Compositional inferences from modeling}\label{regolithModels}

Having described the approach, terminology, and caveats generally associated with spectral modeling in Section \ref{subsec:caveats}, we summarize here models of ring composition to date, as well as their primary conclusions.

\begin{figure}
    \centering
    \resizebox{4.5in}{!}{\includegraphics{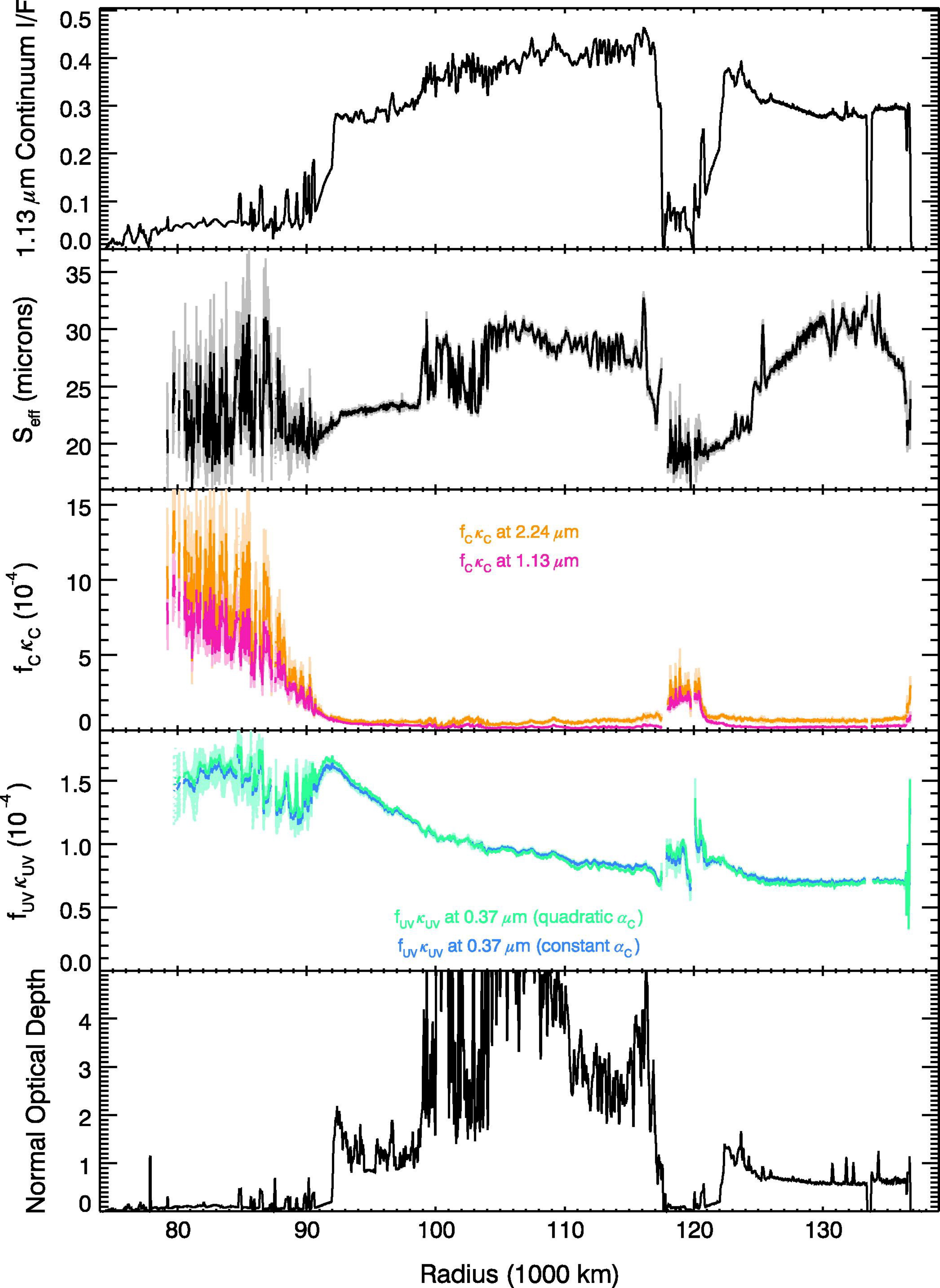}}
    \caption{Radial profiles of spectral parameters for the rings of Saturn derived from selected Cassini-VIMS observations and the 1D light-scattering model of \cite{Shkuratovetal1999}.  
    The upper panel shows the low-phase reflectivity of the rings at the continuum wavelength of 1.13~$\mu$m,
    The second panel shows the effective scattering length in the regolith, which may be loosely interpreted as an average `grain size'.
    The third and fourth panels show the volume-weighted absorption coefficients of the continuum (or neutral) and UV absorbers, respectively. The lower panel shows an optical depth profile of the rings obtained from an occultation of the star $\gamma$~Crucis, binned to 10~km resolution. From \cite{Hedmanetal2013}.}
    \label{fig:spectral_profiles}
\end{figure}

In one of the first attempts to apply a physics-based light-scattering model to the VIMS data, \cite{Hedmanetal2013} modeled a selected subset of VIMS spectra using the 1D reflectance model of \cite{Shkuratovetal1999}. 
Their principal results, shown in Figure~\ref{fig:spectral_profiles}, indicate that while effective scattering lengths (a surrogate for grain size) are larger in the outer A and B rings than in the C ring or Cassini Division, the volume fraction of the UV absorber is in fact similar in the Cassini Division to that in the A and outermost B ring.  The volume fraction of the UV absorber then increases monotonically inwards across the B ring, before decreasing again in the outer C ring.  
The above behavior supports the idea that the UV absorber is uniformly mixed with the water ice and is thus probably embedded within the icy grains, but also suggests that its volume mixing ratio generally increases with proximity to Saturn. In the Hedman models this component also appears to have a local maximum in the middle C ring, although some caution is appropriate here, as described below. 
\par
The model of \citet{Hedmanetal2013} requires all the observed spectral properties to correspond to one of their three components: water ice, UV absorber, or neutral absorber. If the reality is more complex than three components, this approach may lead to analytical artifacts. 
For example, Fig.~14 of \citet{Cuzzietal2002} shows very similarly increasing short-wavelength reddening inwards to the B ring edge, but nothing similar in the mid-C ring. 
Given the likelihood of excess non-icy material of {\it some kind} in the mid-C Ring (see Section \ref{bulk}), inclusion of an additional component in the model to account for rocky material may be more appropriate. 
Similarly, the \citet{Hedmanetal2013} model ascribes a considerable amount of fine-scale photometric structure in the mid-B ring to grain size variations that affect spectral slopes. However, reality is more complex than any model, and there may well be other radial variations on those scales that became convolved into the grain size parameter. 
\par

\par
The models shown in Figure~\ref{fig:spectral_profiles}
confirm the much higher concentration of a neutral absorber within the Cassini Division and especially the 

C ring, where its volume fraction is $5-10$ times higher than in the A and B rings. It is noteworthy that the abundance of the neutral absorber seems to increase monotonically inwards across the C ring, whereas the microwave measurements show a peak in the abundance of the nonicy material near the mid-C ring (Section \ref{bulk}), much like that of the UV absorber shown here. These two different radial trends may further support different origins for the UV and neutral absorbers, and a link between the UV absorber and the non-icy material from microwave observations. The models of \cite{Hedmanetal2013} shown in Figure~\ref{fig:spectral_profiles} do not attempt to identify either the UV absorber or the neutral absorber, only their spatial variations.
\par

\cite{Clarketal2012} pointed out the similiarity in shape and central wavelength of the strong water ice band at $\sim3.0~\mu$m, and in the visible wavelength spectral slope, between the dark regions on Iapetus and other satellites with properties of the C ring and Cassini Division. They suggested on the basis of this similarity that the best overall spectral fit, including the UV slope and infrared water bands, may be obtained with nano-scale inclusions (i.e. an intramixture) of iron and/or iron-oxides, rather than the so-called `ice-tholins'. However, significant overall spectral and albedo differences between the satellites and the rings preclude any obvious similarities in the composition of even the C Ring, much less the A and B rings, with these dark regions \citep[e.g. figures 15-34 and 15-37 in][]{Cuzzietal2009}. In addition, recent spectral observations of the satellites in question actually find better fits from carbon-tholin mixtures than iron-hematite mixtures (see chapter by Ciarniello et al, this volume). 
More recently, spectral fits to UV-visible STIS data from the Hubble Space Telescope that bridge the gap between the Cassini UVIS and VIMS datasets indicate that a good match to the steep UV slope seen in the A and B ring spectra can be obtained with tholins\footnote{
Physically, the decrease in reflectivity is attributed to a $\pi-\pi^\ast$ transition in the carbon rings common in tholins and other PAHs \citep{JaffeOrchin1962,  Clar1964, Birks1970, Salamaetal1996,  Mallocietal2004}.} 
embedded in fine-grained water ice \citep{Cuzzietal2018}.
Attempts to fit this region of the rings' spectra using nanophase iron-bearing materials were found to be unsatisfactory.
These models are discussed further below, and illustrated in Figure~\ref{fig:cuzzietal_HST_spectral_fits}.

\par

\begin{figure}
    \centering
    \resizebox{4.5in}{!}{\includegraphics{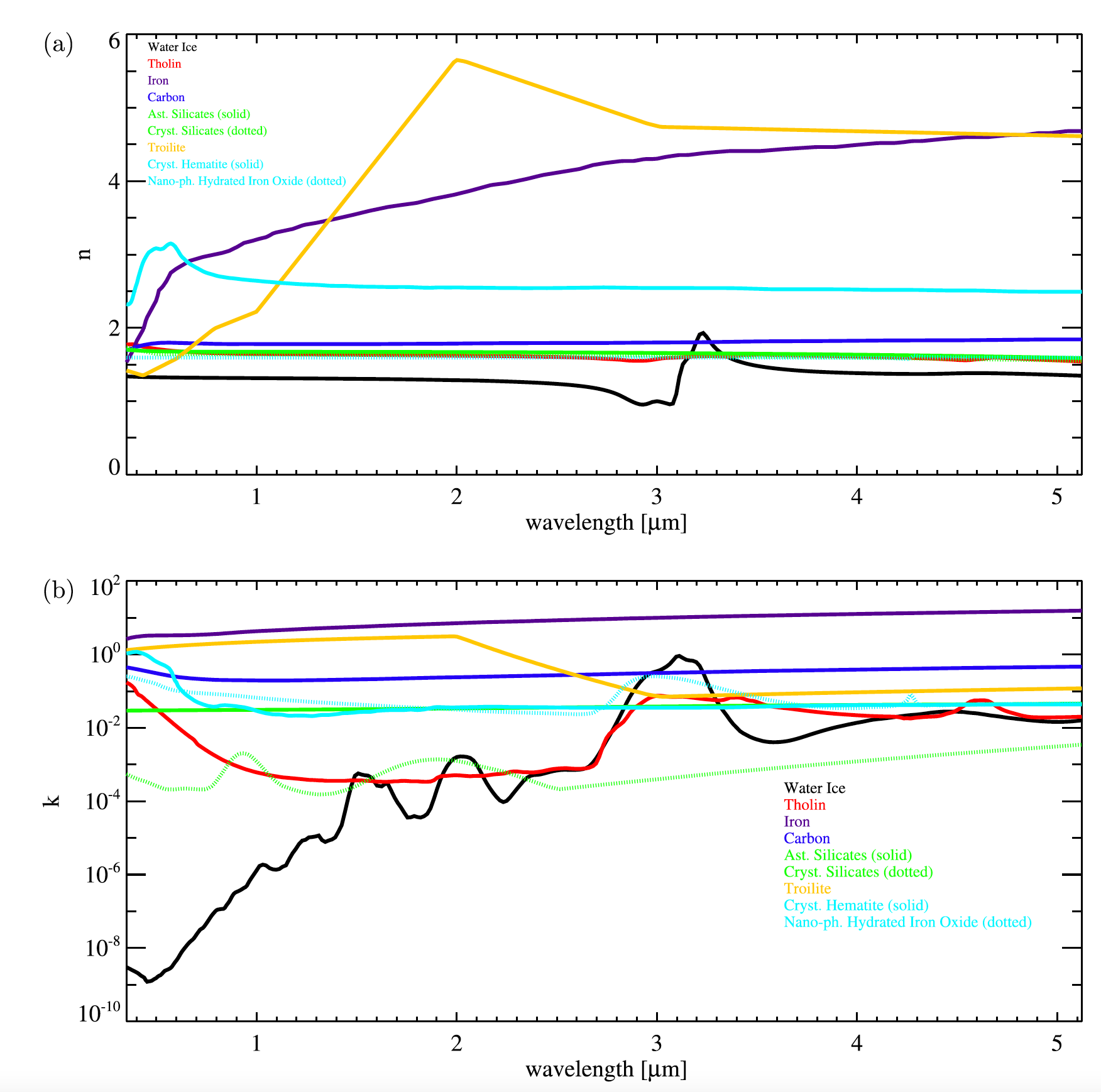}}
    \caption{Real (top) and imaginary (bottom) part of the refractive index of different
materials used for ring spectral modeling. 
Note the decreasing absorption coefficient of water ice at shorter wavelengths, while those
of tholins, hematite, and (to a lesser extent) carbon are rapidy increasing.
From \cite{Ciarnielloetal2019}.}
    \label{fig:ciarniello_absorption_coeffs}
\end{figure}

Additional clues as to the composition of the non-icy component(s) of the main rings are provided by more sophisticated modeling of the VIMS spectra of Saturn's rings and moons based on Hapke's theory of light-scattering in planetary regoliths \citep{Hapke2012}.
\cite{Ciarnielloetal2019} showed that the main ring spectra can be reproduced by intramixtures of water ice grains with inclusions of organic materials providing the UV absorption \citep[Titan tholins;][]{Imanakaetal2004, Imanakaetal2012, Westetal2014}, intimately mixed with sub-micron water ice grains and a variable abundance of other compounds contributing as neutral absorbers, either in pure form (amorphous carbon or amorphous silicates) or embedded in water ice (nanophase hydrated iron oxides, carbon, silicates, crystalline hematite, metallic iron, troilite). Corresponding optical constants are shown in  Figure  \ref{fig:ciarniello_absorption_coeffs}.

Given the lack of diagnostic VIS-IR features, the nature of the neutral absorber and its corresponding absolute volumetric abundance cannot be unequivocally determined remotely.  
The latter in fact depends on the particular compound selected and on the corresponding grain size. 
In this respect, the model VIS-IR spectrum of opaque and spectrally-bland materials is virtually independent of the grain size, provided they are large enough to completely absorb light passing through them. 
This prevents a robust grain size characterization, thus potentially biasing the inferred abundance of the corresponding compound.  
\par
Nonetheless, assuming a common composition and similar properties of the neutral absorber across the rings, results from spectral modeling can still be interpreted in relative terms. 
Keeping this in mind, \cite{Ciarnielloetal2019} determined the abundance of the non-icy materials across regions of the main rings characterized by different optical depths, under the assumption that the neutral absorber is represented by 10 µm-sized amorphous carbon grains. 
For the A and B rings the best unmixing fits are achieved by assuming that water ice grains intramixed with tholins are intimately mixed with submicron (0.2 $\mu$m) water ice grains and 10~$\mu$m sized carbon grains. 
\par
\begin{figure}
    \centering
    \resizebox{4.6in}{!}{\includegraphics{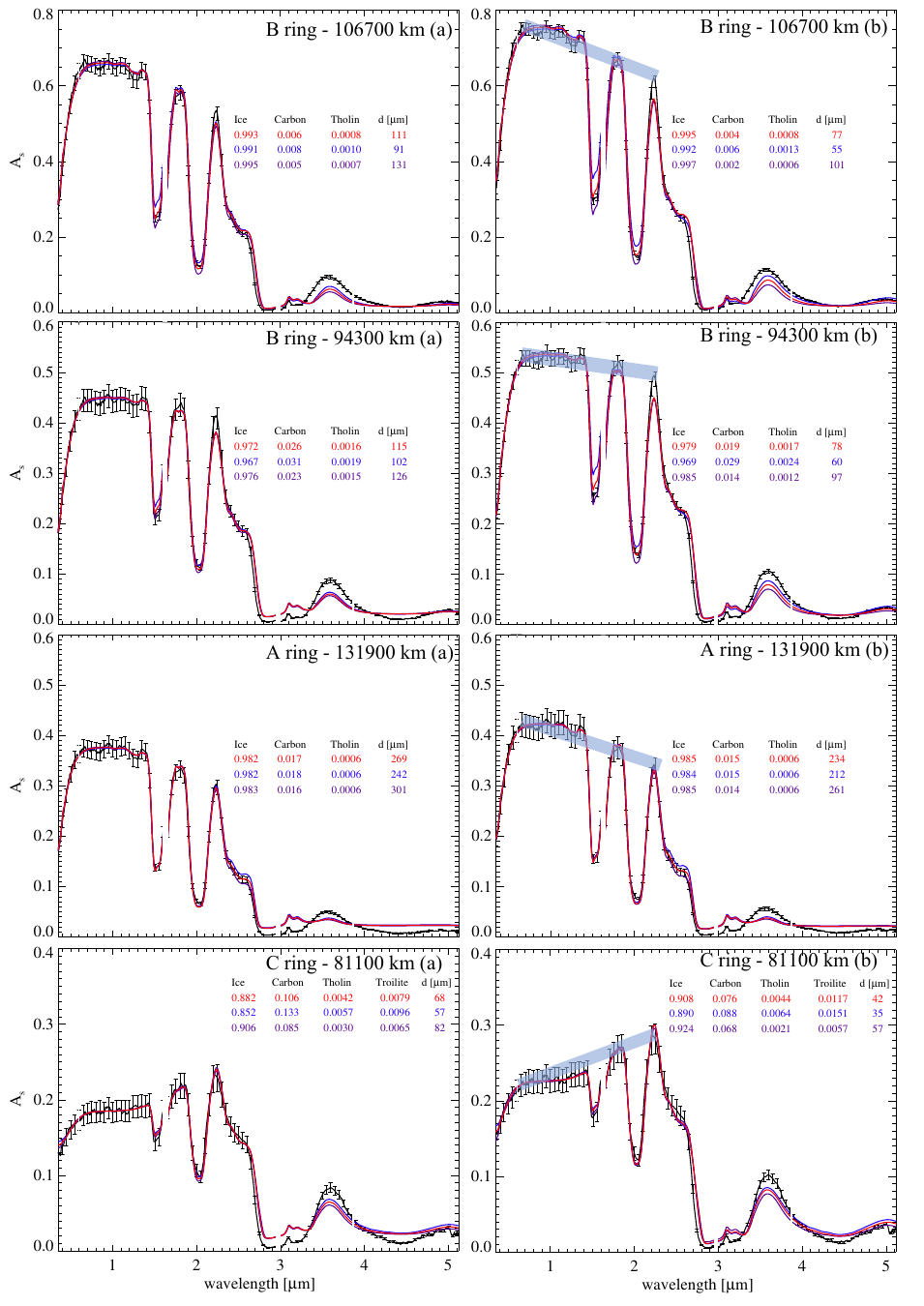}}
    \caption{Bond albedo spectra of Saturn's main rings derived from Cassini VIMS observations (black lines) and the best-fitting spectral models (red curve: best fit; blue and violet curves: models with accuracy comparable to the best fit). For each ring region two solutions for the Bond albedo spectra are shown (a,b), as resulting from ring photometric modeling performed in \cite{Ciarnielloetal2019}.  
    Missing parts of the spectrum indicate instrumental order sorting filters. 
    Note the positive continuum slope in the C ring between 0.6 and $2.3~\mu$m, compared with the flat or blueish slope in the A and B rings, characteristic of mostly pure water ice. Adapted from \cite{Ciarnielloetal2019}.}
    \label{fig:ciarniello_spectral_fits}
\end{figure}
\begin{figure}
    \centering
    \resizebox{4.5in}{!}{\includegraphics{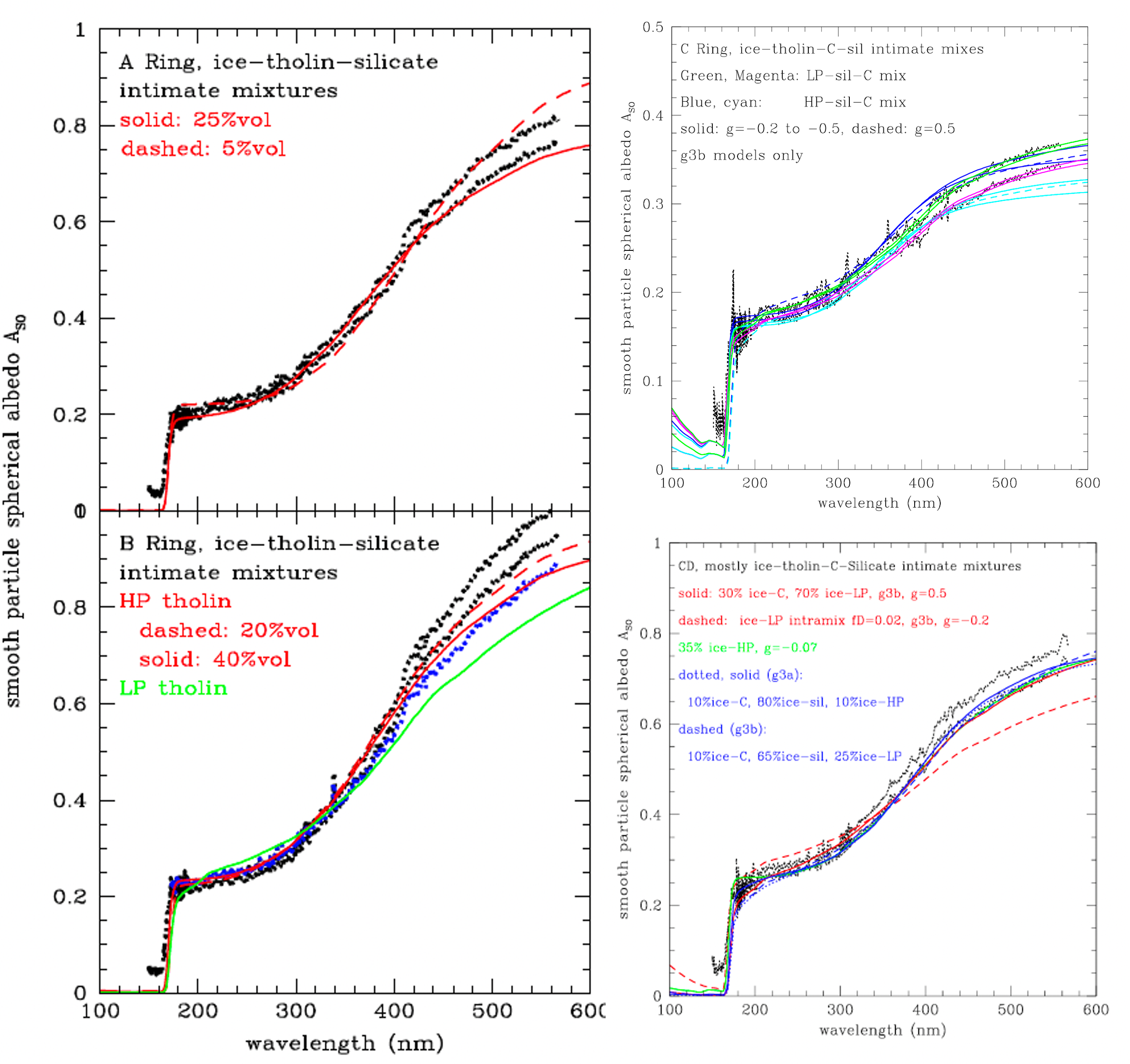}}
    \caption{``Smooth particle" Bond albedo of Saturn's main ring (A, B, C, CD) particles derived from HST observations (heavy and light dotted curves, representing two different particle shadowing parameters) along with the best-fitting spectral modeling results (colored curves).Several variants of Hapke models are shown, to provide a sense of potential model-based uncertainties. The parameter $g$ is the assumed regolith grain scattering asymmetry, with negative values referring to backscattering grains and positive values to forward scattering grains. Models denoted g3a and g3b are slightly different ways Hapke theory can be implemented. For details of the models see \citet{Cuzzietal2018}.
   }
    \label{fig:cuzzietal_HST_spectral_fits}
\end{figure}

For the C ring it is necessary to add a further population of water ice grains containing troilite inclusions (intramixture) in order to match the higher broad-band absorption and to introduce a positive continuum spectral slope across the whole VIS-NIR spectral range.
Their findings are summarized in Table \ref{tbl:ring_models} and in Fig. \ref{fig:ciarniello_spectral_fits}, and indicate that the content of tholin increases inwards  across the main rings, while the fraction of darkening materials (amorphous carbon and troilite) is anti-correlated with optical depth. (See Section \ref{subsec:caveats} however, for a cautionary note about a possible role for on-particle shadowing in lowering apparent particle albedo.)

\begin{table}[h]
\footnotesize
\begin{center} 
\caption{\textbf{Inferred composition of Saturn's main rings from spectral modeling of VIMS data by \cite{Ciarnielloetal2019}}. 
The models assume water ice grains with embedded tholin (ice-tholin  intramixture) in intimate mixture  with sub-micron water ice grains ($0.2~\mu$m size) and amorphous carbon grains ($10~\mu$m size). 
The C ring model requires an additional population of water ice grains with embedded troilite, whose size is assumed equal to that of the ice-tholin population (indicated as icy-grain size). 
The possible ranges in the total percentage of volumetric abundance of the different constituents are reported. }
\label{tbl:ring_models} 
\begin{tabular}{cccccccc} 
\toprule
Ring & Radius & $\tau$ & Carbon & Tholin & Troilite & Water & Icy-grain \\
region& [km] & & & & & ice & size [$\mu$m]\\
\midrule
outer A ring & 131900 & 0.6 & 1.39-1.78 & 0.061-0.063 & - & 98.15-98.54 & 212-301\\ 
mid B ring & 106700 & 4.6 & 0.24-0.81 & 0.13-0.57 & - & 99.09-99.70 &  55-131\\
inner B ring & 94300 & 1.4 & 1.42-3.08 & 0.12-0.24& - & 96.73-98.46  & 60-126\\ 
mid C ring & 81100 & 0.1 & 6.76 - 13.30 & 0.21-0.64 & 0.57- 1.5 & 85.18-92.45& 35 - 82\\ 
\botrule
\end{tabular}
\end{center}
\end{table}

\par
These results are consistent with the trends found by \cite{Hedmanetal2013} for the UV and the neutral absorber, respectively (see Figure~\ref{fig:spectral_profiles}).
Also, the inferred size of the icy grains tends to be smaller in the C ring, in agreement with the scattering lengths or regolith grain sizes estimated by \cite{Hedmanetal2013} and \cite{Filacchioneetal2014}.
\par 
\par

The models discussed above were based entirely on VIS-IR data obtained by the Cassini-VIMS instrument that span the wavelength range $0.35 - 5.2~\mu$m.
By extending such physical models for the ring particle reflectivity into the UV it is possible to derive further clues about the rings' composition, complementary to those obtained in the VIS-IR region.
The near-UV spectral range, in fact, is very sensitive to the presence of chromophores, or UV-absorbers, including tholins, silicates, nanophase iron and amorphous carbon mixed with the dominant water ice.
\par
\cite{Cuzzietal2018} fitted high-quality HST spectra of different ring regions  using \cite{Hapke1993} theory, with additional corrections made to take into account the shadowing effects caused by the particles' roughness \citep{Cuzzietal2017}. 
Lacking such a correction, the shadows caused by the irregular surfaces of the ring particles darken the particle and can lead typical smooth particle models  to overestimate the actual fractional abundance of darkening material. The main ring spectra were modeled with combinations of intimate and intramixture modalities, of 10~$\mu$m size grains made of water ice, amorphous carbon, tholins and silicates. 
Tholins appear only within a fraction ($4-50\%$) of the water ice grains, arguing for a heterogeneous mixture. Here, HP and LP refer to High Pressure- and Low-Pressure formation regimes \citep{Imanakaetal2004, Imanakaetal2012,Westetal2014}. The LP tholins tend to have more aromatic bonds and the HP tholins tend to have more aliphatic bonds \citep{Cuzzietal2018}. 
In general neither iron nor hematite, in nanophase form or in larger grains, was found to provide satisfactory spectral fits to the HST data for the A or B rings. 
Examples of the best fits from \cite{Cuzzietal2018} are shown in Fig.~\ref{fig:cuzzietal_HST_spectral_fits}, while the corresponding model parameters are listed in Table \ref{tbl:cuzzietal_HST_models}. 
   
\begin{table}[h]
\footnotesize
\begin{center} 
\caption{\textbf{Total fractional abundances (F) of tholin, carbon and silicate by ring region.} The total average abundance $F_j$ for component $j$ is given for different ring regions. $F_j$ is the
product of the volume fraction of grains containing component $j$ , multiplied by the
volume fraction $f_j$ of pollutant intramixed within those grains. The range shown
is inferred from best-fitting models across a
wide array of different Hapke-based models that all fit comparably well. This systematic sensitivity testing
provides a sense of the uncertainties in the spectral modeling results. Non-asterisked values come
from the bulk of the simulations in which the grain size for each population is allowed
to vary within a plausible range between 3–10 $\mu$m diameter. The grain size is constrained by the location of the UV edge \citep{Bradleyetal2010}.
Asterisked models assume all constituents have the same
grain size; in general constant grain size models provide worse fits to the data. 
From \cite{Cuzzietal2018}.}
\label{tbl:cuzzietal_HST_models} 
\begin{tabular}{lccc} 
\toprule
Ring & $F_{\rm tho}$ & $F_{\rm carb}$ & $F_{\rm sil}$  \\
\midrule
A & $(4-10) \times 10^{-3}$  & $(2-4) \times 10^{-4}$ & $(0.5-7) \times 10^{-5}$ \\ 
A$^*$ & $9 \times 10^{-3}$  & $1.5 \times 10^{-4}$ & $1.6 \times 10^{-3}$ \\ [0.4em]
B & $(2.5-3) \times 10^{-3}$  & $(0.5-3) \times 10^{-5}$ & $(0.5-3) \times 10^{-5}$ \\ 
B & $1.2 \times 10^{-2}$  & $(2-6) \times 10^{-6}$ & $(0-1) \times 10^{-6}$ \\ 
B$^*$ & $8 \times 10^{-3}$  & $2.5 \times 10^{-7}$ & $1.9 \times 10^{-3}$ \\ [0.4em]
C & $(2-10) \times 10^{-3}$ & $(5-15) \times 10^{-3}$ & $(1-30) \times 10^{-4}$ \\
C$^*$ & $8 \times 10^{-3}$  & $2.6 \times 10^{-2}$ & $2.5 \times 10^{-4}$ \\ 
C$^{'*}$ & $7 \times 10^{-3}$  & $4.5 \times 10^{-2}$ & $2.0 \times 10^{-4}$ \\ 
\botrule
\end{tabular}
\end{center}
\end{table}

 Two fits are given for the B Ring, where we see that the tholin abundance can exceed one percent or so only if the carbon and silicates are both reduced to extremely low levels. 
For the A ring, the upper limit of about one percent tholin also occurs only in models with almost no carbon. 
The main difference between fits with variable (A, B) and constant (A$^*$, B$^*$) regolith grain sizes is that the latter have more silicates --- but still only a fraction of a percent.
For the C Ring, the variable grain size model (non-asterisked) has much more silicate than do the corresponding A and B Ring models. 
Underneath the nominal C ring fit, we show two alternate fits with constant grain sizes for comparison with the A$^*$ and B$^*$ models: one for the nominal rough surface (C$^*$), and one where the particles are assumed to be smooth (C$^{'*}$). 
The latter needs more carbon ({\it i.e.,} darkening agent) than does the nominal case. 
\par
As noted above, the radial distribution of the neutral absorber ({\it i.e.,} carbon or silicates) supports an exogenous origin by meteoroid or cometary bombardment \citep{CuzziEstrada1998}, which more effectively contaminates regions of low surface mass density, while the radial trend of the UV absorber (tholin) may reflect an intrinsic origin in the proto-Saturnian nebula \citep{Cuzzietal2018}. Alternatively, it may be derived from different parts of different precursor moons (see Chapter 12 by Crida et al), or perhaps reflect indirectly the influence of Saturn's magnetic field.
\par

\section{Microwave observations of main ring bulk composition}\label{bulk}

The UV, visible and near-IR spectra, of course, only sample the composition of the outermost layers of the decimeter to meter-sized particles in the main rings, which may differ from the composition of their interiors depending on how effective the mixing processes are \citep{ElliottEsposito2011}.  In order to probe the interiors of the ring particles, and thus the bulk of the ring material, we must turn to the longer wavelengths of the rings' microwave emission spectrum, which penetrate many meters of solid ice because of its very low microwave absorptivity. 
\par
Early ground-based measurements in the 1970s showed that the rings' brightness temperature was quite low at wavelengths longer than a few mm, suggestive either of a very low absorption coefficient (and hence low emissivity) of the ring material, or of average particle sizes less than $\sim1$~cm. 
The rings' very strong S-band radar reflectivity effectively ruled out the latter possibility, leaving the former. With its unusually low radio opacity, pure, cold water ice soon emerged as the most likely candidate \citep{Pollacketal1973, 
PettengillHagfors1974, Cuzzietal1980}, subsequently  supported by the strong water ice bands seen in the near-IR spectrum \citep{ClarkMcCord1980}. 

Interferometric observations by \cite{Grossman1990} suggested a maximum non-icy mass fraction assumed to be silicates in the A and B rings of $\sim1\%$, a value that is much lower than the bulk composition of any of the mid-sized saturnian satellites. This surprising result was not challenged or improved upon until the Cassini mission 25~years later.  
The assumption that silicates are the only nonicy constituent is one of convenience; the microwave absorption properties of all other candidates are less well known. So, the microwave observations should be interpreted as constraining the total amount of nonicy material (silicates, organics, etc). 
\par
The radiometer channel of the Cassini RADAR instrument \citep{Elachietal2004} observed Saturn’s rings at a wavelength of 2.2~cm between 2004 and 2008 with resolutions of order $2000$~km, providing an ideal window through which to study the non-icy material fraction within the bulk of the ring mass. 
At 2.2~cm, the absorptivity of water ice is negligible compared to that of most non-icy materials, and thus the intrinsic thermal radiation from the ring layer is dominated by its non-icy components.
\par
\begin{figure}
    \centering
    \resizebox{4.5in}{!}{\includegraphics{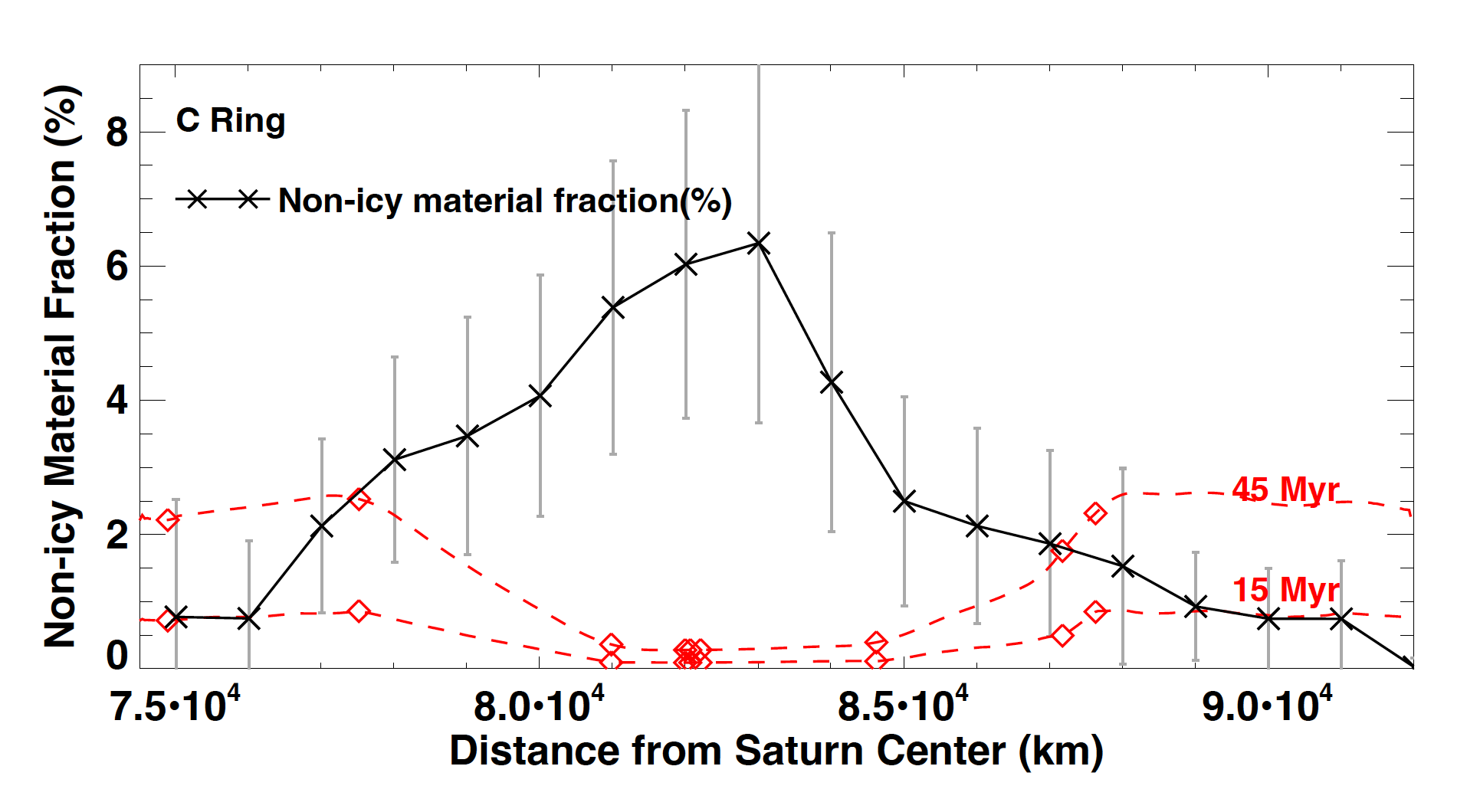}}
    \caption{The black curve shows the non-icy material fraction in the C ring derived from Cassini-RADAR observations. Red dashed curves show the predicted evolution of the non-icy material fraction over 15 and 45 Myr, assuming that meteoroid bombardment is the only source of contamination and that the meteoroid flux has remained constant over the past few tens of Myr. Red diamonds indicate the positions where opacity measurements have been made using density waves. The broad minimum in the predicted nonicy fraction is caused by the small opacities (high surface mass densities) observed in that part of the C Ring \citep{Baillieetal2011, HedmanNicholson2014}. Figure from \cite{Zhangetal2017C}.}
    \label{fig:cringsilicate}
\end{figure}

The Cassini radiometer observations show an exceptionally high brightness when the C ring is in front of Saturn, at near-zero azimuthal angles. Besides directly-transmitted planetary thermal radiation, a large portion of the scattered radiation by the ring particles must be in the forward direction, suggesting a high porosity of 70-75\% for the C ring particles. 
The ring particles could achieve such a quasi-equilibrium, highly  porous regolith as a result of a balance between frequent fragmentation and excavation of debris due to micrometeoroid bombardment, ballistic transport to relocate the debris more gently on other particle surfaces, and inter-particle collisions to compact the particles. 
Most regions in the C ring contain about 1-2\% silicates by volume \citep[Figure~\ref{fig:cringsilicate}; ][]{Zhangetal2017C}.
These results are consistent with an initially nearly pure-ice ring system that has been continuously contaminated by in-falling micrometeoroids over tens or hundreds of Myr \citep{CuzziEstrada1998, Kempfetal2023}, and assuming that the C ring optical depth and surface density has not changed significantly during that time. This absolute time scale is inversely proportional not only to the flux at infinity, but also to the amount of gravitational focusing by Saturn that the micrometeoroids experience before encountering the rings. See Chapter 11 for a more extensive discussion.
\par

An enhanced abundance of non-icy material is found in the middle C ring ( Figure~\ref{fig:cringsilicate}). When assumed to be mixed volumetrically (“intramixed”) 

with the water ice, this enhanced contamination reaches a maximum concentration of 6–11\% silicates by volume at a ring radius of $\sim83,000$~km. It is quite probable, however, that this non-icy material is not well mixed with the water ice, but is instead present as large chunks in the cm to m size range, coated with porous icy mantles. This could naturally account for the low opacity found in the middle C ring \citep{HedmanNicholson2013, Zhangetal2017C}. 
The significantly higher concentration in the middle C ring cannot be explained by direct meteoroid deposition alone. A possible scenario is that the entire C ring has been continuously polluted by meteoroid bombardment since it first formed, while the middle C ring was further contaminated by an incoming Centaur, a rocky object torn apart by tides and ultimately broken into pieces that currently reside in the middle C ring. 
If correct, the spatial extent of the enhanced non-icy material fraction suggests that the Centaur was likely to have been captured and integrated into the rings as recently as $\sim10-20$~Myr ago. 
\par
\citet{Zhangetal2017AB} found non-icy volume fractions of $\sim0.3-0.5\%$ in the inner and outer B ring and as little as $\sim0.1-0.2\%$ in the central region. 
For the A ring interior to the Encke gap, the derived non-icy volume fraction is $\sim0.2-0.3\%$ everywhere.
Finally, the Cassini Division has a non-icy fraction of $\sim1-2\%$, similar to most regions in the C ring. 
The overall pollution exposure times calculated for the A and B rings and the Cassini Division from microwave constraints range from $\sim30-150$~Myr, in line with the $\sim15-90$~Myr derived for most regions in the C ring. 
\par

\section{D ring}\label{dring} 

\begin{figure}
    \centering
    \resizebox{4in}{!}{\includegraphics{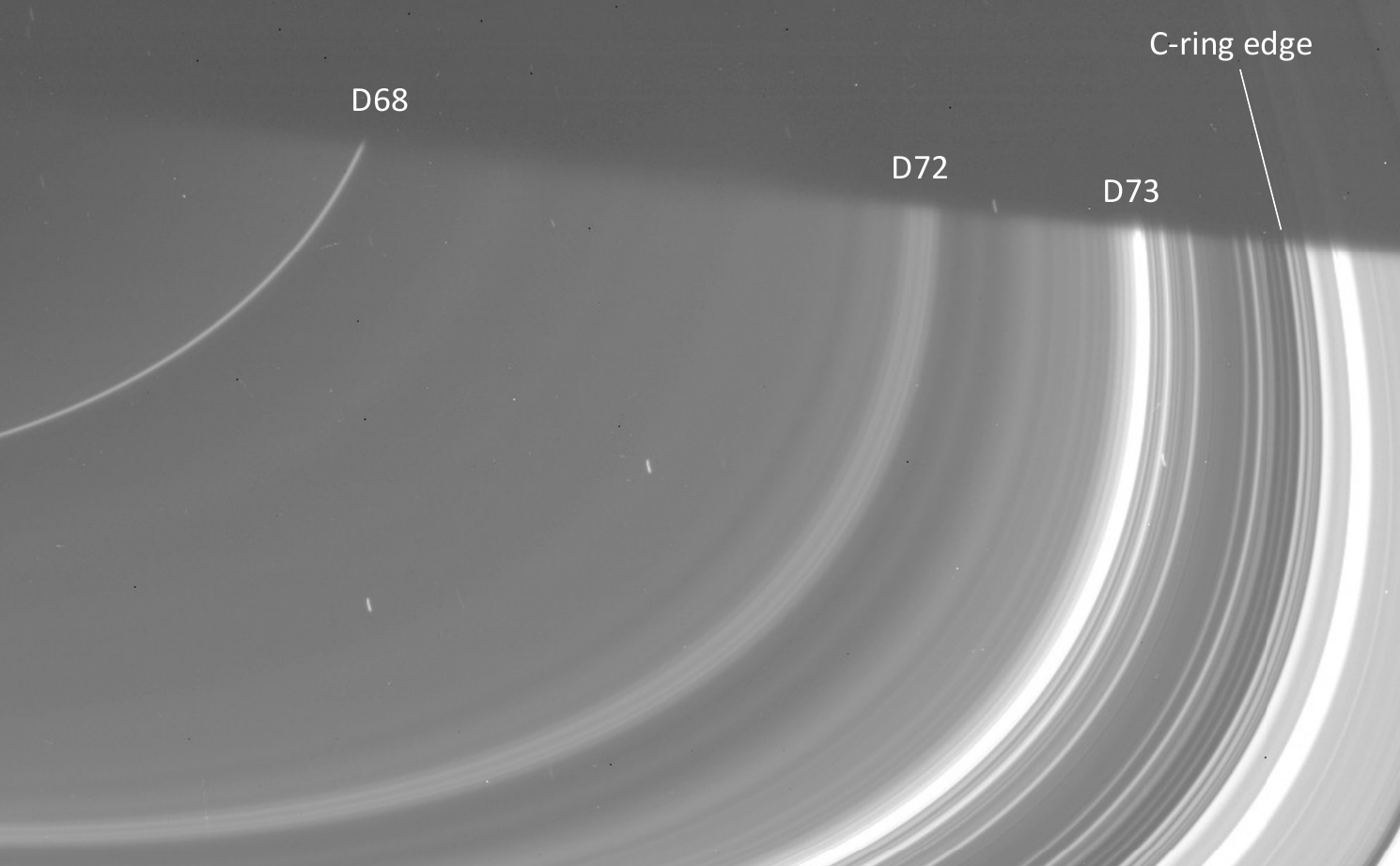}}
     \resizebox{4in}{!}{\includegraphics{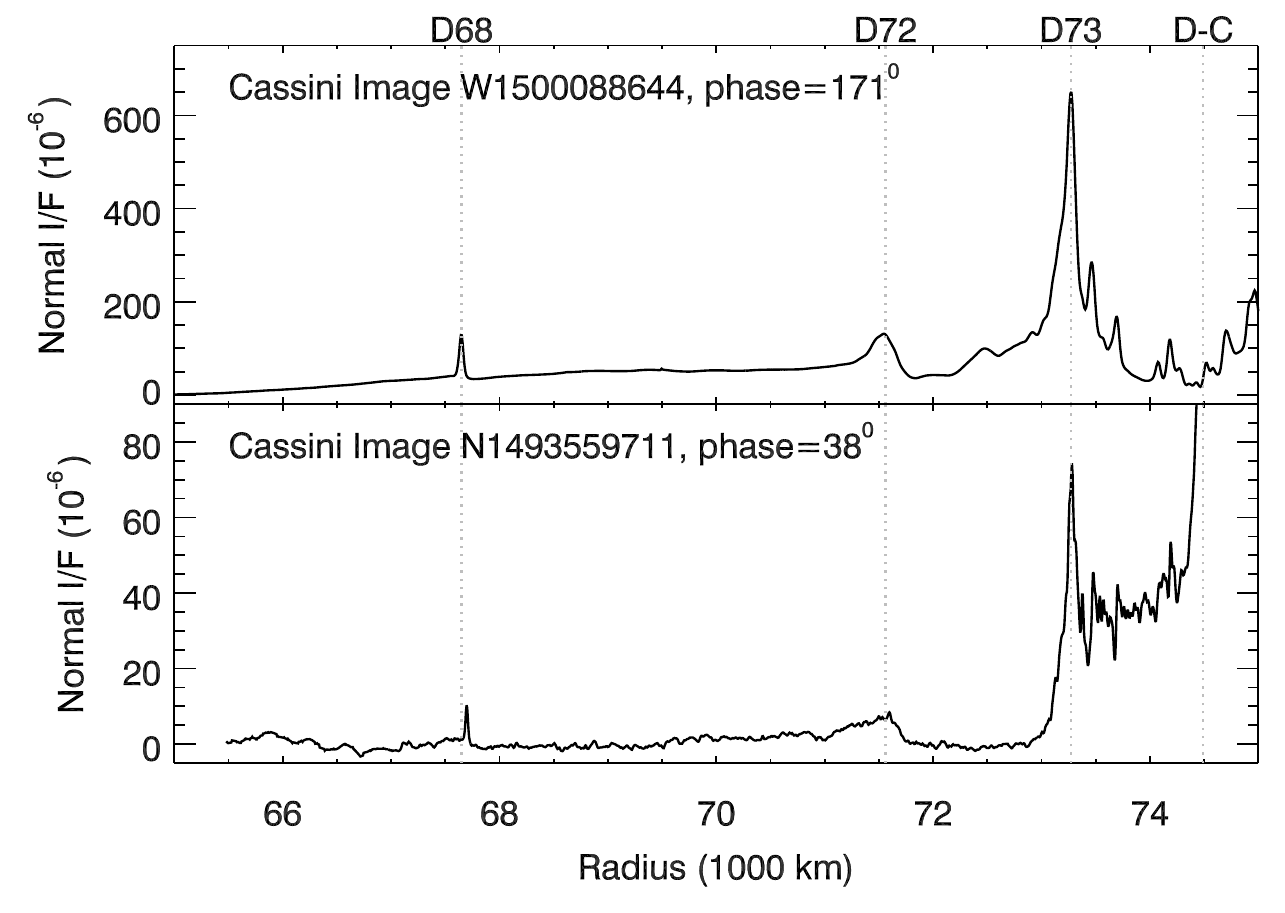}}
    \caption{Overview of the D-ring's structure. The top image shows a Cassini image of the D ring obtained at a phase angle of 154$^\circ$ (N1541397571), while the bottom image shows radial brightness profiles of the ring from selected images obtained at high and low phase angles \citep[adapted from][]{Hedmanetal2007b}. In all cases, the locations of the three named ringlets and the inner edge of the C ring are marked.}
    \label{fig:dringov}
\end{figure}

Having reviewed the physical and chemical composition of the main rings, we now turn to the innermost D ring. The D ring is a relatively tenuous but complex ring that extends from the main rings down almost to the planet's upper atmosphere. Any material coming from the main rings and entering the planet's equatorial atmosphere must pass through this region, so the structure and composition of this ring is relevant to connecting the composition of the main rings with the material flowing into the planet. 

Figure~\ref{fig:dringov} provides a general overview of the D-ring's structure \citep{Hedmanetal2007b}. The most prominent features in this ring are three ringlets designated D68, D72 and D73 after their distance from Saturn's center in thousands of km. Exterior to D73 there is a region that contains a collection of additional ringlets when viewed at high phase angles, but appears instead as a faint shelf of material when viewed at low phase angles. Meanwhile, D72 and D68 are surrounded by a broad, diffuse sheet of material that is only clearly visible at high phase angles. This diffuse sheet of material gradually fades interior to D68 with a scale length of order 1000 km and no distinct inner edge.

\subsection{Constraints on particle size and composition}\label{dringcomp}

Remote sensing data provide some constraints on the composition and typical particle size in the D ring, and are the only data source for the bulk D ring properties. The fact that all parts of the ring become brighter at higher phase angles implies that the majority of the visible particles are less than 100 microns across \citep{Hedmanetal2007b, HedmanStark2015}. Larger particles may exist in the region exterior to D73, which is both unusually bright at low phase angles and has a measurable opacity in some stellar occultations \citep{Hedmanetal2007b}. The sudden appearance of clumps in the D68 ringlet near the end of the Cassini mission provides indirect evidence that larger objects may reside within the ringlet  \citep{Hedman2019, AHearnetal2021}, but these objects have not yet been directly observed. Thus far, no evidence for larger particles has been uncovered elsewhere in the D ring. At the same time, sub-micron particles are very inefficient at scattering light at visible wavelengths, so there are at present no strong constraints on the distribution of such tiny particles within the D ring, a fact that should be kept in mind when assessing the {\it in situ} data from CDA, MIMI and INMS described in Section  \ref{eqInflow}.

\begin{figure}
    \centering
    \resizebox{2in}{!}{\includegraphics{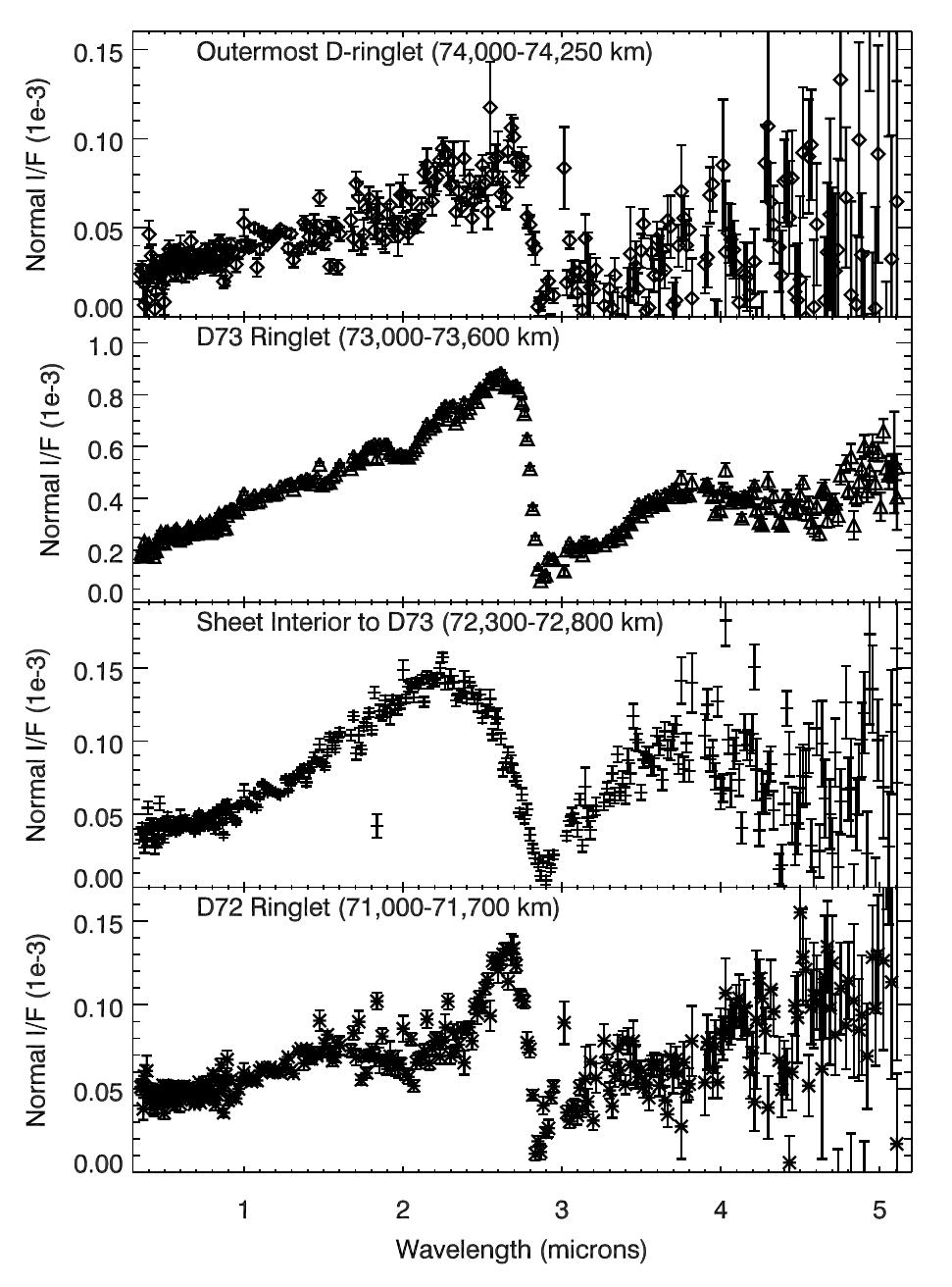}}
     \resizebox{2in}{!}{\includegraphics{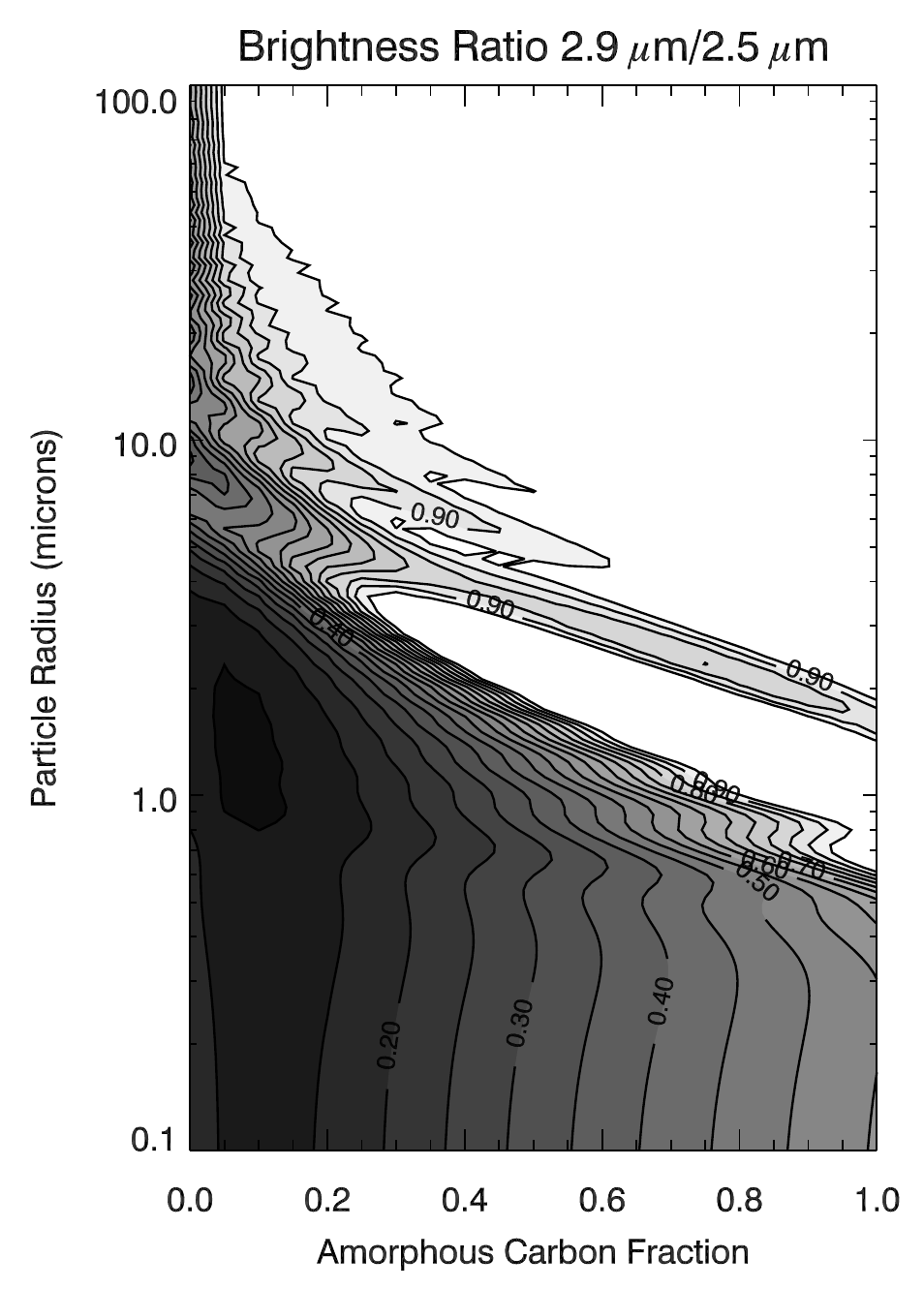}}
    \caption{Compositional information from near-infrared D-ring spectra. The left image shows spectra of various regions in the D ring derived from VIMS observations obtained at a phase angle of 167$^\circ$  \citep{Hedmanetal2007b}. Note that all these spectra show a strong band at 3 $\mu$m due to the fundamental water-ice absorption band. These spectra have ratios for particle brightness at 2.5 $\mu$m and 2.9 $\mu$m that range between R = 0.04 and R = 0.24. The right panel shows as contours the predicted values for R as a function of particle size and composition assuming the particles are spheres with optical constants given by a mixture of water ice and amorphous carbon.}
    \label{fig:dringspec}
\end{figure}

The small size of the visible particles in the ring limits the amount of compositional information that can be extracted from the remote sensing data. Fortunately, near-infrared spectra of the D ring obtained by VIMS do provide some useful constraints on the composition of this material. As shown in Figure~\ref{fig:dringspec},  many different parts of the D ring show a clear and deep absorption band around 3 microns due to water-ice. This indicates that water-ice is a major component of the D ring, as it is for the rest of the rings. The question then becomes how much non-icy material could be in this region. 

It is reasonable to expect that the D ring contains a higher fraction of non-icy material than the main rings because the observed spectral trends across the main rings clearly indicate that the concentration of non-icy materials become higher closer to the planet \citep{Hedmanetal2013, Filacchioneetal2014}. We can place a rough constraint on the amount of non-icy material in the outer and middle D ring based on the observed depth of the very strong 3~$\mu$m ice band. This particular band is visible at high phase angles because the real index of water-ice approaches unity, thus inhibiting surface scattering. This dip becomes weaker if non-icy material is added to the particles because this tends to increase the real index away from unity. 
\par
Figure~\ref{fig:dringspec} shows the predicted ratio of brightness between 2.9~$\mu$m and 2.5~$\mu$m for dielectric spheres computed from Mie theory as a function of particle size and composition for various mixtures of water ice and amorphous carbon. Note that the optical constants used for these calculations are obtained using measurements for water ice and amorphous carbon by \cite{Mastrapaetal2009} and  \cite{RouleauMartin1991}, respectively, and the effective optical constants for the mixtures are computed using formulae from \cite{Cuzzietal2014b}.  These calculations indicate that adding even 20 volume\% of amorphous carbon to ice is enough to significantly change the depth of the ice band; this limit scales roughly proportionally with the real index of the contaminant at 2.9~$\mu$m. Thus the strength of these ice bands suggests that the material in at least the outer and middle D ring is still predominantly ($\ge 80$ vol.\%) water ice. However, the ice-band strengths in the innermost D ring have not yet been analyzed in detail. 

\subsection{Evidence for disturbances}\label{dringchange}

Another important aspect of the D ring relevant for understanding the flow of material between the rings and the planet is that unlike the main rings, the D ring is time variable on time scales of years to decades \citep{Hedmanetal2007b}. Between the Voyager flybys in 1980/81 and the arrival of Cassini in 2004 the structure of the middle D ring changed dramatically, with the brightest feature in the entire ring during the Voyager encounters \citep{Showalter1996} turning into a much more subtle feature 25~yrs later. During the Cassini mission, various parts of this ring were perturbed by sudden events, either impacts or temporal anomalies in the planet's magnetosphere \citep{Hedmanetal2007b, Hedmanetal2009a, Hedmanetal2015, HedmanShowalter2016}.
\par
Finally, the innermost narrow feature in the ring, D68, developed a collection of bright clumps towards the end of the Cassini mission \citep{Hedman2019, AHearnetal2021}. This localized brightening is most easily explained as the result of disruption of source bodies trapped in this region, releasing additional dust into the system. These sorts of changes could well have influenced the rate at which material flows from the rings into the planet.

\section{Inflowing material}\label{inflowIntro}

During Cassini’s Grand Finale orbits, the spacecraft passed between the inner edge of the rings and Saturn’s upper atmosphere, to a depth of 1360 km above the 1 bar level. Initial expectations were that only Saturn’s atmosphere would be detected by the suite of {\it in situ} instruments during these proximal orbits. As the spacecraft completed the proximal orbits, it became evident that an abundance of material was measured by the Ion Neutral Mass Spectrometer \citep[INMS,][]{Waiteetal2004}, the Magnetospheric Imaging Instrument \citep[MIMI,][]{Krimigisetal2004}, and the Cassini Dust Analyzer \citep[CDA,][]{Sramaetal2004} at or near the equatorial plane flowing into the atmosphere from the rings \citep{Waiteetal2018, Mitchelletal2018, Hsuetal2018}. This chemically complex material included grains from nanometers to tens of nanometers in size, with mixing ratios that increased relative to the H$_{2}$ and He atmosphere with increasing altitude \citep{Perryetal2018}, supporting its origin exogenous to Saturn's atmosphere. Here, we review the current understanding of the material that has been observed flowing into Saturn's atmosphere from the rings, which breaks into two categories: ring rain at mid-latitudes coming in along magnetic field lines, and the equatorial inflow observed by Cassini.

\subsection{Charged ring rain at mid-latitudes}\label{ringRain} 

The first evidence for the flow of material from Saturn’s rings into the atmosphere was measurement of ionospheric electron density profiles by Pioneer and Voyager \citep{Klioreetal1980,Tyleretal1981}, which indicated a paucity of electrons with respect to the modeled densities \citep{Waiteetal1979,AtreyaWaite1981}. These previous chemical models required high electron densities to reproduce observations of long-lived H$^+$ ions. The addition of charge-exchange reactions with water as an additional sink for H$^+$ ions reduced the required electron densities and reconciled theory and data, leading to the identification of “ring rain” \citep{ConnerneyWaite1984}. Further analysis and modeling of this phenomenon identified that the main influx was due to movement of charged material along magnetic field lines, with deposition at mid-latitudes especially in the southern hemisphere \citep{NorthropHill1982, Connerney1986, Ip1983}. 
\par
More recently, ground-based observations revealed that H$_3^+$ ion emission in Saturn's atmosphere is latitudinally variable, with a symmetric pattern offset from the equator that maps along magnetic field lines to areas of the rings with greater optical depth, which was interpreted as additional evidence of ring rain \citep{Odonoghueetal2013}. \cite{Mooreetal2015} reproduced the observed H$_3^+$ observations with an atmospheric chemistry model that includes a shifting balance between two loss pathways for H$^+$: reaction with water or reaction with vibrationally excited H$_2$. This leads to a complex relationship between H$_3^+$ and H$_2$O. Charge exchange with vibrationally excited H$_2$ produces H$_3^+$, and competes with the water pathway for removal of protons, subsequently decreasing the electron density. This in effect slows the rate of loss of H$_3^+$ by electron recombination. Combined with constraints on the electron density \citep[][]{Klioreetal2009} and haze latitudinal distribution \citep{Connerney1986}, both of which are removed by water, \cite{Odonoghueetal2013} and \cite{ODonoghueetal2019} utilized the H$_3^+$ distribution to infer the presence and abundance of ring rain.
\par
Notably, \cite{Cravens1987} identified quenching with atomic H as a major loss pathway for vibrationally excited H$_2$, and photoionization of atomic H is a major source for H$^+$ \citep{ConnerneyWaite1984}. Atomic H may play an important role in the inflow of ring material, including water, at the equator \citep[][ see also Section \ref{eqInflow} below]{Mitchelletal2018}. Observations of Saturn's Lyman-alpha brightness between 5 and 35°N suggest that the mid-latitude abundance of atomic H may be higher than current model predictions by a factor of two to three \citep[][]{Ben-Jaffeletal2023}. Better understanding of the latitudinal distribution and source of atomic hydrogen may therefore be an important step forward for understanding the Saturn-rings system.

The Cassini Grand Finale orbits provided a unique opportunity to directly measure the mass flux and composition of ring rain particles, which includes the small fraction of high-speed ejecta from micrometeoroid impacts across the rings that can be charged and travel along magnetic field lines \citep[e.g.,][]{NorthropConnerney1987,Ipetal2016}. Measuring the flux of these nano-grains plays an important role in constraining the rings' age, origin, and their remaining lifetime (discussed in more detail in Chapter 12 by Crida et al.), as well as for understanding certain observed properties of Saturn's atmosphere itself.
 
\par During Cassini's Grand Finale, the CDA instrument detected inflowing grains with peaks in distribution in the northern and especially southern mid-latitudes when the instrument was pointed in the plasma-ram direction, as well as at the equator when the instrument was pointed towards the Keplerian-ram direction \citep{Hsuetal2018}. Mapping of this distribution along magnetic field lines to the B and C rings was interpreted as evidence that this material constitutes direct observation of charged ring rain. 

One of the interesting aspects of this observation is that in general, many of the detected particles are relatively water ice-poor compared to the composition of the main rings from which they originated, with between 8 and 33 weight \% silicate grains reported in the CDA data \citep{Hsuetal2018}. There are a few ways in which this may be explained: (a) The detected grains could be biased towards non-icy material by the nature of the high-velocity initial impact into the rings that charged the grains, perhaps because the ice resident in the ring particles is more easily vaporized, possibly recondensing back onto shadowed ring surfaces, or is reduced to grains too tiny for CDA sampling; (b) impact ejecta grains retaining ice are larger and more closely restricted to low latitudes that were not as well sampled by CDA; or (c) photosputtering of icy-silicate ring ejecta readily converts the water ice into vapor as the grains flow into the planet's atmosphere at high latitudes. 

These possibilities may be assessed by a parallel analysis of the total amount of water that is entering the planet's midlatitudes. CDA measured a total flux of nanometer-sized grains arriving at Saturn consistent with a mass-deposition rate (MDR) of $\sim 300-1200$ kg s$^{-1}$, 70\% of which fell near the midplane, and the remaining 30\% ($\sim 100-370$ kg s$^{-1}$) at the so called ring rain latitudes \citep{Hsuetal2018}. This water-depleted inflow is insufficient to explain the amount of charged water products needed ($\sim 430-2800$ kg s$^{-1}$) to account for the observed latitudinal pattern of H$_3^+$ emission in Saturn's atmosphere \citep{ODonoghueetal2019}. This may suggest that the water has been liberated from the observed grain population, and that this missing water is still somehow making it into the atmosphere, perhaps in vapor form. If one were to reconstitute the CDA grain population with sufficient water so that it were similar in composition to the inner edge of the C ring ($\sim 6\%$ non-icy by mass), this would give a lower bound of an additional $\sim 470-1750$ kg s$^{-1}$ of water \citep{DurisenEstrada2023} which is more consistent with the \cite{ODonoghueetal2019} result. 

This water liberation may be explained by photosputtering of the icy material {\it en route} from the rings to the planet (S. Hsu, personal communication 2018). This rate can be simply estimated from the far-UV flux at Saturn (between 10-1200 \AA) of approximately $10^{-3}$ J m$^{-2}$s$^{-1}$, scaled from 1 AU \citep{Colinaetal1996, Chadney2017}. A 20 nm radius grain, the best-fit size for CDA data \citep[][]{Hsuetal2018}, then receives about $5 \times 10^{-18}$ J s$^{-1}$. The latent heat of water is 2800 J g$^{-1}$ so the energy needed to photosputter away a 20 nm radius grain is about $0.9 \times 10^{-13}$ J, giving an exposure time of about five hours, comparable to the time the grain takes to wander along field lines into the planet's upper atmosphere at ring rain latitudes \citep[and personal communication]{Hsuetal2018}. Thus it is possible that water ice is photosputtered off of icy ejecta grains, and that the water vapor subsequently becomes charged and follows along with the solid grain residuals along magnetic field lines into Saturn's ring rain latitudes.   

\subsection{Equatorial inflow}\label{eqInflow}
In addition to the flow of charged material inward from the rings along magnetic field lines, neutral material from the inner edge of Saturn’s D ring can be decelerated by impacts with Saturn’s extended atmosphere, and subsequently deorbited \citep{Mitchelletal2018, Perryetal2018}. The highest influx of material observed during the Grand Finale was centered around the equator and measured by INMS, which detected masses across its full range from 1 to 99 u \citep{Waiteetal2018, Milleretal2020, Seriganoetal2020, Seriganoetal2022}. This dataset provides a new window into the composition of Saturn’s rings, but there are some challenges to its interpretation.
Cassini’s relative velocity compared to the inflowing ring material was approximately 30 km s$^{-1}$. This high velocity may have led to impact fragmentation of material entering the instrument, such that the measured masses may represent impact products of the ring materials rather than direct measurement of unaltered ring materials. 
\par
Furthermore, data on neutrals from INMS in this time period were collected using the closed source, which has a higher sensitivity but can lead to artifacts caused by interactions of incoming material with the instrument. 
Such effects include a time delay in detection of materials that adsorb more strongly to the surface, such as water. 
This effect leads to compositional fractionation, which requires recreation of the original composition via integration of the data from each orbit into a single composite mass spectrum \citep{Mageeetal2009}. 
Finally, while ionization fragmentation can yield unique signatures for many compounds, convolution of those signatures leads to degeneracies that cannot be resolved with the unit mass resolution of INMS. 
With these caveats in mind, some important conclusions can still be drawn about the composition of the inflowing ring material. 

\par

\subsubsection{Composition}\label{eqCompositon} 

\par
\begin{table}
\begin{center} 

\caption{Composition of inflowing ring material in weight percent. Values presented here compare deconvolutions from \cite{Waiteetal2018}, \cite{Milleretal2020}, and \cite{Seriganoetal2022} and are scaled to account for the effects of diffusive fractionation in Saturn's atmosphere following methods from \cite{Waiteetal2018}. Organics (''Org") are held constant for comparison.}
\label{tbl:influx_composition_summary} 
\begin{tabular}{lrrccccc}\\
\toprule
Reference & CH$_4$ & H$_2$O & NH$_3$ & CO & N$_2$ & CO$_2$ & Org \\[.2em]
\midrule
\cite{Waiteetal2018} & 16\% &	24.2\% &	2.4\% &	20\% &	20\% & 0.5\% &	37\%  \\
\cite{Milleretal2020} & 9.0\% &	17.9\% &	3.6\% &	15.7\% &	16.4\% &	0.4\% &	37\% \\
\cite{Seriganoetal2022} & 12.6\% &	16.6\% &	2.2\% &	15.2\% &	13.7\% &	2.7\% &	37\% \\

\botrule
\end{tabular} 
\end{center}

\end{table}
\par
The INMS measurements included gas, both from the rings and from  the nominal H$_2$ and He atmospheric gas component, and volatilized fragments generated by dust impacting at over 30 km s$^{-1}$ into the instrument’s titanium antechamber surface. 
Unexpectedly, there was a high abundance of material heavier than He with identifiable compounds (CH$_4$, CO$_2$, CO/N$_2$/C$_2$H$_4$, H$_2$O, NH$_3$), as well as a host of organic fragments. Together, these compounds comprised a measured mass density of 1.6$\times10^{-16}$ g cc$^{-1}$, as compared to the mean atmospheric mass density in this region of 4.7$\times10^{-15}$ g cc$^{-1}$ \citep{Waiteetal2018}. This material was measured at altitudes above the turbopause, and the presence of these heavy masses with increasing mixing ratios at higher altitudes indicates that they originate from material falling into Saturn's atmosphere rather than being atmospheric constituents \citep[][]{Perryetal2018}. 
\par
Overall, the inflowing equatorial material was found to be quite water-poor in comparison to remote sensing constraints. By mixing ratio, water is still the single most abundant compound, comprising approximately 25 molar\% of the material measured by INMS \citep{Milleretal2020,Seriganoetal2020}, or an estimated 20 volume\% assuming all components have approximately the same porosity (Table \ref{tbl:influx_composition_summary}). This is notably lower than the main ring composition, which is greater than 90 volume\% water (see Section \ref{mainRingsWater} and Table \ref{tab:ring-comp-summary}). However, the D ring composition is also water-poor compared to the main rings (Section \ref{dringcomp}), and the overall trend suggests enrichment of non-water material moving inward through the rings. Other abundant compounds in the equatorial inflow include organics with fragments dominated by C$_4$ compounds and comprising between 10 and 20 weight\% (g per 100 g), CH$_4$ comprising approximately 10 to 16 weight\% of the INMS sample, NH$_3$ at 2 to 3 weight\%, CO$_2$ at $<$ 3 weight\%, and CO, C$_2$H$_4$ and/or N$_2$ with a combined 20 to 30 weight\%.\footnote{The INMS data cannot resolve CO, C$_2$H$_4$ and N$_2$. The confinement of the 28 u peak around the equator in high altitude proximals provides some evidence that more refractory C$_2$H$_4$ contributes to the 28 u counts \citep[see Fig. 1b of ][]{Perryetal2018}. Whether the remainder of 28 u material is dominated by CO or N$_2$ depends on the original source for the material. CO is much more abundant in cometary sources, for example.}
\par
\textbf{Evidence for a gas-grain mixture.}
\par
Based on comparisons of proximal orbits that crossed the ring plane (e.g. Prox290-292) with the final plunge (Prox293) that did not pass through the ring plane, \cite{Milleretal2020} concluded that the inflowing material was latitudinally fractionated, with more refractory components depleted at higher latitudes (see their Figure 3). They interpreted this as inflow of the more volatile components CH$_4$ and CO (or N$_2$) as gas, and the remaining, refractory components as grains, and estimated a gas to dust molar ratio of ~0.7.

\begin{figure}
    \centering
    \resizebox{4in}{!}{\includegraphics{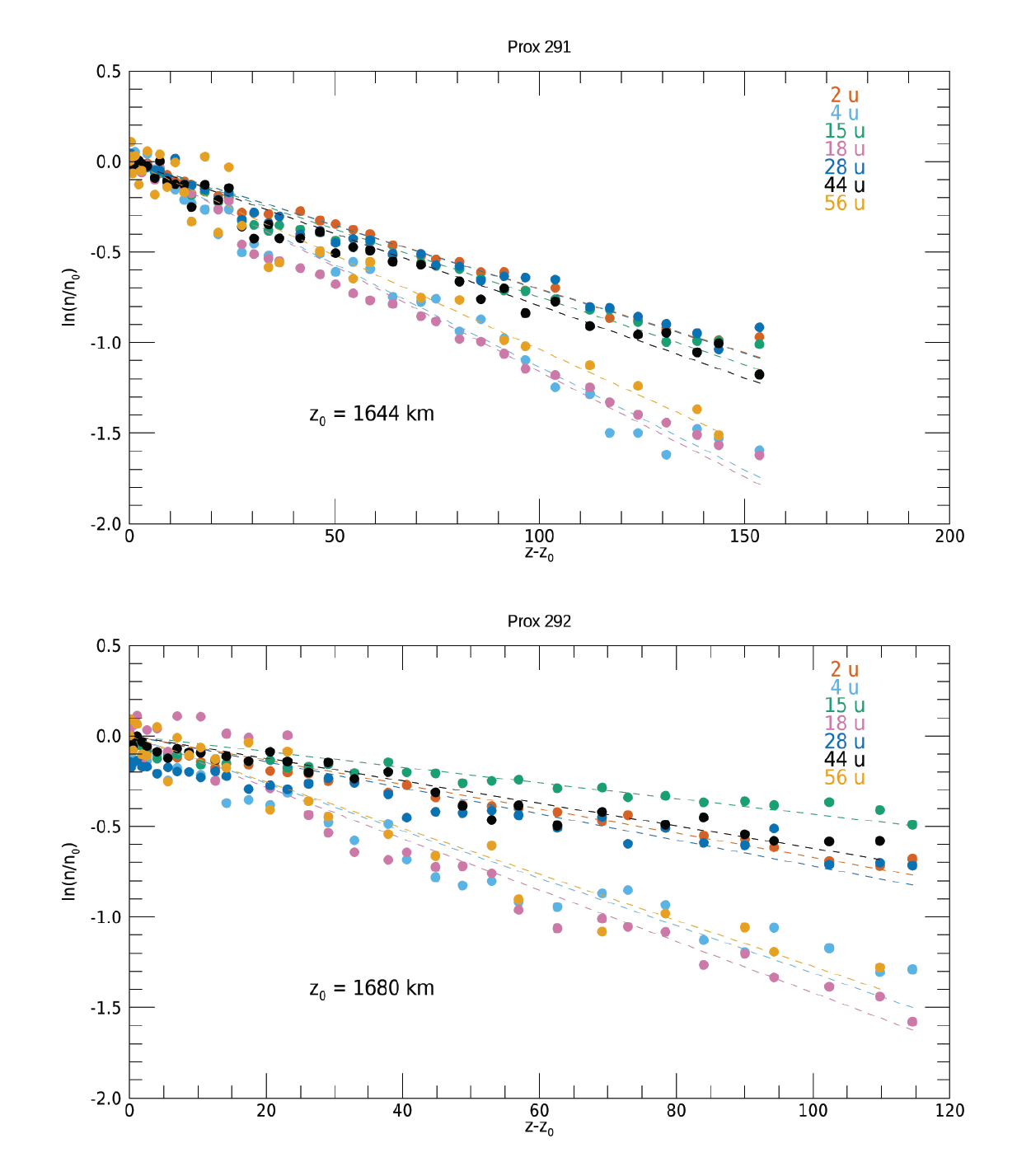}}
    \caption{Calculated scale heights for marker masses for atmospheric H$_2$ (2 u) and He (4 u), as well as for inflowing CH$_4$ (15 u), H$2$O (18 u), CO/N$_2$/C$_2$H$_4$ (28 u), CO$_2$ (44 u), and organics (56 u). Symbols show measured density values, and dashed lines show linear fits to the data. The different scale heights for H$_2$ and He are evidence that the measurements are made above the turbopause where diffusion velocity dominates. The different scale heights between volatile and refractory inflowing masses suggests that refractory materials are falling more rapidly because of contributions from grains.}
    \label{fig:scaleheight}
\end{figure}

\par
New analysis presented in this work of the scale heights of Saturn's atmospheric constituents and the major classes of inflowing material provides additional evidence for a gas-grain division (Fig. \ref{fig:scaleheight}). Here, 15 u represents CH$_4$, 18 u represents H$_2$O, 28 u represents the combination of CO, N$_2$, and C$_2$H$_4$, 44 u represents CO$_2$, and 56 u is representative of organics. The scale height $H$ and pressure $P$ are described by

\begin{equation}
H = \frac{kT}{mg}
\end{equation}
\begin{equation}
P = P_0 \exp\left(-\frac{z-z_0}{H}\right)
\end{equation}

where $k$ is the Boltzmann constant, $T$ is the temperature, $m$ is the compound mass, $g$ is the gravitational acceleration,  and $z$ is the altitude. While scale heights calculated from the slope of the dashed fits in Fig. \ref{fig:scaleheight} match within formal uncertainty (which is not shown for visual clarity), two separate trends or groups  are visible. The scale heights for H$_2$ and He differ by a factor of 2 ($\sim140$ km versus $~70$ km), matching the 2$\times$ mass difference between the compounds. This provides confirmation that the measurements are taken above the turbopause, where the scale height for upward-moving material becomes mass-dependent. Masses in the downward-flowing material that have been associated with the gas phase \citep{Milleretal2020} have scale heights that are most similar to the bulk atmosphere (H$_2$). This suggests diffusive coupling of these gases with the atmosphere; that is, the downward velocity of the gases is controlled by the rate of diffusion through H$_2$ because the gas-phase inflow is strongly coupled with the atmosphere. Masses that have been associated with grains of less-volatile material  \citep[18 u and 56 u;][]{Milleretal2020} have scale heights that are smaller than the bulk atmosphere. This may indicate that these masses are not yet coupled to the bulk atmosphere at the altitudes measured, and are falling at higher velocities.

\par
The presence of inflowing grains is further supported by analyses of the Langmuir probe (LP) dataset. \cite{Johanssonetal2022} modeled secondary emission produced by high velocity impacts of water vapor molecules on the Langmuir probe, and derived a median mixing ratio for gaseous water molecules of $~10^{-5}$. This value is approximately an order of magnitude lower than the water mixing ratio reported by \cite{Waiteetal2018}, suggesting that the majority of water measured by INMS may be present as grains.
\par
Ionospheric chemistry also points towards the presence of grains in the equatorial inflow. Photochemical modeling identifies loss processes for H$^+$ and H$_3^+$ via reaction with a heavy compound M, which may be  CH$_4$ or H$_2$O \citep{Cravensetal2019}. The modeled mixing ratio for M, f$_M$, is the same order of magnitude as the mixing ratio measured by INMS for water. At higher latitudes, f$_M$ increases, while the equatorial region is dominated by a second heavy compound R that reacts with H$_3^+$ but not H$^+$. N$_2$ and CO are suggested as candidates for R. While the photochemical model of \cite{Cravensetal2019} therefore also supports latitudinal fractionation, the higher abundances of volatile CO and/or N$_2$ at the ring plane are contrary to the findings of \cite{Milleretal2020}. However, the modeled reactions are for gas-phase compounds, and the depletion in f$_M$ at the equator may be explained by retention of most water in grains at lower latitudes, while CO and/or N$_2$ may be in the gas phase at all latitudes.
\par
This explanation is consistent with further modeling by \cite{Vigrenetal2022}, who consider the role of grain charging in this environment. Their model results require higher mixing ratios of M-type molecules and lower mixing ratios of R-type molecules than reported by \cite{Milleretal2020}. Similar to \cite{Cravensetal2019}, their photochemical model considers reactions of gas-phase water with H$_3^+$ and H$^+$, but does not include solid-phase water in icy grains. \cite{Vigrenetal2022} report that the number density of grains required by their model is approximately equal to the difference between the number of positive ions, n$_i$, and the number of electrons, n$_e$. This difference is greater at lower latitudes, which may be suggestive of an increased grain population near the equator. They suggest an upper limit of $\sim40$ ppm for M, and compare this to the 200 to 400 ppm abundances reported by \cite{Milleretal2020}. This may suggest a molar gas to dust ratio for M molecules on the order of 40:200 (0.2) to 40:400 (0.1) if the grains are dominated by water ice, somewhat lower than the value of 0.7 estimated by \cite{Milleretal2020}.

Modeling of Saturn's atmospheric chemistry by \cite{Mosesetal2023} suggests that the physical state of the inflow as a gas-grain mixture rather than as gas only is significant for interpretation of the inflow timescales; see Section \ref{inflowTimescale}. Additional work to understand the distribution of gas and grains in the inflowing material is needed.

\begin{table}[htbp]
\centering
\caption{Comparison of ring compositions via in situ and remote sensing measurements. Remote sensing values summarize the upper and lower limits for non-water and water ice components respectively. In situ compositions include the average value from deconvolution results in \cite{Seriganoetal2022} and \cite{Milleretal2020}. Other ices includes NH$_3$ and CO$_2$. Gas includes CH$_4$, N$_2$, and CO. Rocky material includes silicates and troilite.}
\label{tab:ring-comp-summary}
\begin{adjustbox}{width=\textwidth}
\begin{tabular}{@{}llllllllllll@{}}
\toprule
& \multicolumn{5}{c}{In Situ} & \multicolumn{6}{c}{Remote Sensing} \\
\cmidrule(lr){2-6} \cmidrule(lr){7-12}
& \multicolumn{3}{c}{INMS inflow} & \multicolumn{2}{c}{SUDA inflow} & \multicolumn{1}{c}{D ring$^{a}$} & \multicolumn{1}{c}{C ring edges$^{b}$} & \multicolumn{1}{c}{C ring mid$^{c}$} & \multicolumn{1}{c}{B ring$^{d}$} & \multicolumn{1}{c}{CD$^{e}$} &\multicolumn{1}{c}{A ring$^{d}$}   \\
& molar\%$^{f}$ & weight\%$^{g}$ & volume\% & weight\%$^{g}$ & volume\% & volume\% & volume\% &volume\% & volume\%&volume\% &volume\% \\
\midrule
water & 24 & 16 & 31 & 67--92 & 87--97 & \textgreater{}80 & \textgreater{}92 & \textgreater{}75 & \textgreater{}95 & \textgreater{}92 &\textgreater{}97 \\
organics & 34 & 50 & 55 & -- & -- & \textless{}20 & \textless{}6 & \textless{}14 &\textless{}4 & \textless{}6 & \textless{}2 \\
other ices & 11 & 8 & 14 & -- & -- & -- & -- & -- & -- & -- & -- \\
gas & 30 & 26 & -- & -- & -- & -- & -- & -- & -- & -- & -- \\
rocky & -- & -- & -- & 8--33 & 3--13 & -- & \textless{}2 & \textless{}11 & \textless{}1 &\textless{}2  &\textless{}1  \\
\bottomrule
\end{tabular}
\end{adjustbox}
\begin{tablenotes}\footnotesize
        \item[a]\, see Section \ref{dring}
        \item[b]\, \citep{Cuzzietal2018, Zhangetal2017C}; away from the central, anomalous peak (see figure \ref{fig:cringsilicate})
        \item[c]\, \citep{Zhangetal2017C,Ciarnielloetal2019}
        \item[d]\, \citep{Cuzzietal2018, Zhangetal2017C,Ciarnielloetal2019}
        \item[e]\, quantitative constraints from \cite{Zhangetal2017C}, plus similarity to C ring from \cite{Hedmanetal2013}
        \item[f]\, effectively molecules per 100 molecules
        \item[g]\, grams per 100 grams
    \end{tablenotes}
\end{table}

\subsubsection{Comparison of mass deposition rate derivations}\label{eqMassFlux} 
\par
In situ measurements by Cassini captured instantaneous compositional snapshots of parcels of Saturn’s atmosphere along each proximal orbit. To determine the mass deposition rate (MDR) required to produce these measured compositions, the downward velocity of ring material traversing through each parcel and the total surface area represented by each parcel must be constrained. To date, there are five reported calculations of a large equatorial mass influx of ring material based on data from Cassini's Ion Neutral Mass Spectrometer \citep[INMS;][]{Waiteetal2004}, including two different methods and three independent calculations \citep{Waiteetal2018, Perryetal2018, Yelleetal2018, Seriganoetal2020, Seriganoetal2022}. The results are summarized in Table~\ref{tbl:mass_influx_summary}. Calculations based on measurements by the MIMI Ion Neutral CAmera (INCA) of the equatorial influx \citep{Mitchelletal2018}, and by CDA of grains at the equator as well as at mid-latitudes are also included \citep{Hsuetal2018}. Ground-based  measurements from the Keck telescope constraining the mass influx of charged water at mid-latitudes from modeling of H$_3^+$ densities are provided for comparison \citep{ODonoghueetal2019}.

\par
\begin{sidewaystable}
\begin{center} 

\caption{Mass deposition rates (MDR) calculated for inflowing material measured during Cassini's Grand Finale. Inflows are categorized as mid-latitude ring rain ("Mid") or as equatorial inflow ("Eq").}
\label{tbl:mass_influx_summary} 

\begin{tabular}{p{0.15\linewidth}p{0.1\linewidth}p{0.1\linewidth}p{0.15\linewidth}p{0.05\linewidth}p{0.3\linewidth}}\\
\toprule
Reference & Dataset & Material included & MDR (kg/s) & Latitude & \multicolumn{1}{c}{Notes}  \\
\midrule
\cite{Waiteetal2018} & INMS & Volatiles + nanograins & 4800 - 45000 & Eq. & \\
\midrule
\cite{Perryetal2018} & INMS & Volatiles + nanograins & 4000 - 30000 & Eq.& Altitudes and velocities comparable to MIMI \citep{Mitchelletal2018}\\
\midrule
\cite{Perryetal2018} & INMS & Volatiles + nanograins & 10000 - 200000 & Eq. & Sedimentation velocities, and mid-altitudes comparable to \cite{Waiteetal2018, Yelleetal2018}\\
\midrule
\cite{Yelleetal2018} & INMS final plunge & CH4 & 1300 - 11000 & Eq. & Here we assume 10$^{10}$ km$^2$ surface \newline area for comparison\\
\midrule
\cite{Seriganoetal2020} & INMS & CH$_4$, H$_2$O, NH$_3$ & 1390 - 9580 & Eq. & Following \cite{Yelleetal2018} \\
\midrule
\cite{Seriganoetal2022} & INMS & All compounds & 19700 - 80600 & Eq. & \\\midrule
\cite{Mitchelletal2018} & MIMI & Particles 10,000 to 40,000 u & 5.5 & Eq. & Some evidence for flow of charged material on magnetic field lines  \\
\midrule
\cite{Hsuetal2018} & CDA & Grains & 224 - 854 & Eq.& \\
\hline
\hline
\cite{Hsuetal2018} & CDA & Grains & 100 - 370 & Mid & \\
\midrule
\cite{ODonoghueetal2019} & Keck telescope & Water & 432 - 2870 & Mid & Based on models of observed H$_3^+$ densities \\
\botrule
\end{tabular} 
\end{center}

\end{sidewaystable}

\par
MIMI detected high energy impacts of large molecules in a tight band with a half width of +/- 1.4 degrees centered around the equator, when the spacecraft was moving through the upper atmosphere at a relative velocity of greater than 30 km s$^{-1}$. Particles were found to have a limited mass range greater than 1 $\times10^{4}$ u. Assuming a mass density of 1 g cc$^{-1}$ for the grains, the inferred particle sizes range from 1.6 to 2.5 nm. MIMI's INCA sensor detected a dust number density of 0.1 cm$^{-3}$ \citep{Mitchelletal2018}. Utilizing terminal velocity from their model of collisional deceleration of dust grains (2.5 km s$^{-1}$ at an altitude of 3000 km), and assuming a grain size of 15,000 u, \cite{Mitchelletal2018} scale the measurement for regions within 1° latitude of the equator to calculate a grain MDR of 5 kg s$^{-1}$.
\par
The MIMI value is several orders of magnitude lower than the MDR calculated from INMS data, and the values do not appear to be reconcilable via reasonable adjustment of the model parameters. The MIMI influx calculation scales approximately linearly with grain mass. Use of the terminal velocity yields the upper limit for mass influx, and choice of a different altitude consistent with the same atmospheric model would lead to lower values for the MDR. Imposing a 10$\times$ increase on the atmospheric H density leads to a ~3$\times$ increase in the mass influx. The most likely explanation may be measurement by the two instruments of different grain populations based on size \citep{Perryetal2018}, discussed further below.

\par

INMS data are dominated by signal at the lowest altitudes, where ring material is diffusively coupled to the atmosphere, e.g. below 1700 km. Therefore, in the \cite{Waiteetal2018} calculations of influx rate, diffusive velocities\footnote{ ``Diffusive velocities" are defined from a simplified diffusion equation, which is easily seen to be just a rearrangement of terms in equation \ref{eq:simplediffeq} to $F_i=D_i(X_i N_{H_2}/H_i)$ where the term in parentheses approximates a vertical gradient. The same equation can then be written as $F_i=(D_i/H_i)X_i N_{H_2} \equiv v_D N_i$ where $v_D=D_i/H_i$ is thereby defined as the diffusive velocity, multiplying the local number density.} are used rather than terminal velocity. Measurements are still made above the turbopause, where diffusion dominates over turbulent mixing; this is evidenced by the different scale heights measured for H$_2$ and He \cite[Fig. \ref{fig:scaleheight}; see also Fig. 3 of ][]{Waiteetal2018}.

\cite{Waiteetal2018} base their velocities off of bounding limits for the downward velocity of methane, with the lower limit (45 m s$^{-1}$) given by the limiting flux approximation, and the upper limit (100 m s$^{-1}$) adapted from a simple hydrostatic model \citep{Belletal2014}. Velocities for different compounds are then scaled using diffusion coefficients \citep{MasonMarrero1970} for each of the major molecular components identified by INMS. These velocities are much lower than the terminal velocities utilized by \cite{Mitchelletal2018}. Note that using higher velocities at the same altitudes would imply higher mass influxes to explain the observed mixing ratios for inflowing material. The \cite{Waiteetal2018} calculation uses a wider spread of the material than \cite{Mitchelletal2018} at 4° north and south of the equator, based off of the half-width for INMS counts. The mass influx numbers calculated over three separate orbits ranged from 4800 to 45,000 kg s$^{-1}$ of infalling material at masses heavier than the H$_2$ and He atmospheric compounds \citep{Waiteetal2018}. The mass influx variation from orbit to orbit differs by ~2.5, while differences from the two bounding cases for velocity are a factor of ~3.8; uncertainty in velocity contributes more strongly to the range of values reported.

Similar influx estimates (10,000 to 200,000 kg s$^{-1}$ for the mid-altitude range) were determined by \cite{Perryetal2018}, with velocity estimates calculated based on Epstein drag:
\begin{equation}
v_s=\frac{0.36\rho_pg(2a)}{m_{H_2} n_{H_2} v_{H_2}},
\end{equation}

where $v_s$ is the sedimentation velocity, $\rho_p$ is the particle mass density, g is the gravitational acceleration, a is the particle radius, and $v_{H_2}$, $n_{H_2}$, and $m_{H_2}$ are the mean thermal speed, number density, and mass for atmospheric H$_2$. They calculate velocities at 1700 km of between 3 and 30 m s$^{-1}$. \cite{Perryetal2018} also report that use of INMS data at altitudes (3000 km) and terminal velocities (1 to 3 km s$^{-1}$) matching the method and region used for the MIMI MDR result in a $\sim50\%$ reduction in the calculated INMS MDR. However, this method still results in an exceptionally high inflow rate, and at the higher altitudes it is likely that material is undercounted in the INMS dataset because of the lower signal to noise ratio.
\par
\cite{Perryetal2018} concluded that the MDR discrepancy between MIMI and INMS was best explained by the size distribution of the infalling material, with a peak near a mass of 500 u and INMS measuring the smallest grains. Such small grains may be more likely to avoid sublimation \citep{Hamiletal2018}, consistent with the minimal effects observed on atmospheric chemistry \citep{Mosesetal2023}. These data imply an extremely steep size distribution if the INMS and MIMI grains are derived from the same population. However, the physical process of deorbiting ring material via collisions with atmospheric gases \citep[][]{Mitchelletal2018, Perryetal2018} is likely to fractionate the original grain population as a function of grain radius $a$. While grains with a larger radius are more likely to have collisions with atomic H (an $a^2$ dependence), the greater momentum of grains with higher mass may require more collisions to deorbit (an $a^3$ dependence). Further work is needed to understand the relationship between size distribution of the measured material, the size distribution of the source material in the D ring, and any connection to the main rings e.g. by ballistic transport \citep[][]{DurisenEstrada2023}. 
\par
The \cite{Yelleetal2018} work draws on a different atmospheric model, based on \cite{KoskinenGuerlet2018}, and utilizes the methane mixing ratio in the deep atmosphere reported from occultations \citep{KoskinenGuerlet2018} as well as mixing ratios calculated from the INMS data to construct and solve a diffusion profile of the form:

\begin{equation}\label{eq:simplediffeq}
X_i=(H_i F_i)/(N_{\rm{H}_2} D_i ),
\end{equation}
where $X_i$ is the mixing ratio of compound $i$, $H$ is the scale height, $F$ is the flux, $N_{\rm{H}_2}$ is the density of H$_2$, the dominant atmospheric component, and D is the diffusion coefficient for compound $i$. They utilize the same source for coefficients for diffusion through H$_2$ gas as \cite{Waiteetal2018}: \cite{MasonMarrero1970}. Their separate analysis confirms that the region sampled is dominated by diffusion, and calculates similar mass influxes independent of the underlying atmospheric model.
\par
\cite{Seriganoetal2020} and \cite{Seriganoetal2022} build on this approach to calculate the MDR through the conversion 
\begin{equation}
{\rm{MDR}} = F_i m_i\times 2\pi R_{\rm{S}}^2 \theta,
\end{equation}
where $m$ is the molecular mass, and $\theta$ is the latitudinal width over which the influx is spread. For $\theta$, they utilize a latitudinal width of 16°, chosen to match the drop in signal for minor constituents at $\pm 8$° from the equator. They follow the same formulation for F$_i$ as \cite{Yelleetal2018}, adding H$_2$O and NH$_3$ \citep{Seriganoetal2020}, as well as deconvolved compounds across the INMS mass range \citep{Seriganoetal2022}. Their results suggest that the MDR may reach 74,000 kg s$^{-1}$.

\par
In conclusion, the inferred mass deposition rate for the INMS dataset is measured to be of the order of 10$^4$ kg s$^{-1}$. This rate is dependent on the measured number densities of the infalling material, as well as the downward velocities imposed. Analysis via two different atmospheric models plus separate consideration of Epstein drag suggests that velocities are on the order of $<100$ m s$^{-1}$. The use of higher velocities, such as the thermal velocity of the bulk atmosphere, results in higher influxes. The MDR of fine-grained material therefore appears to be exceptionally high. However, modeling of ring dynamics suggests that ballistic transport in the rings and, to a lesser degree, mass loading could naturally lead to inward flow of the inner B ring and C ring with an MDR on the order of 10$^4$ kg s$^{-1}$ \citep{DurisenEstrada2023}. The high mass inflow may therefore be representative of the dynamical evolution of the rings. Alternatively, the phenomenon may be transient, and Cassini may have observed an unusually high rate as discussed below in Section \ref{inflowTimescale}; see also discussion of changes in the D ring in Section \ref{dringchange}. 

\par
\subsubsection{Atmospheric observational inferences and their implications for time variability}\label{inflowTimescale} 
\par
\cite{Mosesetal2023} place constraints on the duration and/or properties of the equatorial ring inflow based on comparisons between Saturn’s observed atmospheric composition and the modeled impact of the inflow on atmospheric chemistry. They demonstrate that the non-detections of HCN and HC$_3$N in Saturn's stratosphere from infrared observations, and the low inferred stratospheric abundance of CO$_2$ and H$_2$O from infrared, ultraviolet, and sub-millimeter observations, are inconsistent with coupled ion-neutral photochemical models that adopt ring-inflow compositions from either \cite{Seriganoetal2022} or \cite{Milleretal2020}. \cite{Mosesetal2023} conclude that the large inflow rates inferred from INMS measurements must be a recent phenomenon (i.e., $\lesssim$ 4 months to 50 years since the onset of such large ring inflow rates), or the ring material must consist primarily of small grains that do not ablate in the atmosphere at the depths perceived by CIRS. Even a $\lesssim$50-year-old continuous ring-inflow event would provide observational signatures of excess HC$_3$N, HCN, and CO$_2$ in Saturn's stratosphere, unless the material flowing in from the rings is primarily present as unablated grains smaller than $\sim$50 nm for entry velocities of 25 km s$^{-1}$ \citep{Hamiletal2018} \footnote{For grains with a radius of 10 nm, slightly smaller than the CDA best-fit size of 20 nm \citep{Hsuetal2018}, and density of 917 kg/m$^3$, ablation occurs at velocities above 15 km/s \citep{Hamiletal2018}. These velocities are significantly higher than the velocities calculated by MIMI \citep{Mitchelletal2018}, which reach a maximum value of 2.5 km/s, and INMS \citep{Perryetal2018, Yelleetal2018, Waiteetal2018}, which cite velocities on the order of 30 to 100 m/s for diffusively coupled material.}, with only the most volatile compounds (CO, N$_2$, CH$_4$) being present as vapor; furthermore, this volatile vapor component must spread globally before reaching the top of the stratosphere to prevent observational consequences in the stratosphere. Inflow of gas-phase CO and/or N$_2$ in particular cannot be reconciled with longer inflow timescales. The effects of assuming 28 u instead represents gas-phase C$_2$H$_4$ have not yet been examined.
\par
Of the three atmospheric products that constrain the inflow timescale (HC$_3$N, HCN, and CO$_2$), HCN is the longest-lived, persisting for $\sim2000$ years at 1 mbar and $\sim30$ years at 1 $\mu$bar. Therefore, even if the inflow of material is a transient but repeating phenomenon, the times when this process is inactive may be more common than when it is active. The possible division of inflowing material into gas-phase and solid-phase components is consistent with compositional constraints discussed in Section \ref{eqCompositon}. Saturn's atmospheric chemistry  places time constraints on the inflow of gas-phase material, but does not directly constrain the inflow timescale for solid-phase material unless it ablates. The short timescales for the present gas inflow rate may be tied to recent changes in the D ring (Section \ref{dringchange}).

\section{Ring atmosphere}\label{ringAtmosphere} 
The exposure of the ring particles to photons, charged particles, and meteoroids produces a toroidal gaseous envelope of neutrals that is referred to as the ring atmosphere. Atoms and molecules are subsequently scattered from this atmosphere, creating a system-wide neutral corona that extends into Saturn’s atmosphere and out into the magnetosphere well beyond the A ring \citep{Johnsonetal2006, Tsengetal2010, Tsengetal2013Seasonal, Cuzzietal2009}. This  extended ring atmosphere is superimposed on the toroidal atmospheres produced by outgassing from Enceladus, atmospheric escape from Titan, and, to a lesser extent, by  sputtering of the icy satellites and the grains in the tenuous E, F and G rings, as seen in modeling of their  H$_2$ and H contributions to the atmosphere over the ring plane \citep{Tsengetal2011b, Tsengetal2013Hydrogen}. Although the presence of an ambient gas with molecules precipitating into Saturn’s atmosphere was predicted prior to Cassini's arrival \citep{Carlson1980}, the simple ion mass spectrum obtained during the outbound Saturn Orbital Insertion (SOI) trajectory was surprising and is the only in situ measurement over the B ring. It was the least contaminated mass spectrum obtained by the Cassini Plasma Spectrometer \citep[CAPS; ][]{Youngetal2004} during the mission and consists of narrow peaks of O$^+$ and O$_2^+$ \citep{Tokaretal2005}. These data are the basis of a model for the production, composition and spatial distribution of the ring atmosphere, dominated by O$_2$ \citep[e.g.][]{Johnsonetal2006}. That species other than oxygen must be present below the CAPS detection limits is suggested by reflectance observations of the rings (see Sections \ref{nonWater} and \ref{regolithModels}) and, indirectly, by plasma observations outside the rings discussed below.

\begin{figure}
    \centering
    \resizebox{4.5in}{!}{\includegraphics{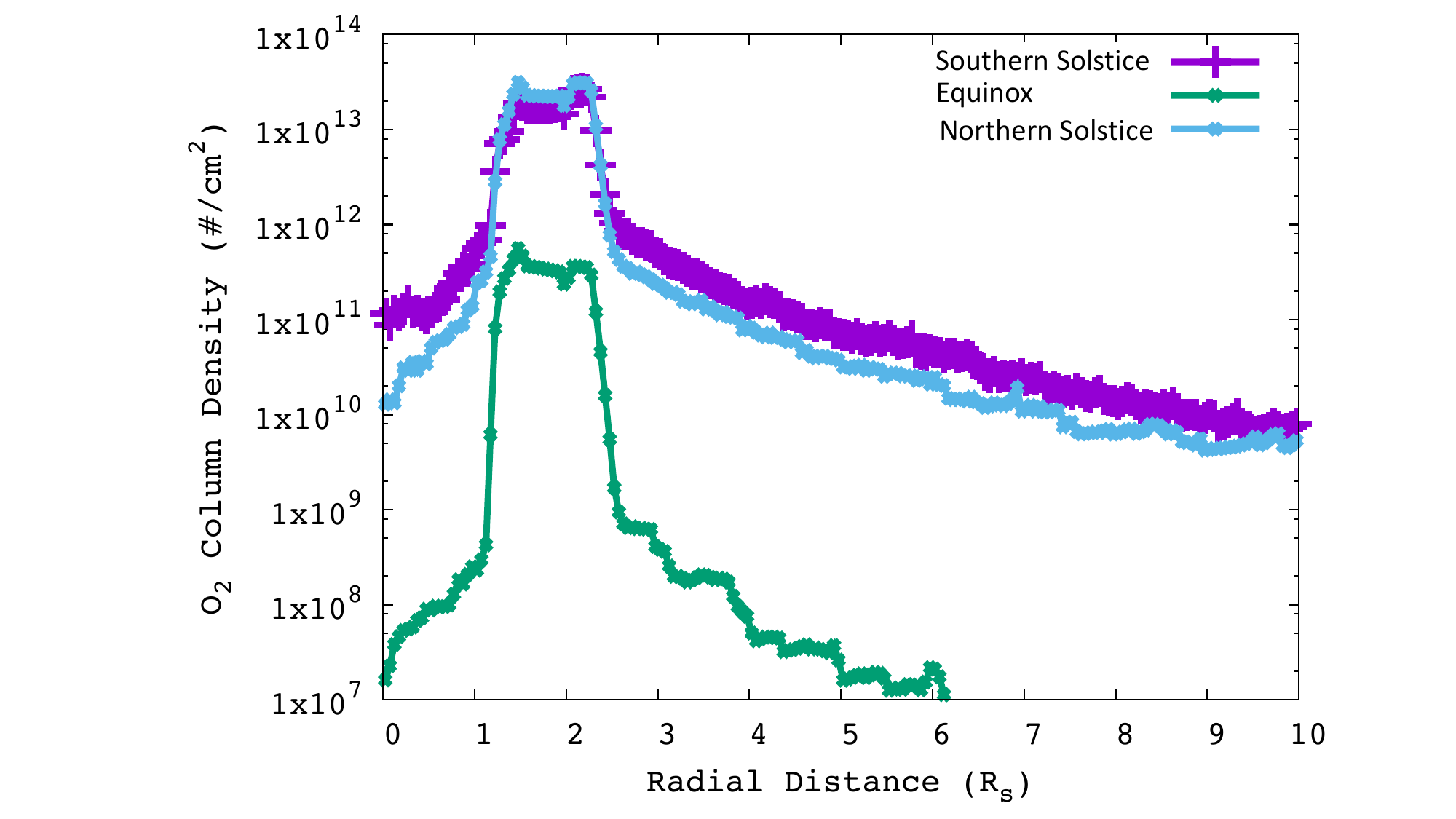}}
    \caption{Radial distribution of the modeled neutral O$_2$ column density for illumination of the A, B and C rings at three seasons with their minimum photodecomposition O$_2$ source rates: Southern Solstice (SOI, ~2x10$^{27}$ O$_2$ s$^{-1}$), Equinox (~2x10$^{25}$ O$_2$ s$^{-1}$) and Northern Solstice (Grand Finale, ~2x10$^{27}$ O$_2$ s$^{-1}$). The respective injection rates into Saturn's atmosphere are  ~5x10$^{26}$ O$_2$ s$^{-1}$ (~27kg s$^{-1}$), ~5x10$^{23}$ O$_2$ s$^{-1}$ (0.03kg s$^{-1}$), and  ~2.5x10$^{26}$ O$_2$ s$^{-1}$ (~27kg s$^{-1}$). \citep{Tsengetal2010,TsengIp2011,Tsengetal2013Seasonal}.}
   
    \label{fig:Ring_O2_neutral}
\end{figure}

\begin{figure}
    \centering
    \resizebox{5.0in}{!}{\includegraphics{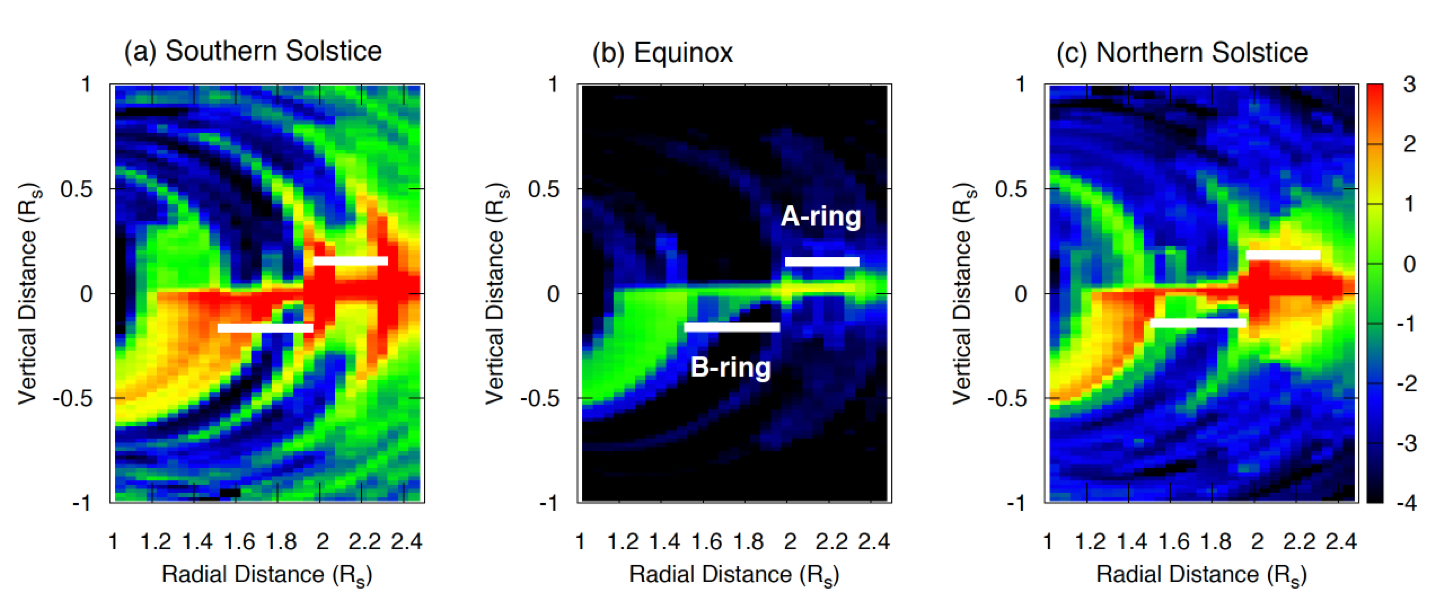}}
    \caption{O$_2^+$ ion density (cm$^{-3}$) in log10 scale as indicated by the color bar at right for the three seasons and neutral O$_2$ source rates in Fig~\ref{fig:Ring_O2_neutral}: (a) Southern Solstice (SOI), (b) Equinox and (c) Northern Solstice (Grand Finale). 
 \citep{Tsengetal2010,TsengIp2011,Tsengetal2013Seasonal}. }
    \label{fig:Ring_O2_ion}
\end{figure}

\par
During Cassini’s outbound SOI trajectory the CAPS instrument detected oxygen ions north of the B ring, through the Cassini division, out to the inner edge of the A ring  \citep{Tokaretal2005}. Oxygen ions were again detected when CAPS was turned back on beyond the outer edge of the A ring as Cassini moved into the magnetosphere \citep{Elrodetal2012,Elrodetal2014}. Although other water products were seen by INMS closer to the ring plane over the A ring \citep{Waiteetal2005}, the dominance of oxygen ions was roughly consistent with our understanding of the radiation effects at the icy Jovian satellites \citep{Johnsonetal2004}. That is,  when ice is the dominant surface constituent, radiolysis and photolysis result in the ejection of water products, a significant fraction of which are O$_2$ and H$_2$. These volatiles do not recondense at the expected ring particle temperatures, resulting in residence times  over the B-ring of the order of a half year for O$_2$ and an order of magnitude larger for H$_2$ \citep{Johnsonetal2006}. 
Because the inflowing magnetospheric plasma is suppressed at the edge of the A ring, photolysis is the dominant source of  decomposition of the ice particles. Since the southern side of the ring plane was illuminated at SOI, the detection of a significant component of oxygen ions north of the ring plane indicates that a toroidal O$_2$, and concomitant H$_2$, atmosphere permeates and co-orbits with the ring particles with an O$_2$  scale height $\gtrsim 0.02$ \Rss. Their lifetimes are primarily determined by photoionization and photodissociation producing O, O$^+$, O$_2^+$ and H, H$^{+}$ and H$_2^+$. Although ionization predominantly occurred south of the ring plane at SOI, the CAPS detection of O$^+$ and O$_2^+$ north of the plane indicates that the B ring is not fully opaque to the UV and/or to the flow of ions along field lines through the plane \citep{Johnsonetal2006}. Of course, the estimates of the source rates and the lifetimes needed to interpret the CAPS data depend strongly on the uncertain fate of the chemistry initiated by the interaction of the radiation products with the ring particles, a fraction of which are charged. In the following the fate of ionization and dissociation products is discussed using a rough estimate of their recycling on the ice grains  \citep{Johnsonetal2006}. Using this estimate, ~270kg/s of O$_2$ is emitted from the A, B and C rings at solstice initiated by the photolysis of ice (Fig~\ref{fig:Ring_O2_neutral}).
\par
The motion of the ionic products is especially interesting as the magnetic equator is $\sim 0.046$ \Rss  north of the ring plane and the magnetic field co-rotation speed falls below the neutral orbital speed at $R \lesssim 1.86$ \Rs. Due to this geometry, ions formed and then 'picked up' (accelerated)  by the rotating field south of the ring plane and inside of $\sim  1.86$ \Rss  precipitate into Saturn’s southern hemisphere \citep{Luhmannetal2006,Bouhrametal2006} as seen in Fig~\ref{fig:Ring_O2_ion}. This accounts for a molecular ion contribution to ring rain \citep[see more detail in Fig. 4 in ][]{Tsengetal2010}. This contribution consists of on the order of $~0.5$ to 1 percent of the oxygen neutral source rate from the ring particles, up to $\sim0.3$ kg s$^{-1}$, suggesting a minor contribution relative to the grain deposition rates of \cite{Hsuetal2018} and \cite{ODonoghueetal2019}.
\par
Pick-up ions that do not precipitate into Saturn  will oscillate along the field lines about the magnetic equator prior to impacting a ring particle. Such ions have a centrifugal scale height of O$_2^+$ varying from $\lesssim 0.1$-to-$0.16$ \Rss across the region where CAPS ring atmosphere data were acquired ($\sim1.8-2.08 R_S$). Therefore, although neutral molecules, including O$_2$, have not been directly measured, the observed population of ions oscillating about the magnetic equator with lifetimes limited by impact with ring particles indicates the presence of a neutral source that we call the ring atmosphere.
\par
This picture of a ring atmosphere dominated near the ring plane by O$_2$ is strongly supported by the ion temperatures measured by CAPS \citep{Johnsonetal2006}, which are approximated by the sum of the ions' energy at formation and their pick-up gyro energy. Newly formed O$_2^+$ and O$^+$, which are accelerated by Saturn's rotating magnetic field, are produced with very different initial speed distributions. Dissociative photoionization of O$_2$ produces an O$^+$ with a significant excess energy due to the kinetic energy associated with loss of the O atom:

\[
\mathrm{O}_2 + \mathit{hv} \rightarrow \mathrm{O} + \mathrm{O}^+ + \mathit{e}^{-}
\]

In contrast, O$_2^+$ is formed with very little excess energy because of the low mass of the electron that is lost:

\[
\mathrm{O}_2 + \mathit{hv} \rightarrow \mathrm{O}_2^+ + \mathit{e}^{-}
\]

\par
Since the pick-up contribution to the ion temperature depends on the relative speeds between the neutrals that orbit at the Keplerian orbital velocity, and ions confined by the rotating magnetic field orbiting at the magnetic field co-rotation velocity, the O$_2^+$ temperature was seen to go through a minimum at $\sim 1.86$ \Rss, where the Keplerian and co-rotation velocities are roughly equal \citep[see Fig. 1b of ][]{Johnsonetal2006}. That is, because of the low excess energy of formation of O$_2^+$, its temperature is dominated by the pick-up energy, whereas the O$^+$ temperature is dominated by its excess energy and does not exhibit as strong of a radial influence from the pick-up energy. The radial temperature trends of O$_2^+$ and O$^+$ therefore support their relationship to neutral O$_2$.
\par
Because the distance of the magnetic equator above the ring plane is of the order of the neutral scale height, those ions moving about the magnetic equator can collide with and charge exchange with neutrals, producing hot neutrals with energies determined primarily by the ion's speed. These scattered neutrals create a highly extended component of the ring atmosphere indicated by the radial column densities in Figure \ref{fig:Ring_O2_neutral}. It is seen that a fraction of the scattered O$_2$, as well as H$_2$ not shown here, precipitates into Saturn’s atmosphere as neutrals with estimated rates at three phases given in the caption. The absolute magnitude of the ring atmosphere density and, hence, these precipitation rates, depends on the still-uncertain recycling of hydrogen and oxygen on the ring particles. 
\par
Although the much lighter H$_2$ dominates O$_2$ in the extended ring atmosphere it is not the dominant source of H$_2$ or H over the ring plane. Modelling of significant contributions to hydrogen in this region indicate it also comes from the Saturn's extended atmosphere, escape from Titan's atmosphere and the Enceladus torus \citep{Tsengetal2011b,Tsengetal2013Hydrogen} each with very different scale heights. This toroidal atmosphere produces the Lyman-alpha flux in this region \citep[e.g.][]{Ben-Jaffeletal2023} as well as the hot proton plasma detected by Cassini at the inner \citep{Kollmannetal2018} and outer \citep{Cooperetal2018} edges of the main rings. However, the observed seasonal and longitudinal dependence of the Lyman-alpha flux  \citep { Ben-Jaffeletal2023} are suggestive of a ring source of  hydrogen. These data have also been used to help constrain hydrogen recycling on the ring particles. Of particular interest is the role of the incident H and H$^+$ on the particles of the inner D-ring ringlets that has yet to be modeled. 
\par
The extension of the neutral ring atmosphere beyond the edge of the A ring (R$_S$ = 2.27) seen in Figure \ref{fig:Ring_O2_neutral} contributes to the local plasma out to large distances. As Cassini did not fly over the rings again until the end of the mission, by which time the CAPS instrument was unfortunately shut off, the model of the O$_2$ ring atmosphere was further confirmed by analysis of the plasma data outside the A ring  \citep[e.g.][]{Elrodetal2014, Elrodetal2012, Christonetal2013}. That is, since the oxygen plasma density roughly tracks the local neutral density, the plasma outside the A ring exhibited a variability consistent with a steep drop in the solar flux onto the ring plane as Saturn approached equinox, as indicated by the drop in the source rate in Figure \ref{fig:Ring_O2_neutral}. Post equinox, when the magnetic equator is above the sunlit ring plane, the same processes occur. Therefore, the tilt of the ring plane to the sun leads to an extended ring atmosphere that varies with inclination to the solar illumination \citep{Tsengetal2010}.
\par
To explain the radiation processing by the magnetospheric plasma adsorbed at the outer edge of the A ring, \cite{TsengIp2011} suggested that the erosion of the icy ring particles in this region might be sustained by water from Enceladus. Refreshed ice grains in this region were also suggested to account for the observed delay in the post-equinox repopulation of O$_2^+$ plasma \citep{Christonetal2013}. Using a direct simulation Monte Carlo model and a plume outflow rate of 10$^{28}$ molecules H$_2$O s$^{-1}$, \cite{CassidyJohnson2010} find a mass deposition rate from Enceladus' neutral cloud onto the rings of approximately 100 kg s$^{-1}$, which is equivalent to a quarter of the mass of the main rings in approximately 1 Gyr if the plume rate were constant. (Note that the active lifetime of the Enceladus plume is at present poorly constrained.) However, using a similar plume rate \cite{SmithRichardson2021} find a much lower water product deposition rate. Further research into contributions from Enceladus to the rings is needed, including how such contributions would affect regional composition and albedo in the rings.
\par
Even though INMS detected other water products closer to the ring plane during SOI \citep{Waiteetal2005} when CAPS was briefly off, the absence of non-ice products was striking. This is especially the case given the material from Enceladus that is deposited onto the ring particles \citep{JuracRichardson2007,Farrelletal2008,CassidyJohnson2010}. This material is likely processed in two ways. Due to collisions of ring particles, contaminants more refractory than H$_2$O can be dispersed into the ring particles, depleting surface concentrations. In addition, any desorbed volatiles, or surface species made volatile by reactions with the water products, can be scattered into the outer magnetosphere or into Saturn's atmosphere as discussed above for the photolytically produced O$_2$.  That the latter process occurs is confirmed by the seasonal variation of ions of mass 28 as well as O$_2^+$ observed by the MIMI CHEMS (Charge-Energy-Mass Spectrometer) sensor at suprathermal energies \citep{Christonetal2014}. Further analysis of CHEMS data presented at the Cassini Science Seminar in 2018 by Hamilton and colleagues showed that C$^+$ and CO$^+$ also exhibit a seasonal variation suggestive of a photolytic ring source that is seen to peak in fall of 2014 and persist through to the Grand Finale \citep[][]{Hamiltonetal2018}. This was interpreted to be due to CO$_2$ deposited on the rings from the Enceladus torus. The observed lifetimes of CO$^+$ and CO$_2^+$ seen in the plasma appears to correlate with their photoionization rates producing CO$^+$ and CO$_2^+$ over the rings, which are scattered into the outer magnetosphere as neutrals by charge exchange with the dominant O$_2$.
\par
Although the D-ring particles are also dominated by ice so that photolysis  can contribute water products to the ring atmosphere, they contain a higher concentration of more refractory materials. Whereas the plasma flux onto the other rings is highly suppressed, \cite{Kollmannetal2018} examined the hot plasma quenching going from the outer regions of Saturn's atmosphere onto the inner edge of the D-ring. In this way they were able to estimate the extended scale height of Saturn's atmosphere as well as sense the effect of the ambient gas-phase species over the inner ringlets, which are exposed to hot neutrals and energetic ions. Surprisingly, their estimated flux, ~10$^{16}$ H$_2$O/m$^2$ is remarkably similar to that estimated by \cite{Tsengetal2010} for the contribution to the column density in this region of O$_2$ formed from the A, B and C rings.  Therefore, the D-ring contribution to its local atmosphere might indeed be small, in spite of any chemistry induced in the particles by the incident protons and hydrogen. These interactions should produce a not-yet modeled gas component originating from this region that would also contribute to Saturn's atmosphere. If this D ring atmosphere is related to the CH$_4$-rich inflowing gas measured by INMS during the Grand Finale, then its composition must be quite different than the ring atmosphere over the main rings.

\subsection{Possible ring atmosphere connection to the inflowing material}\label{bulk-volatile-loss} 

\par

We have seen that the composition of the inflowing material differs substantially from the observed composition in the main rings and in the D ring, namely in how water-poor the inflowing material is (Table \ref{tab:ring-comp-summary}). Here, we consider whether this difference in composition is intrinsic to the material, or whether the initial source composition may be identical and the observed differences could be caused by different evolutionary processes occurring in Saturn's ring environment. In the latter case, more volatile compounds may be lost via such processes, and the material observed in situ may represent the remaining residue. This volatile component may contribute to the ring atmosphere, including in the D ring region. We consider this component here. Since the D ring bulk composition is poorly constrained (Section \ref{dringcomp}), we assume that it is related to the main ring composition with a greater abundance of non-icy material.
\par
A large mass of material more volatile than water (e.g. CH$_4$, CO) was observed by INMS, and the latitudinal variation suggests it is present as a volatile component rather than as an impact fragment from more refractory material \citep{Milleretal2020, Seriganoetal2020}. However, compounds more volatile than water are not observed in the main rings (e.g. Section \ref{otherIces}), and should be lost more quickly than water ice in the absence of a lag deposit. Therefore, to calculate the missing water mass, we make the simplifying assumption here that all non-water masses detected by INMS can be attributed to impact fragments of refractory organic material related to the neutral and/or UV absorbers in the main rings (Sections \ref{neutralAbsorber} and \ref{UVAbsorber}). In this simplified scenario, there are two observed components (water and organic products), and the missing mass of water is the mass required to reconcile the relative abundance of these two components with remote observations of the D ring or main rings.
\par
The range in observed inflowing mass from INMS is bracketed by values in \cite{Seriganoetal2020} (see Table \ref{tbl:mass_influx_summary}), between 2080 and 74000 kg s$^{-1}$. The observed inflowing composition was approximately 16 to 24 wt.\% water (Table \ref{tbl:influx_composition_summary}). If we take 24 wt.\% as representative, this implies that the inflowing non-water mass is 1,581 to 56,260 kg s$^{-1}$. For the source material, we take the lower limit of 80 vol.\% water ice in the bulk D ring composition (Figure \ref{fig:dringspec}), and assume a density ratio for water/organics of 0.9/1.4 \citep{ReynardSotin2023} i.e., equal porosity for the two compositions. This corresponds to 72 wt.\% water ice in the bulk D ring. In order to reconcile the observed composition with the bulk D ring, between 3,566 and 126,857 kg s$^{-1}$ of water must be volatilized or lost in the ring environment. Estimates of the ring atmosphere production rate for the main rings (Figure \ref{fig:Ring_O2_neutral}) peak at $2\times10^{27}$ O$_2$ molecules s$^{-1}$, which corresponds to approximately 120 kg s$^{-1}$ of water ice, a value that is significantly lower than the required water mass loss ($3,566-126,857$ kg s$^{-1}$). The ring atmosphere in the D ring region has not yet been modeled, but as discussed above, estimates by \cite{Kollmannetal2018} of plasma quenching suggest that the D ring atmosphere flux may be relatively small. If true, then this estimate of 120 kg s$^{-1}$ may be an upper limit. Steady ongoing production of the water-poor inflow composition is therefore not realistic, since the water loss rate required to keep pace with the mass deposition rate ($3,566-126,857$ kg s$^{-1}$) is much greater than the expected rate of water volatilization in the D ring region.

\par
 An alternative explanation is that ring "residue" may accumulate, since the observed inflow may be a transient process (Section \ref{inflowTimescale}). From \cite{Mosesetal2023}, the time constraints on the inflow are between 4 and 600 months. This time range, together with the water mass loss rate from above ($3,566-126,857$ kg s$^{-1}$) implies a total water mass loss between $4\times10^{10}$ and $2\times10^{14}$ kg. We can generate a rough estimate for the total mass of the ring atmosphere by taking an upper limit on the column density, $1\times10^{14}$ cm$^{-2}$ (Figure \ref{fig:Ring_O2_neutral}), and the surface area of the rings, $4\times10^{20}$ cm$^{2}$ (Table \ref{surfaceDensity}). This yields on the order of $4\times10^{34}$ molecules of O$_2$, or $2\times10^{9}$ kg. For the extended ring atmosphere from 2.5 to 10 R$_S$, we calculate an additional $5\times10^{8}$ kg using an upper limit column density of $1\times10^{12}$ cm$^{-2}$ (Figure \ref{fig:Ring_O2_neutral}). We therefore estimate a total ring atmosphere mass of $\sim3\times10^{9}$ kg. The estimated water loss that would be needed to reconcile the inflowing composition with the bulk D ring composition ($4\times10^{10}$ and $2\times10^{14}$ kg) is likely orders of magnitude greater than the mass of the ring atmosphere. We would therefore expect that this magnitude of water loss may be significant, and have a noticeable effect on the ring system. 
\par
While the total required water mass loss appears large in comparison to the estimated ring atmosphere mass, it may be possible that the water loss has occurred over a sufficiently long timescale that it can be reconciled with the expected ring atmosphere production and loss rates. As calculated above, the production rate for the ring atmosphere is equivalent to on the order of 120 kg of water s$^{-1}$. To remove $4\times10^{10}$ to $2\times10^{14}$ kg of water into the ring atmosphere would then require between 10 and 50,000 years. This is rapid on geologic timescales, and further suggests that the observed inflow may be transient if the material is derived from the bulk D ring. This additionally assumes concentration of residue material at the inner edge of the D ring. The overall contribution of the D ring region to the ring atmosphere has not yet been modeled but is expected to be small in comparison to the main rings. Loss rates for the ring atmosphere into Saturn's atmosphere are $\lt30$ kg s$^{-1}$ (Figure \ref{fig:Ring_O2_neutral}), and transfer of water vapor products outward to the magnetosphere may be required instead if the INMS material does represent D ring residue. While we have assumed here that all inflowing material may be represented by a combination of water and complex organic material, this simplified scenario does not explain the presence of CH$_4$ and CO or N$_2$ in the vapor phase (Section \ref{eqCompositon}).
\par
A related scenario may involve water loss from a subset of the D ring material, and selective deorbiting of that residue into Saturn's atmosphere. The "tholin" material used for spectral fitting in \cite{Cuzzietal2018} produces the best fits when it is concentrated into a subset of grains that comprises 2-40\% of the overall population of grains. We suggest that these grains may be volatilized more quickly due to increased absorption of sunlight by the darker organics, and subsequently higher temperatures for organic-bearing grains. Water loss may reduce the size of these organic-rich grains, which in turn may increase the likelihood that they fall into Saturn's atmosphere since small grains may be decelerated more rapidly by impacts with Saturn's extended atmosphere \citep{Mitchelletal2018}. The INMS material is hypothesized to be very small in size \citep[e.g.][]{Perryetal2018}, and if the inflowing material selectively draws from the organic-rich material in the D ring, such a scenario may reduce the water loss mass required to reconcile INMS observations with the bulk D ring composition.

\section{Relationship between the rings and ring moons}\label{ringMoon} 
While inwardly connected with Saturn through the observed mass inflow discussed above, from a compositional standpoint the main rings also appear to be outwardly linked with Saturn's ring moons, and in particular with the innermost ones. These include small satellites orbiting within (Pan, Daphnis) and just outside the A ring (Atlas), shepherding the F ring (Prometheus and Pandora), and the co-orbitals Janus and Epimetheus (orbiting at $\sim$151,000 km).

As discussed in greater detail in \textcolor{orange}{Ciarniello et al., Chapter 11 this volume}, Cassini remote sensing observations at UV-VIS-IR wavelengths show significant spectral similarities between the main rings and these ring moons. Both are characterized by the dominance of water ice features with absorption bands at 1.5, 2 and 3 $\mu$m, and by UV absorption at wavelengths shorter than $\sim$ 0.55 $\mu$m, indicative of the presence of non-icy material. Although such spectral features can also be observed on the rest of Saturn's icy moons, spectral indicator analysis \textcolor{orange}{(see Section 6 in Ciarniello et al., Chapter 11 this volume)} reveals similar trends between the main rings and inner ring moons, for which the intensity of the UV absorption is in general positively correlated with the albedo and the water ice band depths. This is not the case for the mid-sized icy satellites, thus suggesting a similar compositional paradigm for the rings and the inner ring moons in which the UV absorber is embedded in water ice grains, with an additional endmember as a broad-band absorber (see Section \ref{RRTModels}).

Further evidence of a compositional link between the main rings and ring moons is provided by the radial trend of the ring moon spectral shapes, which become progressively similar to that of the A ring moving inward from Janus and Epimetheus, which appear the most polluted, to Pan. In fact, this radial trend may result from the accumulation of nearby A ring particles on the ring moon surfaces. This accumulation is more effective for the inner moons, and counteracts the contamination from meteoritic bombardment that modifies the surface composition inherited from the rings. An additional contribution may also come from a seasonal flow of neutral species from the ring atmosphere that are scattered  beyond the edge of the rings by molecules and ions generated from the rings by photoionization and subsequently accelerated by the magnetic field \citep{Tsengetal2010}.
\par
Ring moons are thought to originate within or in the vicinity of the A ring, accumulating ring material on their external layers in the process, and then moving outward due to the gravitational torque of the rings \textcolor{orange}{(see Section 5 of Ciarniello et al., Chapter 11 this volume and references therein)}. As a consequence, Janus and Epimetheus, orbiting at larger distance from Saturn with respect to the other ring moons, are likely older and may have endured a longer exposure to meteoritic bombardment. In this respect, it is interesting to note that these two moons appear spectrally similar to the C ring, the most contaminated of the main rings, thus pointing to a similar role for the pollution mechanism. However, we note that contrary to the main rings, ring moons are affected by additional exogenous processes such as contamination from E-ring grains and the effects of charged particle fluxes \textcolor{orange}{(see Section 8 of Ciarniello et al., Chapter 11 this volume)}, thus requiring particular care in comparisons with contamination/alteration processes in the main rings.

\section{Conclusions}\label{conclusions} 

\par
From this retrospective look at the Cassini dataset and at previous remote observations of the Saturnian ring system, a few themes for the ring composition emerge. Overall, the bulk composition of Saturn's rings is overwhelmingly dominated by water ice. This is especially true in the optically thick B ring, which dominates the ring mass. However, remote sensing constraints suggest that even the 'water-poor' D ring may be as much as 80 vol.\% water. This composition is consistent with fairly minimal signs of pollution or contamination of the rings by external impactors. While microwave observations suggest that some silicate material may remain 'hidden' from surficial spectroscopy in the C ring, in general the dynamic evolution of the rings suggests they are well-mixed and this composition is representative of the bulk. Thus, the effective exposure age of the rings to micrometeoroid flux may be relatively young, on the order of 10$^8$ years.
\par
While the composition is dominated by water throughout, compositional trends are observed with both radial distance (Table \ref{tab:ring-comp-summary}) and optical depth. The abundance of non-water components tends to increase going from the A ring towards the D ring. This trend mirrors the inward increase in temperature seen by VIMS and CIRS data on the sunlit side. Similarly, the abundance of small particles increases from the inner A ring to the D ring. The Cassini Division deviates from these trends, and instead points to the influence of optical depth on the ring composition. Namely, optical depths are lower in the C ring and Cassini Division, and as a result these regions are more susceptible to pollution from external impactors. Microwave data show that the composition of the middle C ring may also be silicate-rich, with icy mantles concealing interiors that may point towards the recent break up of a small rocky body.
\par
Data collected during Cassini's Grand Finale of material from the rings support this trend of enhanced abundance of non-water components moving inwards towards Saturn. However, the degree of enhancement from INMS measurements on the smallest size fraction is extreme. This material is comprised of small grains as well as gas that includes a 28 u component (likely dominated by CO with some C$_2$H$_4$ and possibly some N$_2$) and methane. Such highly volatile material has not previously been detected in the main rings from remote sensing observations. While inflow of charged 'ring rain' material to the mid-latitudes is likely a steady process, constraints from Saturn's atmospheric chemistry suggest that the more massive equatorial inflow detected during the Grand Finale may represent a recent event. This is consistent with the appearance of bright clumps in the D68 ringlet in 2014, just a few years before the measurements.
\par
The composition of this material may be partially reconciled with the bulk D ring composition via photolysis and sublimation processes, which are known to generate the ring atmosphere. While these processes in the D ring region have not been specifically modeled, estimates for the local gas density from plasma flux onto the rings suggest that D ring contributions to the ring atmosphere may be relatively small. If the equatorial inflow does represent evolved material from the main rings, the timescale for devolatilization must be much longer than the inflow timescale ($~10^5\times$) and the ring 'residue' must accumulate and be effectively fractionated by the inflow process. The simplified scenario presented here may broadly allow reconciliation of the inflow composition with the bulk composition, but it does not explain the presence of the vapor component in the INMS inflow. This processing of the ring material may be related to the ring 'cleaning' suggested by \cite{Cridaetal2019}. 
\par
Ballistic transport indicates mass inflow of material from across the rings with total inward fluxes consistent with the equatorial inflow abundances. However, additional work is needed to explore the relationship of these processes to the compositionally- and size-fractionated material in the D ring region. Similarly, future work should examine how the Grand Finale inflow fits within the broader context of ring evolution, and the degree to which these processes are stochastic or smooth in time.

\backmatter

\section*{Declarations}

\subsection*{Availability of data and materials}
Cassini data used in this work are archived in NASA’s Planetary Data System (PDS) and publicly available.

\subsection*{Competing interests} The authors have no competing interests to declare.
\subsection*{Funding} Nothing to declare.

\section*{Author Contributions}
The idea for this review article came from GF and JC. All authors contributed to the literature search. New data analysis was performed by KB, CF, MMH, JC, KEM and WLT. All authors contributed to drafting and revisions.

\section*{Acknowledgements}
This research was supported by the International Space Science Institute (ISSI) in Bern, through ISSI Workshop "New Vision of the Saturnian System in the Context of a Highly Dissipative Saturn" (Bern, 9-13 May 2022). This research has made use of NASA’s Astrophysics Data System. KEM acknowledges support from SwRI. GF and MC acknowledge support from INAF-IAPS.

\bibliography{references_all}

\end{document}